\documentclass[fleqn,usenatbib]{mnras}

\usepackage[T1]{fontenc}
\usepackage{ae,aecompl,ulem}
\usepackage{lscape} 
\usepackage{graphicx}	
\usepackage{amsmath}	
\usepackage{soul}
\usepackage{subfig}
\usepackage{longtable}
\usepackage{supertabular}
\usepackage{array}
\usepackage{mathabx}
\usepackage[utf8]{inputenc}
\usepackage{fixltx2e}
\usepackage{tablefootnote}
\usepackage{threeparttable}
\usepackage{ulem}
\usepackage{xcolor}

\title[TTV Analyses of 16 Systems]{Looking For Timing Variations in the Transits of 16 Exoplanets}

\author[Yal\c{c}{\i}nkaya et al.]{
S. Yalçınkaya$^{1,2}$\thanks{E-mail: yalcinkayas@ankara.edu.tr}, E. M. Esmer$^{1,2}$, \"O. Ba\c{s}t\"urk$^{1,2}$, A. Muhaymin$^{3}$, A. C. Kutluay$^{4}$, 
\newauthor
D. \.{I}. Silistre$^{3}$, F. Akar$^{1}$, J. Southworth$^{5}$, L. Mancini$^{6,7,8}$, F. Davoudi$^{9}$,  E. Karamanl{\i}$^{3}$, 
\newauthor
F. Tezcan$^{10}$, E. Demir$^{1}$, D. Y{\i}lmaz$^{3}$, E. G\"ulero\u{g}lu$^{4}$, M. Tekin$^{4}$, \.{I}. Ta\c{s}k{\i}n$^{3}$, Y. Alada\u{g}$^{11}$, 
\newauthor
E. Sertkan$^{1}$, U. Y. Kurt$^{1}$, S. Fi\c{s}ek$^{12}$, S. Kaptan$^{13}$, S. Ali\c{s}$^{12}$, N. Aksaker$^{11,14}$,
\newauthor
F. K. Yelkenci$^{12}$, C. T. Tezcan$^{10,15}$, A. Kaya$^{3}$, D. O\u{g}lakkaya$^{11}$, Z. S. Ayd{\i}n$^{3}$,
\newauthor
 C. Ye\c{s}ilyaprak$^{10,15}$
\\
\\
$^{1}$ Department of Astronomy \& Space Sciences, Faculty of Science, Ankara University, TR-06100, Ankara, Türkiye \\
$^{2}$  Ankara University, Astronomy and Space Sciences Research and Application Center (Kreiken Observatory), İncek Blvd., \\
TR-06837, Ahlatlıbel, Ankara, Türkiye \\
$^{3}$ Bilkent University, Faculty of Science, Department of Physics, TR-06800 Ankara, Türkiye \\
$^{4}$ Middle East Technical University, Science Faculty, Physics Department, Ankara, Türkiye \\
$^{5}$ Astrophysics Group, Keele University, Staffordshire ST5 5BG, UK \\
$^{6}$ Department of Physics, University of Rome “Tor Vergata”, Via della Ricerca Scientifica 1, I-00133, Roma, Italy\\
$^{7}$ INAF – Turin Astrophysical Observatory, Via Osservatorio 20, I-10025, Pino Torinese, Italy\\
$^{8}$ Max Planck Institute for Astronomy, Königstuhl 17, D-69117, Heidelberg, Germany\\
$^{9}$ Astrobiology Research Unit, Universit\`e de Li\`ege, All\`ee du 6 A\^out 19C, B-4000 Li\`ege, Belgium\\
$^{10}$ Atatürk University, Science Faculty, Department of Astronomy and Space Sciences, 25240, Erzurum, Türkiye  \\
$^{11}$ Space Science and Solar Energy Research and Application Center (UZAYMER), University of Çukurova, 01330, Adana, Türkiye \\
$^{12}$ Department of Astronomy and Space Sciences, Faculty of Science, Istanbul University, 34119 Istanbul, Türkiye \\
$^{13}$ Astronomy and Space Sciences Program, Institute of Graduate Studies in Sciences, Istanbul University, 34116 Istanbul, Turkey \\
$^{14}$ Adana Organised Industrial Zones Vocational School of Technical Science, University of Çukurova, 01410, Adana, Türkiye \\
$^{15}$ Atatürk University Astrophysics Research and Application Center (ATASAM), Yakutiye, 25240, Erzurum, Türkiye \\ 
}

\date{Accepted XXX. Received YYY; in original form ZZZ}

\pubyear{2023}

\begin{document}
\label{firstpage}
\pagerange{\pageref{firstpage}--\pageref{lastpage}}
\maketitle

\begin{abstract}
We update the ephemerides of 16 transiting exoplanets using our ground-based observations, new TESS data, and previously published observations including those of amateur astronomers. All these light curves were modeled by making use of a set of quantitative criteria with the {\sc exofast} code to obtain mid-transit times. We searched for statistically significant secular and/or periodic trends in the mid-transit times. We found that the timing data are well modeled by a linear ephemeris for all systems except for XO-2\,b, for which we detect an orbital decay with the rate of -12.95 $\pm$ 1.85 ms/yr that can be confirmed with future observations. We also detect a hint of potential periodic variations in the TTV data of HAT-P-13\,b which also requires confirmation with further precise observations.   
\end{abstract}




\section{Introduction}
\label{sec:introduction}
Since the observations of the first transits in an exoplanet system \citep{Charbonneau2000}, several questions have arisen regarding their formation, evolution, atmospheric composition, and orbital dynamics. These questions can be further investigated through different observational techniques. For example, radial velocity measurements during transits can be employed to determine the obliquity of a planet's orbit, which in turn can provide important information for improving theoretical models related to orbital evolution \citep{mancini2022}. Occultation observations can provide information about the planet's energy budget \citep{arcangeli2021}, while transmission spectroscopy can reveal its atmospheric composition \citep{maguire2022}. However, these observations require high levels of precision, which can only be achieved by making use of large ground-based or space-borne telescopes. As the observation time for these instruments is in great demand, accurate predictions of transit and occultation times are crucial. Even small uncertainties in transit times can accumulate over time and require updates to the exoplanet's orbital period and reference mid-transit times \citep{mallonn2019}.

Tidal interactions can cause the orbit of the planet to shrink \citep{maci2016}. The period decrease per year may be much smaller than the uncertainty of the mid-transit times, making it difficult to observe. The amplitude of this effect increases over time and may be detected with additional transit (or occultation) observations over a long time range. In addition, transit timing analysis can be used to detect unseen additional bodies in a system that could not be seen with radial velocity (RV) observations due to short phase coverage, stellar activity \citep{2021AJ....162..283T} or if the host star is too faint for precise RV observations (e.g. \citealt{2017Natur.542..456G}). For eccentric systems, the secular motion of the periastron (i.e. apsidal motion, \citealt{1995Ap&SS.226...99G}) is observable with the help of occultation observations \citep{2017AJ....154....4P} and could give insights about tidal effects. To identify these effects using the transit timing variation (TTV) technique, it is essential to have transit timing measurements that cover longer time spans and are well sampled. 

We selected potential periodic TTV targets depending on the known third bodies in the system or depending on their radial velocity residuals.

In the potential TTV group, there are also orbital decay candidates selected based on their stellar and planetary radii, masses and orbital separation and ages to work on systems with maximum tidal interaction potential. There are unitless metrics that are used to select our candidates. Please see \cite{basturk2022}. We observed 38 transit of 16 exoplanets (GJ\,1214\,b, HAT-P-1\,b, HAT-P-10\,b, HAT-P-13\,b, HAT-P-16\,b, HAT-P-22\,b, HAT-P-30\,b, HAT-P-53\,b, KELT-3\,b, QATAR-2\,b, WASP-8\,b, WASP-44\,b, WASP-50\,b, WASP-77A\,b, WASP-93\,b, XO-2\,b) that we selected for their potential to display TTVs and/or large shifts in their observed transit timings.

The transit data that we used for timing calculation were obtained from ground-based telescopes and the Transiting Exoplanet Survey Satellite (TESS) \citep{ricker2015}, and compiled from published observations and open databases\footnote{http://var2.astro.cz/ETD/index.php}$^,$\footnote{http://brucegary.net/AXA/x.htm}. We performed homogeneous transit timing analyses of these systems and updated their ephemeris information. 

This paper is organized as follows. In section \ref{sec:observations_datared}, we describe the telescopes and the detectors we used for transit observations, data reduction, and photometry procedure as well as light curve selection criteria. TTV analyses and our results are presented in section \ref{sec:analysis}. We discuss our findings in section \ref{discussion}. 

\begin{table*}
    \scriptsize
\centering
	\caption{Fundamental stellar and planetary properties and the number of light curves analyzed for each planetary system in our sample.}
	\label{tab:planetdata}
        \begin{threeparttable}
	\begin{tabular}{lcccccccccc} 
		\hline
		Planet & $P_{\rm orb}$ & $M_{\rm p}$ / $M_{\star}$ &  $T_{\rm eff,\star}$ & Database & Literature & Our & TESS & Kepler & Total \\
                Name & [days] & $\times 10^{-3}$ & [K] & LC Number & LC Number & LC Number & LC Number & LC Number &\\ 
		\hline
		        GJ 1214\,b$^{1}$ & 1.5803925(117) & 0.157 $\pm$ 0.019 & 3026 $\pm$ 130 & 6 & 37 & 5 & - & - & 48\\
                HAT-P-1\,b$^{2}$ & 4.46529(9) & 1.12 $\pm$ 0.09 & 5975 $\pm$ 45 & 2 & 7 & 5 & 12 & - & 26 \\
                HAT-P-10\,b$^{3}$ & 3.7224747(65) & 0.83 $\pm$ 0.03 & 4980 $\pm$ 60  & 16 & 7 & 2 & 4 & - & 29 \\
                HAT-P-13\,b$^{4}$ & 2.916260(10) & {1.22} $\pm$ 0.10 & 5653 $\pm$ 90 & 5 & 19 & 3 & 8 & - & 35 \\
                HAT-P-16\,b$^{5}$ & 2.77596(3) & 3.29 $\pm$ 0.13 & 6158 $\pm$ 80  & 35 & 11 & 2 & 14 & - & 62\\
                HAT-P-22\,b$^{6}$ & 3.212220(9) & 0.916 $\pm$ 0.035 & 5302 $\pm$ 80 & 5 & 4 & 2 & 14 & - & 25\\
                HAT-P-30\,b$^{7}$ & 2.810595(5) & 1.242 $\pm$ 0.041 & 6304 $\pm$ 88  & 19 & 5 & - & 24 & - & 48\\
                HAT-P-53\,b$^{8}$ & 1.9616241(39) & 1.093 $\pm$ 0.043 & 5956 $\pm$ 50 & 7 & 3 & 2 & 1 & - & 13\\
                KELT-3\,b$^{9}$ & 2.7033904(100) & 1.278 $\pm$ 0.063 & 6306 $\pm$ 50 & 7 & 1 & 1 & 17 & - & 26\\
                QATAR 2\,b$^{10}$ & 1.3371182(37) & 0.740 $\pm$ 0.037 & 4645 $\pm$ 50 &  8 & 19 & 1 & - & 56 & 84\\
                WASP-8\,b$^{11}$ & 8.158715(16) & 1.030 $\pm$ 0.061 & 5600 $\pm$ 80 &  1 & 2 & - & 6  & - & 7\\
                WASP-44\,b$^{12}$ & 2.4238039(87) & 0.951 $\pm$ 0.034 & 5410 $\pm$ 150 &  13 & 11 & 1 & 8  & - & 33\\
                WASP-50\,b$^{13}$ & 1.9550959(51) & 0.892 $\pm$ 0.080 & 5400 $\pm$ 100 &  19 & 13 & 2 & 20 & - & 54\\
                WASP-77\,b$^{14}$ & 1.3600309(20) & 1.002 $\pm$ 0.045 & 5500 $\pm$ 80 & 13 & 6 & 2 & 32 & - & 53\\
                WASP-93\,b$^{15}$ & 2.7325321(20) & 0.65 $\pm$ 0.06 & 6700 $\pm$ 100 & 5 & 4 & 7 & 6 & - & 22\\
                XO-2\,b$^{16}$ & 2.615857(5) & 0.55 $\pm$ 0.07 & 5340 $\pm$ 32 & 13 & 9 & 3 & 18 & - & 42\\
                \hline
                Total & & & & 174 & 158 & 38 & 184 & 56 & 607 \\ 
		\hline
	\end{tabular}
        \begin{tablenotes}
          \footnotesize{
        \item[1] \citet{2009Natur.462..891C}, $^{2}$\citet{bakos2007},$^{3}$\citet{bakos2009a}, $^{4}$\citet{bakos2009}, $^{5}$\citet{bucchave2010},  $^{6}$\citet{2011ApJ...742..116B}, $^{7}$\citet{2011ApJ...735...24J}, $^{8}$\citet{bonomo2017}, $^{9}$\citet{2013ApJ...773...64P}, $^{10}$\citet{2012ApJ...750...84B},$^{11}$\citet{2010A&A...517L...1Q}, $^{12}$\citet{2012MNRAS.422.1988A}, $^{13}$\citet{2011A&A...533A..88G}, $^{14}$\citet{2013PASP..125...48M}, $^{15}$\citet{hay2016}, $^{16}$\citet{burke2007}, 
          }
        \end{tablenotes}
        \end{threeparttable}
\end{table*}
                
\section{Observations and Data Reduction}
\label{sec:observations_datared} 
\subsection{Observations}
\label{subsec:observations}
Photometric transit observations were carried out with the T100, T80, ATA50, UT50 and (numbers in the name of telescopes come from primary mirror diameters in cm)  CAHA 1.23\,m telescopes. Detailed information about the telescopes and their detectors can be found in \cite{basturk2022}. We also observed a multi-color transit of HAT-P-1 b with the Bonn University Simultaneous Camera (BUSCA) on the CAHA 2.2\,m telescope at the Observatory of Calar Alto (Spain). We made use of the well-established defocusing technique \citep{southworth2009} in order to increase photometric precision. Exposure times were set to acquire at least $\sim 50$ frames per transit. The defocusing amount was determined to keep the detector response within its linearity limits while exposing it for larger durations to increase the Signal-to-Noise ratio (SNR) by reading out from a larger area. In general, we selected the photometric filter that gives the maximum SNR. A detailed log of photometric observations is provided in Table \ref{tab:observations}. 

\subsection{Data Reduction}
\label{subsec:data_reduction}
Data reduction (dark, bias and flat correction) and ensemble aperture photometry were performed using the AstroImageJ (hereafter AIJ) \citep{collins2017} software package. To increase the precision in photometry, we selected every star similar in brightness to the target in the field as a comparison, AIJ allows the user to visually inspect the relative flux of the target for a combination of different comparison stars. After finding suitable comparison stars, we experimented with different aperture sizes for both the stars and the sky background, AIJ also allows users to visually inspect relative flux change due to different aperture sizes. When selecting comparison stars and aperture sizes, our goal was to minimize the red noise, especially in contact times where the flux change is abrupt. Red noise during ingress and egress can change the mid-transit times dramatically, but may not affect the error bar of individual data points which results in an underestimation of the mid-transit time uncertainty \citep{pont2006, gillon2006}. This could lead to a higher reduced chi-square ($\chi_\nu^2$) for linear ephemeris, which could (incorrectly) be attributed to TTV. In order to avoid that, we detrended relative fluxes by using time-dependent variables such as airmass and target position on the CCD in an interactive manner using AIJ.

For TESS observations, we downloaded the two-minute light curves from Mikulski Archive for Space Telescopes\footnote{https://mast.stsci.edu/portal/Mashup/Clients/Mast/Portal.html} (MAST) that are processed by Science Processing Operations Center (SPOC) pipeline \citep{jenkins2016}. SPOC generates presearch data conditioning (PDC) light curves and data validation time series (DVT) light curves using simple aperture photometry (SAP). The PDC\_SAP fluxes are the corrected version of the SAP fluxes from instrumental systematics, outliers and flux contamination from nearby stars. The DVT light curves are created by applying a running median filter to the PDC light curves to remove any long-term systematics and search for transits. We used only the DVT light curves because any signal other than transits will deteriorate the transit profiles, which in turn will increase the uncertainty in the measurement of the mid-transit times. For the case of XO-6 b, \cite{2020AJ....160..249R} has shown that the DVT light curves have least scatter, nevertheless, the transit timings from DVT and PDC light curves are practicaly identical.

 We have TESS light curves from the SPOC pipeline for all the planets in our sample except for HAT-P-53, which was observed by TESS during Sector 17, but light curves were not produced. Therefore we downloaded the Full Frame Images (FFI) that has 30 minutes cadence from TESScut\footnote{https://mast.stsci.edu/tesscut/} and performed aperture photometry with the {\sc lightkurve} package \citep{lightkurve2018} and then detrended the light curve using {\sc keplerspline-v2}\footnote{https://github.com/avanderburg/keplersplinev2} while ignoring the transit profiles. Final light curves were not suitable for individual modeling due to insufficient sampling so we time-folded the data using a period from our preliminary analysis. We assumed the period of HAT-P-53 to be constant during Sector 17 but this enabled us to measure only a single mid-transit time from TESS observations. We included every TESS light curves until the end of the extended mission 2 (Sector 69) to our analysis. 

\subsection{Light Curve Selection Criteria}
\label{subsec:criteria}
The main goal of this work is to search for TTVs in the planetary systems listed in Table \ref{tab:planetdata}. This requires precise and accurate mid-transit times measured from high-quality light curves. For this reason, we used the light curve selection criteria given in \cite{basturk2022} to select suitable light curves. First, we compiled available transit light curves from literature, open databases of amateur astronomers (Exoplanet Transit Database\footnote{http://var2.astro.cz/ETD/index.php}, hereafter ETD and Amateur Exoplanet Archive\footnote{http://brucegary.net/AXA/x.htm}, hereafter AXA) along with our own observations and observations from space telescopes (TESS and Kepler Space Telescope's K2 mission \citealt{keplerk2}). We did not include light curves that have large gaps inside transit profiles or high-amplitude signatures of correlated noise, especially in the ingress or egress segments. Then we modeled light curves with the {\sc exofast} \citep{2013PASP..125...83E} (see Section \ref{sec:analysis} for details) and then calculated photometric noise rate (PNR) \citep{fulton2011} from residuals which indicates white noise. We removed the light curves that have PNR values higher than the transit depth. We binned the residuals between the ingress/egress duration $\pm$5 minutes with 1-minute steps and calculated the well-known $\beta$ values as defined in \cite{2008ApJ...683.1076W} as a red noise indicator. We removed the light curves with the median $\beta$-value larger than 2.5. We also removed the light curves if the transit depth is a 5$\sigma$ outlier for the given planet. When we visually inspect the removed light curves, we find that this criterion is very useful to detect problematic light curves. Qatar 2\,b is an exception because it has K2 light curves with incomparably higher precision than other datasets, affecting the $\sigma$-value dramatically. Thus we did not include depth values from K2 for Qatar 2 in the calculations of its 5$\sigma$ level. 

\begin{table*}
  \scriptsize
\centering
	\caption{The log of photometric observations performed for this study. The dates of the light curves that are eliminated and hence not used in the TTV diagrams are marked and the reasons for their elimination are given in the footnotes.}
	\label{tab:observations}
        \begin{threeparttable}
	\begin{tabular}{cccccccccccc} 
		\hline
		System & Telescope & Date & Start  & End & \textit{Filter} & Exp. Time & Images & PNR & $\beta$ & Mid-Transit & Error \\
                Name & & UTC & UTC & UTC & & [s] & Number &  &  & BJD-TDB & [days] \\ 
		\hline
                GJ1214 & T100 & 2020-06-11 & 19:38:36 & 21:45:17 & R & 55 & 112 & 3.52 & 1.12 & 2459012.360728 & 0.000376 \\
                GJ1214 & T100 & 2020-07-03 & 23:07:27 & 00:58:24 & I & 70 & 81 & 3.00 & 2.08 & 2459034.486291 & 0.000454 \\
                GJ1214 & T100 & 2021-04-23 & 21:59:36 & 00:06:47 & I & 55 & 114 & 2.34 & 1.97 & 2459328.441673 & 0.000261 \\
                GJ1214 & T100 & 2021-07-19 $^2$ & 19:06:03 & 22:04:02 & I & 60 & 142 & 2.60 & 2.00 & 2459415.362554 & 0.000460 \\
                GJ1214 & T80 & 2022-05-12 & 22:42:10 & 00:55:10 & $i^{\prime}$ & 50 & 126 & 2.14 & 1.01 & 2455701.413994 & 0.000224 \\
                HAT-P-1 & ATA50 & 2022-09-15 & 17:42:12 & 01:46:02 & $z^{\prime}$ & 100 & 248 & 1.71 & 1.19 & 2456308.238083 & 0.000464 \\
                HAT-P-1 & CAHA (2.2 m) & 2013-09-02 & 22:46:24 & 3:54:39 & u & 40 & 383 & 2.80 & 0.67 & 2456538.550339 & 0.000526 \\
                HAT-P-1 & CAHA (2.2 m) & 2013-09-02 & 22:46:24 & 3:54:39 & b & 40 & 386 & 1.73 & 0.97 & 2456538.549978 & 0.000314 \\
                HAT-P-1 & CAHA (2.2 m) & 2013-09-02 & 22:46:24 & 3:54:39 & y & 40 & 385 & 1.80 & 0.40 & 2456538.548745 & 0.000327 \\
                HAT-P-1 & CAHA (2.2 m) & 2013-09-02 & 22:46:24 & 3:54:39 & z' & 40 & 381 & 1.30 & 0.76 & 2456538.549052 & 0.000227 \\ 
                HAT-P-10 & T100 & 2013-01-15 & 16:31:21 & 20:12:03 & R & 125 & 84 & 0.59 & 1.35 & 2456308.238083 & 0.000203 \\
                HAT-P-10 & T100 & 2020-10-25 $^4$ &  21:23:46 & 02:21:38 & R & 225 & 71 & 0.76 & 2.60 & 2459148.489796 & 0.000421 \\    
                HAT-P-13 & CAHA (1.23 m) & 2014-01-12 & 0:47:47 & 5:45:06 & R & 130 & 137 & 0.42 & 1.14 & 2456669.654639 & 0.000429 \\
                HAT-P-13 & CAHA (1.23 m)& 2015-03-10 & 20:2:10 & 3:8:21 & R & 245 & 110 & 0.65 & 1.00 & 2457092.509213 & 0.000445 \\
                HAT-P-13 & T100 & 2021-01-03 & 18:34:21 & 01:37:17 & R & 245 & 100 & 0.38 & 1.49 & 2459218.452754 & 0.000670 \\
                HAT-P-16 & UT50 & 2020-10-21 $^1$ & 17:57:12 & 23:07:04 & R & 60 & 268 & 3.80 & 1.40 & 2459144.346886 & 0.000759 \\
                HAT-P-16 & ATA50 & 2020-10-07 & 21:21:53 & 02:17:13 & R & 185 & 93 & 0.76 & 2.02 & 2459130.474612 & 0.000521 \\
                HAT-P-22 & T100 & 2021-02-14 & 18:09:54 & 21:43:00 & R & 190 & 63 & 0.47 & 1.36 & 2459260.310347 & 0.000313 \\
                HAT-P-22 & T100 & 2014-02-17 & 22:50:43 & 03:29:44 & R & 185 & 80 & 0.85 & 1.57 & 2456706.584303 & 0.000661 \\
                HAT-P-53 & ATA50 & 2020-11-08 $^2$ & 15:28:21 & 19:38:48 & R & 170 & 73 & 5.20 & 1.33 & 2459162.237801 & 0.002024 \\
                KELT-3 & T100 & 2014-02-18 & 17:43:44 & 03:09:48 & R & 195 & 153 & 0.49 & 2.23 & 2456707.439231 & 0.000560 \\
                QATAR 2 & T100 & 2019-02-17 & 23:26:03 & 03:32:29 & R & 95 & 131 & 1.60 & 0.97 & 2458897.528172 & 0.000234 \\
                WASP-44 & T100 & 2020-08-27 & 21:47:53 & 01:22:40 & R & 125 & 87 & 1.52 & 1.05 & 2459089.484641 & 0.000706 \\
                WASP-50 & T100 & 2019-10-29 & 22:24:53 & 00:41:01 & R & 95 & 74 & 1.59 & 0.49 & 2458786.470867 & 0.000405 \\
                WASP-50 & T100 & 2020-10-09 & 23:03:08 & 01:53:44 & R & 105 & 79 & 0.96 & 0.65 & 2459132.522846 & 0.000285 \\
                WASP-77 & ATA50 & 2020-10-26 & 22:02:55 & 00:56:44 & R & 120 & 75 & 1.42 & 1.07 & 2459149.480906 & 0.000526 \\
                WASP-77 & ATA50 & 2021-10-16 & 21:08:02 & 01:05:21 & R & 115 & 106 & 1.13 & 1.21 & 2459504.447789 & 0.000380 \\                
                WASP-93 & ATA50 & 2021-09-26 & 16:45:56 & 20:56:25 & R & 160 & 93 & 0.67 & 1.50 & 2459484.305639 & 0.000726 \\
                WASP-93 & UT50 & 2020-12-24 & 17:44:38 & 22:07:22 & R & 140 & 112 & 2.01 & 0.73 & 2459208.322251 & 0.001692 \\
                WASP-93 & T100 & 2020-01-12 & 17:01:44 & 20:55:19 & R & 145 & 88 & 0.78 & 1.61 & 2458861.287821 & 0.000651 \\
                WASP-93 & T100 & 2020-09-22 $^4$ & 20:35:59 & 01:50:32 & R & 200 & 89 & 0.58 & 2.70 & 2459115.413786 & 0.000501 \\
                WASP-93 & T100 & 2019-10-30 & 22:05:33 & 02:48:11 & R & 175 & 91 & 0.41 & 1.72 & 2458787.509779 & 0.000421 \\
                WASP-93 & T100 & 2020-11-24 & 16:35:48 & 20:37:04 & R & 105 & 119 & 1.18 & 1.32 & 2459178.263534 & 0.000768 \\
                WASP-93 & T100 & 2021-07-25 $^2$ & 21:06:17 & 01:28:45 & R & 140 & 101 & 0.96 & 1.61 & 2459421.461017 & 0.000772 \\
                XO-2 & T100 & 2020-01-12 & 22:18:58 & 03:16:03 & R & 175 & 94 & 1.36 & 1.89 & 2458861.528988 & 0.000794 \\
                XO-2 & T100 & 2020-12-25 & 20:07:25 & 01:06:58 & R & 220 & 78 & 0.61 & 2.47 & 2459209.437425 & 0.000405 \\
                XO-2 & UT50 & 2020-12-25 & 20:27:58 & 00:16:06 & R & 120 & 108 & 2.45 & 1.35 & 2459209.437700 & 0.0009460 \\
                
		\hline
        \end{tabular}
        \begin{tablenotes}
          \scriptsize{
          \item[1] Eliminated because it is an outlier on the TTV-diagram.
          \item[2] Eliminated because its depth is out of 5$\sigma$ of the average.
          \item[3] Eliminated because its PNR value is larger than its depth.
          \item[4] Eliminated because its $\beta$-factor is larger than 2.5.
          }
        \end{tablenotes}
        \end{threeparttable}
\end{table*}

\section{Data Analysis and Results}
\label{sec:analysis} 

\subsection{Light Curve Modelling and Measurements of Mid-Transit Times}
\label{subsec:measurements}
We followed the same method given in \cite{basturk2022} to model the light curves and measure the mid-transit times. Briefly, we used {\sc exofast-v1} to model the light curves after converting the observation time to Dynamical Barycentric Julian days (BJD-TDB) and detrending the light curves for the airmass effect that need it with the {\sc AIJ}. We used our scripts to convert the timings to BJD-TDB and calculated the airmass values by using relevant modules and functions of the {\sc astropy} \citep{2013A&A...558A..33A, 2018AJ....156..123A} library. The centers and widths of the priors were automatically selected from the NASA Exoplanet Archive\footnote{https://exoplanetarchive.ipac.caltech.edu/} for the atmospheric parameters of the host stars as a Gaussian priors and the orbital periods of the planets as constant values while uniform priors of the limb darkening coefficients were automatically retrieved from \cite{2011A&A...529A..75C} based on the atmospheric parameters of the host stars and the observed passbands. For the passbands that are not available, we choose the passband that has the closest transmission curve (e.g. we choose \textit{I} for the TESS passband and CoRoT for the clear observations). 

After selecting the light curves as described in \ref{subsec:criteria} and measuring the mid-transit times from the individual transit models using a built-in IDL routine AMOEBA that uses downhill simplex method \citep{10.1093/comjnl/7.4.308} to minimise $\chi{^2}$, we constructed the TTV diagrams and fitted a linear ephemeris using {\sc emcee} \citep{2013PASP..125..306F} following the recipe given in \cite{basturk2022}. We discarded the two-tailed 3$\sigma$ outliers from the linear ephemeris not to bias our final results. These light curves with correlated noise, especially during ingress or egress, may survive the $\beta$ > 2.5 criteria and result in an inaccurate mid-transit time with underestimated error bars. We visually inspected the light curves eliminated based on this criterion and we noticed that this criterion is especially useful for the light curves that come from spectroscopic observations (i.e. white light curve, formed by integrating an observed spectrum over the entire wavelenght coverage) because these light curves usually have very high precision (hence low mid-transit time error) but inaccurate mid-transit times due to heavy detrending. We also noticed that this criterion enables us to detect light curves with incorrectly reported time references. We did not apply two-tailed 3$\sigma$ outlier criteria for the XO-2 system because we detect a statistically significant orbital period decrease. For the Qatar 2 system, we calculated the 3$\sigma$ value without including K2 light curves but a linear ephemeris was fitted to all data points including K2.

\subsection{Ephemeris Corrections}
\label{subsec:ephemeris}
For all systems, we fitted independent linear and quadratic ephemeris using the {\sc emcee} package. We followed the same procedure as described in \cite{basturk2022} for selecting random walkers, burn-in period and Markov Chain Monte Carlo (MCMC) steps for convergence. The median values of posterior probability distributions (PPD) of linear elements; slope and $y$-intercept were added to the reference period ($P_{\rm orb}$) and mid-transit time ($T_{\rm c}$) respectively. The updated linear ephemerides are listed in Table \ref{tab:ephemeris} with their uncertainties calculated from PPD.

\subsection{Transit Timing Analyses}
In order to detect potential secular changes in the orbital periods, we fitted quadratic functions to the TTVs of all planets using the method described in section \ref{subsec:ephemeris}. We compared the quadratic ephemeris with the linear to detect any significant secular change for the planets in our sample. 

In Table \ref{tab:tidal_quality_factors}, we report the Akaike Information Criterion differences ($\Delta$AIC) and Bayesian Information Criterion differences ($\Delta$BIC) values between linear and quadratic ephemeris and the rate of the secular period change calculated from the coefficient of the second-degree term of the quadratic ephemeris. We only consider $\Delta$BIC > 10 as suggested strong evidence by \cite{e0988035-67ff-3d00-8d22-b00bd6518fb4} to favour quadratic over the linear ephemerides. 

\begin{figure*}
\includegraphics[width=0.75\paperwidth]{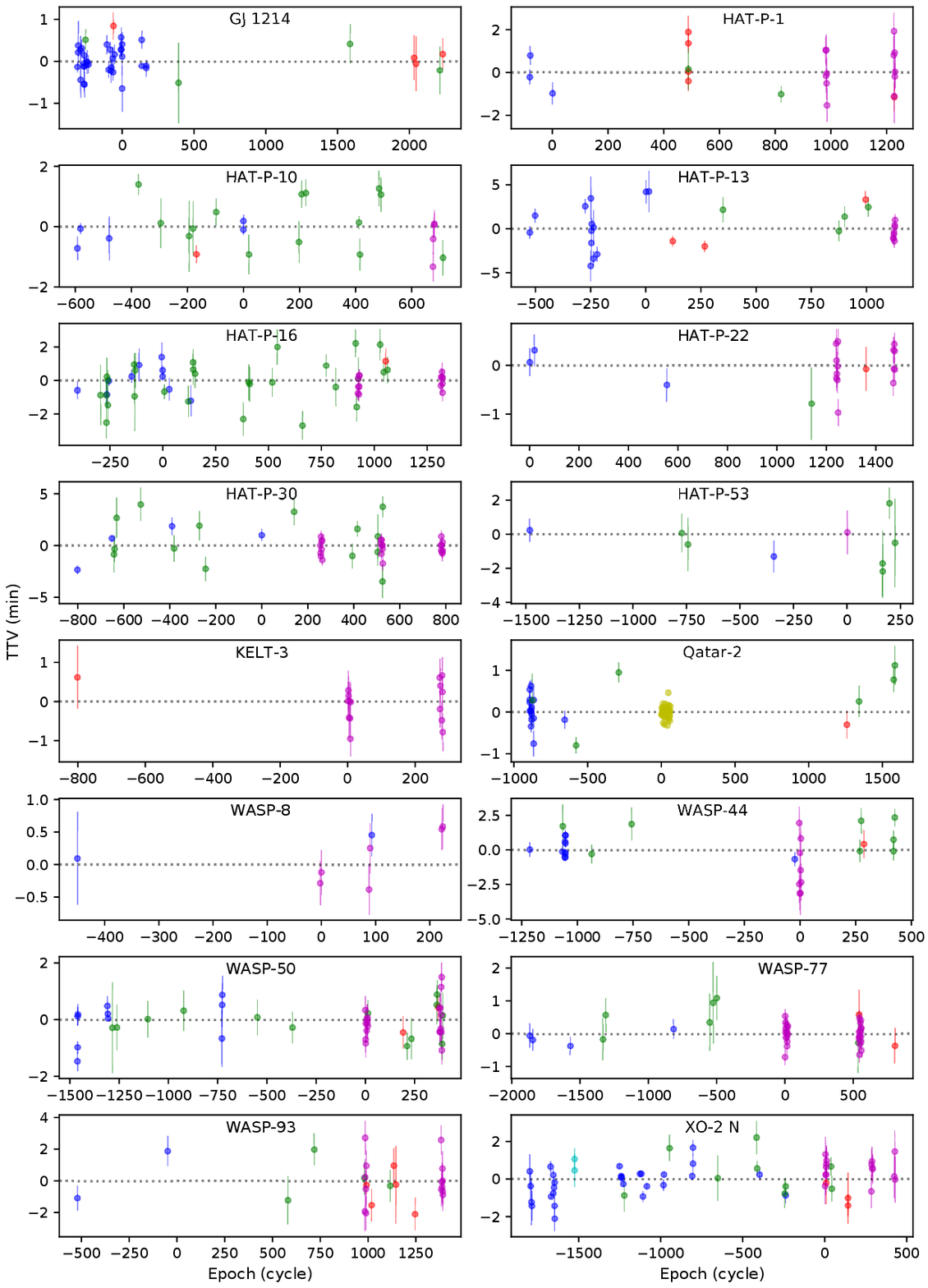}
\caption{Linear residuals of TTV diagrams for all the planets in our sample based on observations from open databases (green), our observations (red), TESS observations (magenta), {\it Kepler} observations (yellow) and light curves published in the literature (blue).}
    \label{fig:ttv_plots_all}
\end{figure*}

After correcting the ephemerides (displayed in Figure \ref{fig:ttv_plots_all}) using the linear coefficients, we performed a frequency analysis to search for potential periodic variations that can be caused by orbital perturbers or the apsidal motion of the planets. We used the {\sc astropy}'s Lomb-Scargle function \citep{2018ApJS..236...16V} to find possible frequencies and their False Alarm Probabilities (FAP). 

\subsubsection{GJ\,1214 System}
GJ\,1214\,b is a sub-Neptune planet ($M_{\rm p}$ = 6.55 $M_\Earth$, $R_{\rm p}$ = 2.678 $R_\Earth$) that orbits an M dwarf star. It has a very high Transmission Spectroscopy Metric (TSM) \citep{kempton2018}, making it one of the most favorable sub-Neptune planets for atmospheric studies \citep{2009Natur.462..891C}. Additional bodies in the system have been searched for using the radial velocity (RV) method with 165 RV points spanning 10 years \citep{clotier2012}, as well as with the transit method using a continuous observing run for $\sim 21$ days from the Spitzer Space Telescope \citep{gillon2014}. Follow-up transit observations have been performed multiple times to investigate TTVs or the atmospheric properties of the planet \citep{kundurthy2011, mooij2012, harpsoe2013, narita2013,caceres2014, nascimbeni2015, 2015MNRAS.453.3875P,2017ApJ...834..151R, 2017A&A...608A.120A,2018A&A...614A..35M,2022A&A...659A..55O, 2022ApJ...939L..11S, 2023A&A...673A.140L, 2023ApJ...951...96G}. We selected the planet for its potential to display TTVs as well as updating its ephemeris for future observation plans especially to understand its atmospheric properties.

We analyzed a total of 48 light curves, including 6 from the Exoplanet Transit Database (ETD), 37 from the literature, and 5 from our observations. However, 5 of the light curves did not meet our selection criteria and were eliminated (as explained in Section \ref{subsec:criteria}). The total data span 10 years of observations, but there was a 5-year gap in the TTV diagram. After analyzing the TTV diagram, we did not detect any significant period change in the GJ\,1214 system.

\subsubsection{HAT-P-1 System}
HAT-P-1\,b is a warm Jupiter with low-density ($M_{\rm p}$ = 0.53 $M_{\rm J}$, $R_{\rm p}$ = 1.36 $R_{\rm J}$) orbiting a G0\,V type star discovered by \cite{bakos2007}. The host star is part of a wide binary system with a companion (HAT-P-1A) of similar effective temperature, making it an excellent comparison star for atmospheric observations in high angular resolution. The planet has a relatively high TSM, which makes it a favorable object for atmospheric studies using ground-based and space-borne telescopes  (e.g. \citealt{montalto2015, wakeford2013}). \cite{bakos2007} suggested a small eccentricity that could be attributed to perturbations by an outer companion, which could be discovered by RV or TTV observations. With follow-up RV observations, \cite{kristo2018} rejected the eccentric orbit and \cite{johnson2008} found that the spin of the orbit of HAT-P-1\,b is aligned with the stellar rotation axis. \cite{winn2007} and \cite{johnson2008} found no significant TTVs in the system.

Here we analyzed 26 transit light curves, 3 of which were eliminated, to update the ephemeris of HAT-P-1\,b. We found no statistically significant periodic or parabolic change in the period analysis. We updated the ephemeris of transit which can be very useful for future atmospheric observations.  

\subsubsection{HAT-P-10/WASP-11 System}
HAT-P-10\,b is a low-mass, hot Jupiter that was independently discovered by \cite{bakos2009a} and \cite{west2009}. Follow-up radial velocity (RV) observations by \cite{knutson2014} revealed a linear trend that suggested the presence of a stellar-mass companion. Adaptive-optic (AO) observations by \cite{ngo2_2015} revealed the existence of a 0.36 \(M_\odot\) companion at a distance of 42 AU ($\sim$0.235\arcsec), which can explain the RV trend. \cite{ngo2016} showed that the companion can not cause Kozai-Lidov migration of the planet, and the eccentricity of the planet is consistent with zero as expected. The Rossiter-Mclaughlin (RM) observations by \cite{mancini2015} indicate that the system is aligned, and this alignment has a primordial origin rather than being due to tidal interactions, owing to the relatively long distance between the star and the planet. Therefore, we do not expect to observe orbital decay in this system. \cite{wang2014} investigated the TTVs to detect any outer companion with the light-time effect (LiTE), but found the orbital period of HAT-P-10 b to be constant. We included this system in our study for the same reasons and studied its TTV diagram with more data spanning a longer baseline. 

We conducted an analysis of 29 transit light curves, consisting of 16 from ETD, 7 from literature, 4 from TESS, and 2 from our own observations. However, we excluded 4 of them and ultimately derived a TTV diagram from 25 mid-transit times that were evenly distributed across a span of 13 years. Our analysis of the TTV diagram did not reveal any significant periodic changes or deviations from a constant period.

\subsubsection{HAT-P-13 System}
HAT-P-13\,b is a warm Jupiter discovered by \cite{bakos2009}, revolving around a Solar-like, metal rich ($T_{\rm eff} = 5653$\,K, ${\rm [Fe/H]}=0.41$) and slightly evolved star. The system consists of at least another planet, HAT-P-13\,c, highly eccentric ($e=0.691$), long period ($P_{\rm c} = 446.27$\,days), massive ($M_{\rm p} \sin{i} = 15.2 \, M_{\rm J}$) outer companion discovered with RV observations. The presence of another outer companion is suggested by the linear trend of RV residuals, as noted by \cite{winn2010} and \cite{knutson2014}. HAT-P-13 was suggested to have a cooler companion ($T_{\rm eff} = 3900$\,K; \citealt{friends3}) blending its lines in its infrared spectrum. However, the AO observations do not reveal a companion \citep{ngo2_2015}, within the limits of the study given in their Figure 4 making the system worthwhile for TTV investigations. \cite{nascimbeni2011} and \cite{pal2011} suggested the system has significant TTV while \cite{southworth2012}, \cite{fulton2011h13} and \cite{sada2016} found the period to be constant by ignoring a single outlier in the TTV diagram. \cite{sunetal2023} detected apsidal motion of the orbit with a $\Delta$BIC = 26.

\begin{center}
    \begin{figure}
    \includegraphics[width=\columnwidth]{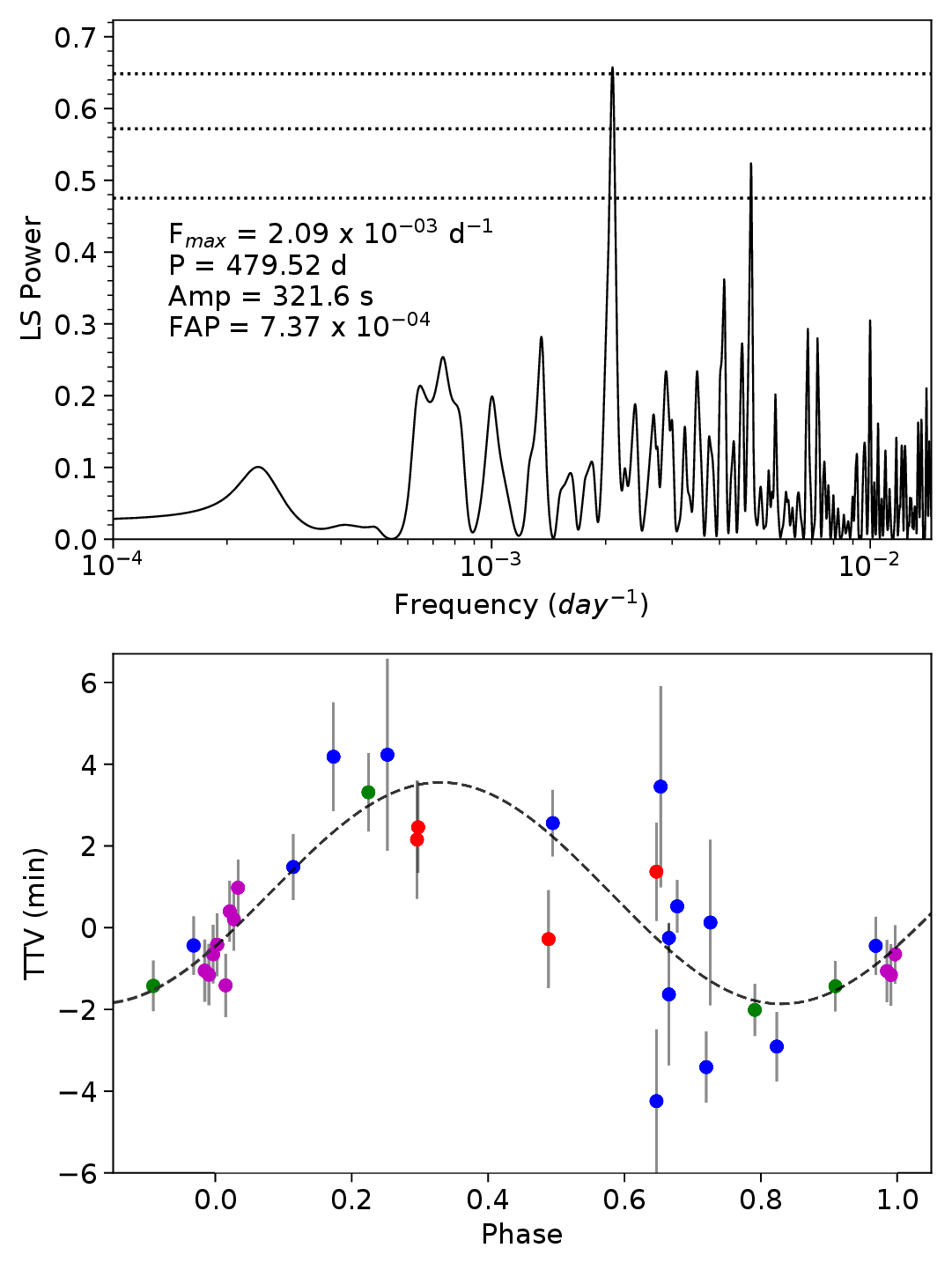}
    \caption{{\it Top}: Lomb-Scargle periodogram of TTV of HAT-P-13 b. The horizontal dotted lines correspond to false alarm probabilities of 0.1, 0.01, 0.001, from bottom to top. {\it Bottom}: phase folded TTV to the frequency with the highest power.}
        \label{fig:hatp13_freq}
    \end{figure}
\end{center}

We used 28 light curves to construct the TTV diagram of HAT-P-13\,b, after eliminating seven of them. We found that the period of HAT-P-13\,b deviates from a constant period. The frequency analysis revealed a peak at 479.52 days with a FAP of 0.0007 and the full TTV amplitude is $\sim$ 321 seconds (see Fig \ref{fig:hatp13_freq}). Assuming that planet \textit{c} is the perturber, and the system is coplanar, the TTV amplitude caused by a planet \textit{c} should be approximately 40 seconds, as previously calculated by \cite{bakos2009}. The RM observations by \citet{winn2010} revealed that the orbit of HAT-P-13\,b is aligned, which supports the coplanar scenario. However, the transit of HAT-P-13\,c has not been observed in long-term observations by \cite{fulton2011h13} and \cite{szabo2010}. Therefore, we conducted a preliminary Newtonian orbital analysis to fit the RVs and TTVs and found that the inclination of the putative planet $c$ must be $\sim 2^\circ$, and its mass should be $\sim 0.4 \, M_{\odot}$ to cause a 321-second TTV. If this is the case, we would expect the impact parameter, $b$, to vary over time, making HAT-P-13\,b's orbit misaligned.

Some of the transit light curves of HAT-P-13\,b exhibit modulations that can be attributed to star spots. This makes it challenging to accurately measure the mid-transit times, which could introduce fallacious TTVs. Additional observations, including upcoming TESS data and new ground-based observations, are needed to determine the true ephemeris of HAT-P-13\,b. We suggest that these light curves require special treatment, such as Gaussian Process \citep{yalcinkaya2021} or spot modeling \citep{mancini2017} for better accuracy.

\subsubsection{HAT-P-16 System}
HAT-P-16\,b is a dense, ($M_{\rm p} = 4.193 \, M_{\rm J}$, $R_{\rm p} = 1.289 \, R_{\rm J}$) hot Jupiter (P = 2.775960 days) orbiting an F8 dwarf, discovered by \cite{bucchave2010}. The planet was found to have a small but statistically significant eccentricity based on its RV observations \citep{bucchave2010, bonomo2017} and its projected spin-orbit angle suggests that it is aligned within the limits of measurement uncertainties (projected spin-orbit angle, $\lambda = -10^\circ \pm 16^\circ$, \citealt{moutou2011}). The small eccentricity could be explained by the young age of the system ($2\pm0.8$\,Gyr: \citealt{bucchave2010}; $0.5\pm0.5$\,Gyr: \citealt{ciceri2013}; $0.8\pm0.2$\,Gyr: \citealt{bonfanti2015}), which might be less than the tidal circularization time. However, the tidal circularization time is linearly dependent on the tidal quality factor ($Q_{\rm p}$, \citealt{2006ApJ...649.1004A}) and it is not well known for hot Jupiters. The stellar age of the cold stars on the main sequence (MS) could not be precisely calculated with the isochrone fitting method because the change in their masses and radii during their MS evolution is comparable to the uncertainties on these parameters. Therefore, it is not possible to deduce the eccentricity of the orbit of HAT-P-16\,b based on a comparison of the tidal circularization time with the stellar age. However, \cite{2010ApJ...720.1569K} measured the $\log{R'_{\rm HK}}$ index as -4.863, which indicates low magnetic activity (e.g. \citealt{1984ApJ...279..763N}), and \citet{ciceri2013} found no starspot induced anomalies in the transit light curves, which is also indicative of low magnetic activity. Hence, the system should not be very young. Using the $\log{(R^{\prime}_{\rm HK})}-$ stellar rotation period ($P_{\rm rot}$) calibration from \cite{2015MNRAS.452.2745S}, we found $P_{\rm rot} \sim 22.5$\,days and $v_{\rm rot} = 2.8$\,km\,s$^{-1}$, meaning that the stellar inclination ($I_{\star}$) is consistent with $90^{\circ}$ within uncertainties (Vsin\textit{i} = 3.5$\pm0.5$, \citealt{bucchave2010}). This result, combined with the RM values, suggests that the orbit of HAT-P-16\,b is well-aligned. \cite{2010ApJ...718L.145W} speculates that hot Jupiter systems may have primordial misaligned orbits, but the tidal dissipation in the convective zones of their host stars can lead to spin-orbit alignment. Considering the relatively high effective temperature of HAT-P-16, the star should have a thin convective zone. The $T_{\rm eff}$ cut-off at which the star will have a negligible convective mass was determined at 6250\,K by \cite{2001ApJ...556L..59P}, while HAT-P-16's $T_{\rm eff}$ is 6158\,K. Then, it should take a few Gyr for HAT-P-16 to diminish the primordial obliquity. On the other hand, using Eq.\,(2) from \cite{2006ApJ...649.1004A} and the limits for Qp between $10^5$ and $10^6$ as given by them, the tidal circularization timescale is only 400 Myr even if the $Q_{\rm p}$ is taken to be $10^6$. Assuming the system is at least a few Gyrs old based on its magnetic activity, the non-zero eccentricity may have been caused by an outer companion \citep{2006ApJ...649..992A}, which may have led to Kozai-Lidov oscillations. \cite{sada2016} searched for TTVs, but did not detect a definitive signal because there were too few observations. \cite{sunetal2023} detected orbital decay with $\Delta$BIC = 167 and apsidal motion with $\Delta$BIC = 317.  We included light curves from several works \citep{2021TJAA....2...28A,bucchave2010,sada2016,ciceri2013,2014NewA...27..102P}, adding up to a total of 62 light curves, nine of which were eliminated, hence we were able to form a TTV diagram covering the widest time range available for analysis. The recently published TESS sector ruled out the orbital decay suggested by \cite{sunetal2023}. We also did not detect any significant cyclic TTV, as suggested by \cite{sunetal2023}, that can be caused by the apsidal motion of the orbit. Although we did not detect any significant cyclic or parabolic changes, we updated the ephemeris for future observations. 

\subsubsection{HAT-P-22 System}
HAT-P-22\,b is a relatively dense ($M_{\rm p} = 2.147 \, M_{\rm J}$, $R_{\rm p} = 1.080 \, R_{\rm J}$), slightly eccentric ($e = 0.0064^{+0.0080}_{-0.0046}$, \citealt{knutson2014}), probably aligned (true spin-orbit angle, $\Psi = 25^\circ \pm 18^\circ$, \citealt{mancini2018}) hot Jupiter discovered by \cite{2011ApJ...742..116B}. Linear trend in the radial velocity residuals has been detected by \cite{knutson2014}, they suggest that this acceleration is an evidence of a presence of at least one additional body in the system. Later on, \cite{friends3} detected a spectroscopic companion with an effective temperature of 4000\,K. However, this companion could not be seen in the AO observations \citep{ngo2_2015}, based on which \cite{friends3} calculated the mass of the potential companion to be $\sim 660 \, M_{\rm J}$ with a maximum separation of 33\,AU. If this companion is responsible for the RV trend, then it should have a face on orbit (e.g. the inclination of the companion's orbit must be close to $0^\circ$). The companion also could have separation larger than 33\,AU but observed at the time when the angular separation is low, which explains the non-AO detection. Based on the mass ratio of the host star and the companion, it is possible for companion to excite the Kozai-Lidov oscillations for HAT-P-22 b from 33\,AU distance (see Fig 5. in \citealt{ngo2_2015}). Moreover, the small eccentricity could be a hint for such oscillation. HAT-P-22 b is also one of the most favorable exoplanet for atmospheric characterization with TSM = 582. We included HAT-P-22 b to our list to attempt to detect TTV and/or update the ephemeris for future observations.

Ground-based photometric follow-up transit observations have been carried out by \cite{hinse2015} and \cite{wang2021}. We used all available observations from the literature, ETD, TESS and our observations, which passed our criteria (six of them were eliminated), to form a TTV diagram of 19 data points spanning a baseline of 13 years. We did not detect any parabolic or periodic changes, and we updated the ephemeris of the exoplanet HAT-P-22 b as a result. 

\subsubsection{HAT-P-30/WASP-51 System}
HAT-P-30\,b is a hot Jupiter ($M_{\rm p} = 0.711 \, M_{\rm J}$, $R_{\rm p} = 1.340 \, R_{\rm J}$), independently discovered by \cite{2011ApJ...735...24J} and \cite{enoch2011}. RV observations show that the planet has a highly oblique (i.e. misaligned) orbit but no potential perturbing companion has been detected in the system through spectral \citep{friends3} or AO observations \citep{ngo2_2015}. \cite{enoch2011} detected a strong Lithium absorption line indicating the system is young (< 1 Gyr). Therefore, it is possible that the planet has not had enough time to damp its obliquity with tidal dissipation \citep{winn2010}. \cite{lu2022} detected TTV for HAT-P-30 b that could be caused by apsidal precession or additional perturbing body. We selected this system to investigate the findings of \cite{lu2022} with the new transit observations.  

We analyzed a total of 48 light curves (6 of them were eliminated), including three TESS sectors and follow-up observations from the literature \citep{wang2021, maciejewski2016} and ETD observations to form the TTV diagram spanning more than 12 years. We were able to accurately update the ephemeris thanks to the multisector TESS observations. However, in contrast to \cite{lu2022}, we did not find any statistically significant TTVs.

\subsubsection{HAT-P-53 System}
HAT-P-53\,b is a hot Jupiter ($M_{\rm p} = 1.484 \, M_{\rm J}$, $R_{\rm p} = 1.318 \, R_J$) that orbits a Sun-like star \citep{hartman2015}. Unfortunately, RV follow-up observations are not sufficiently precise to measure the orbital eccentricity, RM effect or to search for additional bodies even unless they are too massive though \cite{hartman2015} denoted the system's RV can be precisely measured despite the relatively faint host star thanks to its slow rotation and low surface temperature. Although the star rotates slowly, the planet moves rapidly in its orbit. Because of that, HAT-P-53 system is suggested to be a good example for tidal spin-up by \cite{gallet2020}. As the angular momentum transferred from planet's orbit to star, star will rotate faster while planet's orbit shrinks. This effect could be observable with the TTV method if the observed time range is long enough as the amplitude of this effect increases within time. We selected this system to analyze its TTV to attempt to detect such variation.

Photometric transit follow-up observations have been carried out by \cite{kjur2018} and \cite{wang2021}. The system has only 3 transit observations that cover the full transit and does not have 2-minute TESS observations. Nevertheless, we were able to update the ephemeris of the system using the combination of ETD, literature and 30-min TESS data for future observations. We did not detect any deviations from the linear ephemeris in the system in our analysis.

\subsubsection{KELT-3 System}
KELT-3\,b is a hot Jupiter ($M_{\rm p} = 1.477 \,M_{\rm J}$, $R_{\rm p} = 1.345 \, R_{\rm J}$) orbiting a bright, ($V=9.8$ magnitude) late F star discovered by \cite{2013ApJ...773...64P}. A faint nearby star at 3.74 arcseconds angular distance was detected from direct imaging observations  \citep{2015A&A...579A.129W, 2013ApJ...773...64P}. Gaia revealed that this neighbor is actually bound to the system at a linear distance of $\sim 800$\,au. Relatively high surface temperature and brightness of the host star make the KELT-3 system an excellent candidate for probing the atmosphere of its planet in shorter wavelengths with transit observations (e.g. \citealt{2017AJ....153...81C, 2021AJ....162..287C}) or with the occultation observations in longer wavelengths (Emission Spectroscopy Metric, ESM = 170, \citealt{kempton2018}). Despite having a bright host star, the system does not have many follow-up observations in the literature. In fact, our observation is the only one that covers a full transit. \cite{mallonn2019} updated the ephemeris using the transit observations from ETD and observation from \cite{2013ApJ...773...64P}. \cite{wang2021} observed two transits but they were not able to cover the full duration.

We have refined the ephemeris of KELT-3\,b using two sectors of TESS observations and our observation. Based on our criteria, we eliminated the ETD and previous observations from \cite{2013ApJ...773...64P}. However, the TESS observations are relatively precise, thanks to the bright host star, enabling us to correct the ephemeris for future observations of this bright system. 

\subsubsection{Qatar-2 System}
Qatar-2\,b is a short period ($P=1.337$\,d) hot Jupiter ($M_{\rm p} = 2.487 \, M_{\rm J}$, $R_{\rm p} = 1.144 \, R_{\rm J}$) discovered by \cite{2012ApJ...750...84B}. Some of the transit light curves show star-spot occultations by the planet's disk. \cite{mancini2014} observed consecutive transits of QATAR-2 b with ground-based telescopes and they concluded that the planet’s orbital plane is aligned with the stellar rotation by tracking the change in position of one star-spot. Later on, \cite{dai2017} and \cite{mocnik2017} used {\it Kepler} observations and also found that the planet's orbit is aligned. The findings were further confirmed by \cite{esposito2017} based on radial velocity observations during a transit, which revealed a symmetric Rossiter-McLaughlin effect in the prograde direction.

The short period and relatively high mass ratio of Qatar-2\,b make it a potential target to observe an orbital decay (\citealt{dai2017} and references therein), which manifests itself as a parabolic change in the TTV diagram. The amplitude of this effect increases over time, making it detectable with ground-based observations. We observed the transit of Qatar-2\,b after $\sim$1250 epochs later from {\it Kepler} observations but we did not detect statistically significant parabolic change. We also did not detect statistically significant periodic TTV in the system. Although the TTV diagram of Qatar-2 b is not well-sampled, the updated ephemeris precision is the highest among the exoplanets in this study (except WASP-50\,b), thanks to the ultra-precise {\it Kepler} data.

\subsubsection{WASP-8 System}
WASP-8\,b is a warm Jupiter ($P=8.1587$\,d, $M_{\rm p} =2.244 \, M_{\rm J}$, $R_{\rm p} = 1.038 \, R_{\rm J}$), orbiting a bright ($V=9.87$ magnitude) solar-like star, discovered by \cite{2010A&A...517L...1Q}. The planet is very interesting due to its eccentric ($e=0.3044$; \citealt{knutson2014}) and misaligned and retrograde orbit ($\lambda=-143^\circ$; \citealt{bourrier2017}). In the discovery paper, the radial velocity residuals show a linear drift, potentially caused by a companion. The system consists of a physically bound faint M-dwarf, located $\sim 4.5$\arcsec ($\sim 440$\,au) away from WASP-8A \citep{ngo2_2015}. Follow-up RV observations by \cite{knutson2014} revealed that only a part of the observed slope in RV residuals can be due to the presence of WASP-8\,B. Instead, another planet, WASP-8\,c ($P_{\rm c}=4323$\,d, $M_{\rm c}\sin{i_{\rm c}} = 9.45 \, M_{\rm J}$) was found to be responsible for the RV variation. The only photometric follow-up observations were carried out by \cite{borsato2021} with The CHaracterising ExOPlanet Satellite (CHEOPS, \citealt{benz2021}) to improve the precision of ephemeris.  

WASP-8\,b is a promising TTV candidate and has a very high TSM (421), suitable for atmospheric observations. However, its equatorial position ($\delta \sim -35^{\circ}$) and long period (hence long transit duration) make it difficult to observe its full transits. We updated the ephemeris of WASP-8\,b with three sectors of TESS observations and two light curves from previously published observations. This updated ephemeris will be useful for future ground and space-based observations of the system.

\subsubsection{WASP-44 System}
WASP-44\,b is a hot Jupiter ($M_{\rm p} =0.889  \, M_{\rm J}$, $R_{\rm p} = 1.14 R_{\rm J}$) discovered by \cite{2012MNRAS.422.1988A}. \cite{2013MNRAS.430.2932M} found that the radius of the planet is smaller by $10\%$ than first measured and there is no extreme radius variation in the optical wavelengths from multi-band photometry. However, \cite{turner2016} reported the radius of the planet is 1.4$\sigma$ larger in the near-ultraviolet. After $\sim$6.5 years, \cite{addison2019} observed a transit and updated the ephemeris. Similiar to the HAT-P-30 system, WASP-44 system is also exposed to tidal spin-up \citep{gallet2020}. Similiar study has been carried out by \cite{2014MNRAS.442.1844B} and isochrone age was found to be significantly older than gyrochronological age. The angular momentum transfer from planet's orbit to rotation of the star could manifest itself as orbital decay in TTV diagram. We selected this system to attempt to detect such variation.

We analyzed the follow-up transit observations mentioned above, along with our own observations, TESS and ETD data, to update the ephemeris. However, we did not find any evident periodic or secular TTVs in its timing data.

\subsubsection{WASP-50 System}
WASP-50\,b is a hot Jupiter ($M_{\rm p} =1.468  \, M_{\rm J}$, $R_{\rm p} = 1.53 \, R_{\rm J}$), discovered by \cite{2011A&A...533A..88G} revolving on a circular orbit \citep{bonomo2017}. Follow-up photometric transit observations were carried out by \cite{tregloan2013}, \cite{sada2012}, and \cite{sada2018} to update the ephemeris or increase the precision of its transit parameters. \cite{2011A&A...533A..88G} measured the rotation period of WASP-50 from two seasons of WASP photometry as $16.3\pm0.5$\,days; however, \cite{martins2020} found this value to be only 5.488 days from TESS sector-4 light curve. We performed a preliminary analysis of the sector 31 {\sc PDCSAP\_FLUX} of TESS and confirmed the finding by \cite{martins2020}. The measured $\log{R'_{\rm HK}}$ and $P_{\rm rot}$ values by \cite{2011A&A...533A..88G} are in excellent agreement with each other comparing to the empirical values calculated using the $P_{\rm rot}$ - $\log{R'_{\rm HK}}$ relation presented by \cite{2015MNRAS.452.2745S}. Moreover, a rotation rate of 5.488 days indicates a very young age ($\sim 80$\,Myr; \citealt{barnes2007}); however, according to the lithium abundance, the system should be at least $0.6 \pm 0.2$\,Gyr old \citep{2011A&A...533A..88G}. If the true rotation period is 5.488 days, then the lack of lithium suggests that the star could be a good example of tidal spin-up (e.g. \citealt{gallet2020}). \cite{2021ApJ...919..138T} have suggested that even after orbital circularization, the planet's orbit may shrink by transferring angular momentum to its host star and causing its rotation rate to increase. We selected this system to observe such effect via TTV method as it should manifest itself as orbital decay.

We analyzed 45 light curves (9 of them were eliminated) spanning 10 years of best observations available. We did not detect any parabolic TTV but the frequency analysis peaked at 34.45 days with a false alarm probability of 2 per cent. The amplitude of this periodic variation is 57 seconds, which is compatible with our average mid-transit uncertainty. As a result, our findings are inconclusive. Further precise observations are required to confirm this hint of a periodic TTV.

\subsubsection{WASP-77 System}
WASP-77A\,b is a short period ($P=1.36$\,days) hot Jupiter ($M_{\rm p} =1.76 \, M_{\rm J}$, $R_{\rm p} = 1.21 \, R_{\rm J}$) revolving around a G8\,V type, bright ($V=10.12$ magnitude), wide binary with the component WASP-77B at a projected angular distance of $\sim 3.5$ arcseconds \citep{2013PASP..125...48M}. Photometric follow-up transit observations that confirmed the transit parameters of the discovery paper were carried out by \cite{turner2016} and \cite{2020A&A...636A..98C}. The planet has relatively high TSM and ESM (ESM = 333, TSM = 770) and its wide companion (WASP-77B) can be used as a comparison star, making it favorable for atmospheric observations via transmission or emission spectroscopy from the ground \citep{2021Natur.598..580L, 2022AJ....163..159R} or space-borne observations \citep{2022AJ....163..261M}. \cite{gallet2020} suggested the host star might have been affected by tidal-spin-up by its planet WASP-77\,b.

\cite{2020A&A...636A..98C} performed a TTV analysis for WASP-77\,b in a similar way within this work. We added additional transit light curves from TESS sector-31, our observations and the newly available light curves from ETD. As a result, we were able to update the ephemeris with increased precision, thanks to transit light curves covering a longer baseline. As in \cite{2020A&A...636A..98C}, we found no significant secular or periodic TTVs. 

\subsubsection{WASP-93 System}
WASP-93\,b is a hot Jupiter ($M_{\rm p} =1.47 \, M_{\rm J}$, $R_{\rm p} = 1.597 \, R_{\rm J}$) orbiting a fast-rotating ($v\sin{i} = 37 \pm3$\,km\,s$^{-1}$) F4\,V star discovered by \cite{hay2016}. RM observations were attempted twice by \citet{hay2016}; however, the first observation was unable to cover the transit due to ephemeris uncertainty, and the combination of the first and second attempts resulted in inconclusive results due to insufficient RV precision. \cite{2019MNRAS.485.3580G} searched for TTVs in the system using only ETD observations and they did not observe any significant  deviation from linear ephemeris. Although WASP-93\,b has relatively high TSM and ESM, TESS observations do not show significant phase modulations or an occultation signal \citep{2021AJ....162..127W}.

TESS observed WASP-93 during sectors 17, 57, and 58 but, unfortunately, the object was too close to the edge of the camera in sector-58 observations, hence we were unable to use it. We observed 7 transits of WASP-93\,b and used available observations in the literature and ETD to update its ephemeris. The RM observations by \cite{hay2016} show the hint of a retrograde orbit if the transit ephemeris arrived earlier by $\sim$35 min. The timing difference between our ephemeris and the ephemeris from \cite{hay2016} is only $-46 \pm 48$ seconds at the time of the second RM observations. Therefore, we rule out the early transit-retrograde orbit scenario even if the two independent RM observations agree with each other (see Figure 8 in \citealt{hay2016}). We also did not detect any secular or periodic TTV signal.  

\subsubsection{XO-2N System}
XO-2N\,b is a hot Jupiter ($M_{\rm p} =0.57  \, M_{\rm J}$, $R_{\rm p} = 0.98 \, R_{\rm J}$) orbiting around a metal rich, ${\rm [M/H]}=0.44 \pm 0.02$\,dex, wide binary component XO-2N in an aligned orbit \citep{narita2011} discovered by \cite{burke2007}. A binary companion, XO-2S, which is also a metal-rich star, resides $\sim 30\arcsec$ away from XO-2N and has at least two planets discovered with RV observations \citep{desidera2014}. The visual binary components have similar effective temperatures, thus XO-2S is a great comparison star for transmission spectroscopy of XO-2N\,b (e.g. \citealt{sing2012, 2012ApJ...761....7C}). With additional RV observations, \cite{knutson2014} detected that the radial velocity residuals show a linear trend within time, possibly caused by an outer companion. Later on, \cite{2015A&A...575A.111D} revealed that the linear RV residuals are actually only a part of the curve in RV, possibly caused by an outer companion XO-2N\,c or by stellar activity.

\begin{center}
    \begin{figure*}
    \includegraphics[width=2\columnwidth]{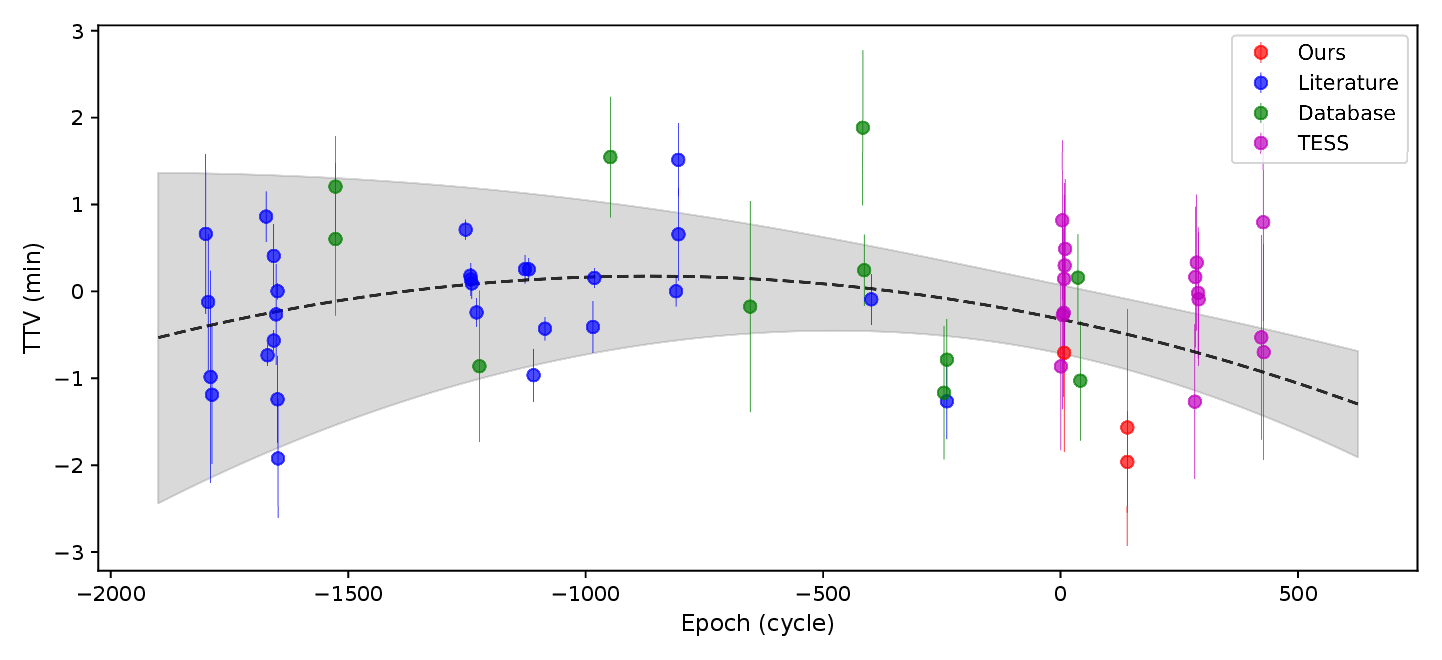}
    \caption{Quadratic TTV model of XO-2N\,b (black dashed line with 3\,$\sigma$ uncertainty shown in grey shaded region). The linear term was subtracted for display purposes.}
        \label{fig:xo2_ttv_quad}
    \end{figure*}
\end{center}

We obtained the transit light curves from several works \citep{burke2007, 2009AJ....137.4911F, kundurthy2011, 2015A&A...575A.111D, 2018IBVS.6243....1M, wang2021}, three sectors of observation (20, 47, 60) from TESS, 13 from amateur astronomers and our three new observations to form the TTV diagram with a total of 42 light curves, spanning almost 16 years. Our analyses indicated that the parabolic ephemeris fit the data better than linear with the $\Delta$BIC of 42.98 and $\Delta$AIC of 45.15, suggesting the orbit of XO-2\,b is decaying with a rate of $-12.95 \pm 1.85$\,ms\,yr$^{-1}$ (Figure \ref{fig:xo2_ttv_quad}). The parabolic ephemeris is as follows:
\begin{align}
T = 2458843.218212(92) + 2.61585862(17)\ E \notag\\
-4.7(8)\times10^{-10}\ E{^2}
\end{align}

 Although the parabolic ephemeris is statistically significant, it does not agree well with the latest TESS observations (sectors 47 and 60). TESS will observe the XO-2N system during cycle 6 of its mission. However, the wide binary component is only $\sim30\arcsec$ away, and a TESS pixel has a $21\arcsec$ field of view. This may add extra red and white noise because the light from XO-2S blends into the aperture selected for XO-2N. Therefore, ground-based observations could be a better option for confirming or discarding the parabolic trend.

\section{Discussion}
\label{discussion}
We constructed the TTV diagrams of 16 exoplanets consisting of the most precise and complete light curves with the longest time span for each of the planets in our sample (Figure \ref{fig:ttv_plots_all}). This allowed us to increase the precision of the orbital period and the accuracy of the ephemeris information for future follow-up observations. Based on the ephemeris information given in Table \ref{tab:ephemeris}, the uncertainty on the predicted ephemeris will provide transit timings with a precision below 5 minutes with an average of 1.8 minutes until 2070 for all the systems except for XO-2N and HAT-P-13 where we detect deviation from the linear ephemeris. 

\begin{table}
\caption{Reference ephemeris information ($T_0$ and $P_{\rm orb}$).}
\label{tab:ephemeris}
\begin{tabular}{lll}  
\hline
\hline
Planet & ~~~~~~~~~~$T_0$ & ~~~~~~~$P_{\rm orb}$ \\
 & ~~~~~~[BJD$_{\rm TDB}$]  & ~~~~~~[days]  \\
\hline
GJ 1214\,b & 2455799.398485(19) & 1.580404418(58) \\
HAT-P-1\,b & 2454363.94778(13) & 4.46530031(18) \\
HAT-P-10\,b & 2456933.615316(50) & 3.72247975(13) \\
HAT-P-13\,b & 2456316.79044(12) & 2.91624121(16) \\
HAT-P-16\,b & 2456204.604299(69) & 2.775967270(90) \\
HAT-P-22\,b & 2454891.67399(13) & 3.21223265(10) \\
HAT-P-30\,b & 2457775.212778(77) & 2.81060070(13) \\
HAT-P-53\,b & 2458771.88605(28)  & 1.96162529(35) \\
KELT-3\,b & 2458872.854373(95) & 2.70339023(43) \\
QATAR-2\,b & 2457218.1101306(63)  & 1.337116440(32) \\
WASP-8\,b & 2458375.23865(11)  & 8.15872562(67) \\
WASP-44\,b & 2458398.69812(13)  & 2.42381131(13) \\
WASP-50\,b & 2458411.093420(55)  & 1.955092447(48) \\
WASP-77A\,b & 2458410.984807(28)  & 1.360028898(50) \\
WASP-93\,b & 2456079.56501(26)  & 2.73253778(25) \\
XO-2\,b & 2458843.218463(69)  & 2.615859462(54) \\
\hline
\end{tabular}
\end{table}
We detect a decrease in the orbital period of XO-2N\,b may have been caused by several events if it is real. As discussed by \cite{Vissapragada2022}, such a decrease can be observed if the system is accelerating towards us, making the transits observed earlier than expected. In that case, the radial velocity residuals should have a slope of $-0.05$\,m\,s$^{-1}$\,d$^{-1}$ ($\dot{\nu}$ = c$\dot{P}/P$, \citealt{Vissapragada2022}) but \cite{2015A&A...575A.111D} reported this value as $+0.0017$\,m\,s$^{-1}$\,d$^{-1}$. \cite{2015A&A...575A.111D} also reported the eccentricity of the planet is consistent with zero. Therefore we did not consider a scenario based on precession. The RV residuals show a parabolic variation that can be caused by a long-period outer companion. This companion may be causing LiTE and changing the orbital period of XO-2\,b within a longer time interval, thus the parabolic TTV could be a part of this periodic variation. We neglect this scenario also because \cite{2015A&A...575A.111D} showed that the parabolic RV variation is due to magnetic activity. Even if the magnetic activity was not the reason for the RV residuals, the phases of LiTE and long-term parabolic RV change do not match. 

We detect a cyclic variation in the period of HAT-P-13\,b which has a semi-amplitude of 160.8 seconds and 479.52\,day periodicity. This variation could be caused by the known outer companion of the system, HAT-P-13\,c ($P = 445.82 \pm 0.11$\,days, $M_{\rm p} \sin{i} = 14.61^{+0.46}_{-0.48} \, M_{\rm J}$, \citealt{knutson2014}). However, in order to cause such a high-amplitude TTV, the orbital inclination of HAT-P-13\,c needs to be $\sim 2^{\circ}$, which translates into a mass of $\sim 0.4 \, M_{\odot}$ using the $M_{\rm p} \sin{i}$ value. \cite{friends3} detected another star in the spectrum of HAT-P-13 that has an effective temperature of $3900^{+300}_{-350}$ ($M_{\rm companion} = 0.6^{+0.086}_{-0.179} \, M_{\odot}$). \cite{friends3} discussed if HAT-P-13\,d, detected by linear drift in the RV residuals, is that spectral companion. But in that case, the inclination of the planet d should be $\sim 5^\circ$, which could lead to Kozai-Lidov oscillations and make the orbit of HAT-P-13\,b misaligned \cite{winn2010}. Our TTV analysis suggests that the observed spectral companion could be HAT-P-13\,c instead. This could explain the non-AO detection by \cite{ngo2_2015} due to short angular separation. However, such a close companion would have catastrophic effects on the system stability aside from making the orbit of HAT-P-13\,b misaligned. 

Nevertheless, in order to assess the timescale of any potential orbital instability of the planet c with an orbital inclination of $2^\circ$, we run an N-body simulation by using {\sc Rebound} code \citep{rebound}, with IAS15 \citep{ias15} integrator. We used the masses and orbital parameters of the planets b and c derived from \cite{winn2010}. We stress that we did not include the potential additional planet (planet d in \cite{knutson2014}), since the orbital parameters and mass of this putative object are currently unknown. Results of our simulation show that the system becomes unstable on the order of 30,000\,yrs. Therefore, we think that the amplitude of the detected TTV signal in our analysis might have been overestimated due to the high scatter caused by correlated noise in the transit light curves.

We do not find any statistically significant periodicities in our timing analysis of any other system. However, we simulated the WASP-8 system in a same way as we did for HAT-P-13 and we found that the TTV due to WASP-8 c should be observable. We assumed the transiting planet WASP-4 b is coplanar with the RV planet WASP-8 c. Using the absolute and orbital parameters from \cite{2010A&A...517L...1Q} and \cite{knutson2014}, we found that the full TTV amplitude of WASP-8 b should be $\sim$40 seconds due to LiTE and $\sim$2 seconds due to gravitational interaction. The typical mid-transit measurement error of TESS data for this system is about 20 seconds, so the TTV of this system can be detected with light curves that has similar precision as TESS. We could not detect this signal due to poor phase coverage. 

\subsection{Discussion on Tidal Quality Factors}
\label{subsec:tidal_quality}
 Although we do not have any observational data on the rotation rates of the exoplanets in our sample, the rotational periods of their hosts are all longer compared to their orbital periods. The energy raised in those tidal interactions can be dissipated in the convective envelope of their host stars, transferring angular momentum from the planet to the star that would cause the star to spin-up while the planet to fall-in \citep{1973ApJ...180..307C, 1996ApJ...470.1187R, 2016ApJ...816...18E, 2023A&A...669A.124H, 2023arXiv230511974W}. In addition to this equilibrium tide, dynamical tide excites internal gravity waves, which dissipate the energy through also the secondary waves it generates via wave-wave interactions \citep{2010MNRAS.404.1849B, 2013MNRAS.432.2339I, 2020MNRAS.498.2270B}. This latter mechanism is especially dominant in a system with a solar-type host and a hot Jupiter-type planet \citep{2016ApJ...816...18E, 2023arXiv230511974W}. 

If we assume the planetary mass to be constant, then the rate of change in the orbital period can be related to the so-called reduced tidal quality factor of the host star by the constant phase lag model of \citet{1966Icar....5..375G} defined as
\begin{equation}
    Q^{\prime}_{\star}= -\frac{27\pi}{2} \left(\frac{M_{\rm p}}{M_{\star}}\right) \left(\frac{R_{\star}}{a}\right)^{5}\frac{1}{\Dot{P}}
\end{equation}
where $M_{\rm p}$ is the mass of the planet, $M_{\star}$ and $R_{\star}$ are the mass and the radius of the host star and $a$ is the semi-major axis of the planet's orbit. The rate of orbital decay, $\Dot{\rm P}$, is derived from the timing analysis, which is twice the value of the quadratic coefficient of the best-fitting parabola. If that best-fitting model is not found to be statistically superior to the linear model, then an orbital decay cannot be argued and the quadratic coefficient can only be used to derive a lower limit for the reduced tidal quality factor, which is the case for all systems in our sample except XO-2N. We derived these limits based on the fundamental parameters of the objects in our sample, which we provide in Table \ref{tab:tidal_quality_factors} together with $\Delta$AIC and  $\Delta$BIC values, indicating the statistical significance of the quadratic model in each of the cases. Positive values for both statistics hint that the quadratic model should be favored.

\begin{table*}
\centering
\caption{Lower limits for the Reduced Tidal Quality Factors ($Q_{\star}^{\prime}$) for the host stars in our sample at 95\% confidence level. $\Delta$AIC and  $\Delta$BIC values were used for comparisons between linear and quadratic models in this order. The quadratic model is favored in the cases where $\Delta$BIC > 10.}
\label{tab:tidal_quality_factors}
\begin{tabular}{lccrrc}  
\hline
\hline \\ [-8pt]
System & $\Delta$AIC & $\Delta$BIC & d$P$/d$E$ (days/cycle) & $\dot{P}$ (ms/yr) & $Q_{\star}^{\prime}$ \\
\hline
GJ\,1214 & -1.58  & -3.32  & $6.63 \pm 11.4 \times 10^{-11}$ &  $1.32 \pm 2.26 $ & $> 2.4 \times 10^{2}$ \\
HAT-P-1 & -1.90  & -3.00 & $-3.63 \pm 9.20 \times 10^{-10}$ & $18.99 \pm 9.83$ & $> 6.2 \times 10^2$  \\
HAT-P-10 & 0.63  & -0.54 & $-9.60 \pm 5.98 \times 10^{-10}$ & $-8.14 \pm 5.07$ & $> 4.1 \times 10^2$  \\
HAT-P-13 & -2.52  & -3.86 & $-1.98 \pm 10.2 \times 10^{-10}$ & $-2.14 \pm 11.10$ & $> 1.1 \times 10^{5}$  \\
HAT-P-16 & -1.62  & -3.59  & $-3.35 \pm 5.01 \times 10^{-10}$ & $-8.93 \pm 10.12$  & $> 2.6 \times 10^4$ \\
HAT-P-22 & 1.07  & 2.02 & $1.14 \pm 0.58 \times 10^{-9}$ & $11.24 \pm  5.65$ & $> 2.4 \times 10^3$  \\
HAT-P-30 & 1.66  & -0.08 & $-1.32 \pm 0.61 \times 10^{-9}$ & $-43.8 \pm  12.12$ & $> 1.6 \times 10^3$  \\
HAT-P-53& -0.42  & -0.61 & $2.0 \pm 1.70 \times 10^{-9}$ & $32.30 \pm 27.55$ & $> 8.2 \times 10^3$   \\
KELT-3 & -0.13 & -0.90 & $2.12 \pm 1.84 \times 10^{-9}$ & $24.72 \pm 21.53$ & $> 8.8 \times 10^3$ \\
Qatar-2 & 7.26 & 4.90 & $2.46 \pm 0.80 \times 10^{-10}$ & $5.81 \pm 1.89$ & $> 1.4 \times 10^4$  \\
WASP-8 & 0.53 & 0.45 & $7.40 \pm 4.76 \times 10^{-9}$ & $28.60 \pm 18.40$ & $> 2.4 \times 10^0$  \\
WASP-44 & 7.27 & 5.94 & $4.78 \pm 1.13 \times 10^{-9}$ & $62.23 \pm 14.69$ & $> 5.5 \times 10^2$  \\
WASP-50 & -0.59 & -2.40 & $-4.42 \pm 3.60 \times 10^{-10}$ & $-7.14 \pm 5.81$ & $> 6.9 \times 10^3$  \\
WASP-77A & 0.15 & -1.63 & $-1.69 \pm 1.20 \times 10^{-10}$ & $-3.93 \pm 2.78$ & $> 7.2 \times 10^4$  \\
WASP-93 & -1.07 & -2.28 & $-7.87 \pm 9.46 \times 10^{-10}$ & $-31.89 \pm 19.03$ & $> 3.7 \times 10^3$  \\
XO-2 & 45.15 & 42.98 & $-10.7 \pm 1.54 \times 10^{-10}$ & $-12.95 \pm 1.85$ & $= 1.5 \times 10^3$  \\
\hline
\end{tabular}
\end{table*}

XO-2N\,b is not one of the prime candidates for orbital decay due to tidal interactions with its host star because it is not a particularly big planet. Although its solar-like host star should have a convective envelope to dissipate the tidal energy, its reduced tidal quality factor should be larger than 550 as derived from the rate of observed period decrease when it is compared with other host stars with similar spectral types and evolutionary history. Infall time calculated from the same rate is $P/\dot{P} \approx 17.45$\,Myr; which is too short compared to age of the system. TESS observations do not follow the orbital decay model found to be statistically superior to the linear model. However, the precision of TESS observations of XO-2N is lower than that achieved for the stars with similar magnitudes. This is because the wide binary component XO-2S is only $\sim 30$\arcsec away from the XO-2N, blending its flux in the TESS aperture, adding extra noise and diluting the transit depth, resulting in high scatter on the TTV diagram. Because of this reason, we encourage observations of the transit of XO-2N\,b with ground-based telescopes in high angular resolution in the future to confirm the orbital decay scenario, which we find unlikely at the moment.

\section*{Acknowledgments}
We gratefully acknowledge the support by The Scientific and Technological Research Council of T\"urkiye (T\"UB\.{I}TAK) with project 118F042. We thank T\"UB\.{I}TAK for the partial support in using the T100 telescope with the project numbers 12CT100-378, 16BT100-1034, 17BT100-1196 and 19AT00-1472. The data in this study were obtained with the T80 telescope at the Ankara University Astronomy and Space Sciences Research and Application Center (Kreiken Observatory) with project number of 22B.T80.10. We thank the observers in AUKR, UZAYMER, ATASAM, TUG, observatories, and all the observers around the world who kindly shared their data with us. This research has made use of data obtained using the ATA50 telescope and the CCD attached to it, operated by Atat\"urk University Astrophysics Research and Application Center (ATASAM). Funding for the ATA50 telescope and the attached CCD has been provided by Atat\"urk University (P.No. BAP-2010/40) and Erciyes University (P.No. FBA-11-3283) through Scientific Research Projects Coordination Units (BAP), respectively. This study was funded by the Scientific Research Projects Coordination Unit of Istanbul University (P.No. FBA-2022-39121). Authors from Bilkent University thank the internship program of Ankara University Kreiken Observatory (AUKR). We thank all the observers who report their observations to ETD and AXA open databases.
L.\,M. acknowledges support from the ``Fondi di Ricerca Scientifica d'Ateneo 2021'' of the University of Rome ``Tor Vergata''.
Based on observations collected at the Centro Astron\'{o}mico Hispano en Andaluc\'{i}a (CAHA) at Calar Alto, operated jointly by Junta de Andalucía and Consejo Superior de Investigaciones Cient\'{i}ficas (IAA-CSIC).
This work includes data collected with the TESS mission, obtained from the MAST data archive at the Space Telescope Science Institute (STScI). Funding for the TESS mission is provided by the NASA Explorer Program. STScI is operated by the Association of Universities for Research in Astronomy, Inc., under NASA contract NAS 5–26555.
We acknowledge the use of public TESS data from pipelines at the TESS Science Office and at the TESS Science Processing Operations Center. Resources supporting this work were provided by the NASA High-End Computing (HEC) Program through the NASA Advanced Supercomputing (NAS) Division at Ames Research Center for the production of the SPOC data products.
This paper includes data collected by the {\it Kepler} mission and obtained from the MAST data archive at the Space Telescope Science Institute (STScI). Funding for the {\it Kepler} mission is provided by the NASA Science Mission Directorate. STScI is operated by the Association of Universities for Research in Astronomy, Inc., under NASA contract NAS 5–26555. 
This work has made use of data from the European Space Agency (ESA) mission {\it Gaia} (\url{https://www.cosmos.esa.int/gaia}), processed by
the {\it Gaia} Data Processing and Analysis Consortium (DPAC,
\url{https://www.cosmos.esa.int/web/gaia/dpac/consortium}). Funding for the DPAC has been provided by national institutions, in particular, the institutions participating in the {\it Gaia} Multilateral Agreement.

\section{Data Availability}
Some of the light curves to derive mid-transit times were downloaded from Exoplanet Transit Database at http://var2.astro.cz/ETD/. All other light curves appearing for the first time in this article are presented as online material. Mid-transit times derived from our own light curves as well as that of other observers are presented as online data sets too. 



\bibliographystyle{mnras}
\bibliography{ttv_16_exoplanets}

\begin{thebibliography}{}
\makeatletter
\relax
\def\mn@urlcharsother{\let\do\@makeother \do\$\do\&\do\#\do\^\do\_\do\%\do\~}
\def\mn@doi{\begingroup\mn@urlcharsother \@ifnextchar [ {\mn@doi@}
  {\mn@doi@[]}}
\def\mn@doi@[#1]#2{\def\@tempa{#1}\ifx\@tempa\@empty \href
  {http://dx.doi.org/#2} {doi:#2}\else \href {http://dx.doi.org/#2} {#1}\fi
  \endgroup}
\def\mn@eprint#1#2{\mn@eprint@#1:#2::\@nil}
\def\mn@eprint@arXiv#1{\href {http://arxiv.org/abs/#1} {{\tt arXiv:#1}}}
\def\mn@eprint@dblp#1{\href {http://dblp.uni-trier.de/rec/bibtex/#1.xml}
  {dblp:#1}}
\def\mn@eprint@#1:#2:#3:#4\@nil{\def\@tempa {#1}\def\@tempb {#2}\def\@tempc
  {#3}\ifx \@tempc \@empty \let \@tempc \@tempb \let \@tempb \@tempa \fi \ifx
  \@tempb \@empty \def\@tempb {arXiv}\fi \@ifundefined
  {mn@eprint@\@tempb}{\@tempb:\@tempc}{\expandafter \expandafter \csname
  mn@eprint@\@tempb\endcsname \expandafter{\@tempc}}}

\bibitem[\protect\citeauthoryear{{Adams} \& {Laughlin}}{{Adams} \&
  {Laughlin}}{2006a}]{2006ApJ...649..992A}
{Adams} F.~C.,  {Laughlin} G.,  2006a, \mn@doi [\apj] {10.1086/506142}, \href
  {https://ui.adsabs.harvard.edu/abs/2006ApJ...649..992A} {649, 992}

\bibitem[\protect\citeauthoryear{{Adams} \& {Laughlin}}{{Adams} \&
  {Laughlin}}{2006b}]{2006ApJ...649.1004A}
{Adams} F.~C.,  {Laughlin} G.,  2006b, \mn@doi [\apj] {10.1086/506145}, \href
  {https://ui.adsabs.harvard.edu/abs/2006ApJ...649.1004A} {649, 1004}

\bibitem[\protect\citeauthoryear{{Addison} et~al.,}{{Addison}
  et~al.}{2019}]{addison2019}
{Addison} B.,  et~al., 2019, \mn@doi [\pasp] {10.1088/1538-3873/ab03aa}, \href
  {https://ui.adsabs.harvard.edu/abs/2019PASP..131k5003A} {131, 115003}

\bibitem[\protect\citeauthoryear{{Alada{\u{g}}}, {Aky{\"u}z}, {Bast{\"u}rk},
  {Aksaker}, {Esmer}  \& {Yal{\c{c}}{\i}nkaya}}{{Alada{\u{g}}}
  et~al.}{2021}]{2021TJAA....2...28A}
{Alada{\u{g}}} Y.,  {Aky{\"u}z} A.,  {Bast{\"u}rk} {\"O}.,  {Aksaker} N.,
  {Esmer} E.~M.,   {Yal{\c{c}}{\i}nkaya} S.,  2021, \mn@doi [Turkish Journal of
  Astronomy and Astrophysics] {10.48550/arXiv.2109.06108}, \href
  {https://ui.adsabs.harvard.edu/abs/2021TJAA....2...28A} {2, 28}

\bibitem[\protect\citeauthoryear{{Anderson} et~al.,}{{Anderson}
  et~al.}{2012}]{2012MNRAS.422.1988A}
{Anderson} D.~R.,  et~al., 2012, \mn@doi [\mnras]
  {10.1111/j.1365-2966.2012.20635.x}, \href
  {https://ui.adsabs.harvard.edu/abs/2012MNRAS.422.1988A} {422, 1988}

\bibitem[\protect\citeauthoryear{{Angerhausen} et~al.,}{{Angerhausen}
  et~al.}{2017}]{2017A&A...608A.120A}
{Angerhausen} D.,  et~al., 2017, \mn@doi [\aap] {10.1051/0004-6361/201730914},
  \href {https://ui.adsabs.harvard.edu/abs/2017A&A...608A.120A} {608, A120}

\bibitem[\protect\citeauthoryear{{Arcangeli}, {D{\'e}sert}, {Parmentier},
  {Tsai}  \& {Stevenson}}{{Arcangeli} et~al.}{2021}]{arcangeli2021}
{Arcangeli} J.,  {D{\'e}sert} J.~M.,  {Parmentier} V.,  {Tsai} S.~M.,
  {Stevenson} K.~B.,  2021, \mn@doi [\aap] {10.1051/0004-6361/202038865}, \href
  {https://ui.adsabs.harvard.edu/abs/2021A&A...646A..94A} {646, A94}

\bibitem[\protect\citeauthoryear{{Astropy Collaboration} et~al.,}{{Astropy
  Collaboration} et~al.}{2013}]{2013A&A...558A..33A}
{Astropy Collaboration} et~al., 2013, \mn@doi [\aap]
  {10.1051/0004-6361/201322068}, \href
  {https://ui.adsabs.harvard.edu/abs/2013A&A...558A..33A} {558, A33}

\bibitem[\protect\citeauthoryear{{Astropy Collaboration} et~al.,}{{Astropy
  Collaboration} et~al.}{2018}]{2018AJ....156..123A}
{Astropy Collaboration} et~al., 2018, \mn@doi [\aj] {10.3847/1538-3881/aabc4f},
  \href {https://ui.adsabs.harvard.edu/abs/2018AJ....156..123A} {156, 123}

\bibitem[\protect\citeauthoryear{{Ba{\c{s}}t{\"u}rk}
  et~al.,}{{Ba{\c{s}}t{\"u}rk} et~al.}{2022}]{basturk2022}
{Ba{\c{s}}t{\"u}rk} {\"O}.,  et~al., 2022, \mn@doi [\mnras]
  {10.1093/mnras/stac592}, \href
  {https://ui.adsabs.harvard.edu/abs/2022MNRAS.512.2062B} {512, 2062}

\bibitem[\protect\citeauthoryear{{Bai}, {Gu}, {Wang}, {Sun}, {Kwok}  \&
  {Hui}}{{Bai} et~al.}{2022}]{lu2022}
{Bai} L.,  {Gu} S.,  {Wang} X.,  {Sun} L.,  {Kwok} C.-T.,   {Hui} H.-K.,  2022,
  \mn@doi [\aj] {10.3847/1538-3881/ac5b6a}, \href
  {https://ui.adsabs.harvard.edu/abs/2022AJ....163..208B} {163, 208}

\bibitem[\protect\citeauthoryear{{Bakos} et~al.,}{{Bakos}
  et~al.}{2007}]{bakos2007}
{Bakos} G.~{\'A}.,  et~al., 2007, \mn@doi [\apj] {10.1086/509874}, \href
  {https://ui.adsabs.harvard.edu/abs/2007ApJ...656..552B} {656, 552}

\bibitem[\protect\citeauthoryear{{Bakos} et~al.,}{{Bakos}
  et~al.}{2009a}]{bakos2009a}
{Bakos} G.~{\'A}.,  et~al., 2009a, \mn@doi [\apj]
  {10.1088/0004-637X/696/2/1950}, \href
  {https://ui.adsabs.harvard.edu/abs/2009ApJ...696.1950B} {696, 1950}

\bibitem[\protect\citeauthoryear{{Bakos} et~al.,}{{Bakos}
  et~al.}{2009b}]{bakos2009}
{Bakos} G.~{\'A}.,  et~al., 2009b, \mn@doi [\apj]
  {10.1088/0004-637X/707/1/446}, \href
  {https://ui.adsabs.harvard.edu/abs/2009ApJ...707..446B} {707, 446}

\bibitem[\protect\citeauthoryear{{Bakos} et~al.,}{{Bakos}
  et~al.}{2011}]{2011ApJ...742..116B}
{Bakos} G.~{\'A}.,  et~al., 2011, \mn@doi [\apj] {10.1088/0004-637X/742/2/116},
  \href {https://ui.adsabs.harvard.edu/abs/2011ApJ...742..116B} {742, 116}

\bibitem[\protect\citeauthoryear{{Barker}}{{Barker}}{2020}]{2020MNRAS.498.2270B}
{Barker} A.~J.,  2020, \mn@doi [\mnras] {10.1093/mnras/staa2405}, \href
  {https://ui.adsabs.harvard.edu/abs/2020MNRAS.498.2270B} {498, 2270}

\bibitem[\protect\citeauthoryear{{Barker} \& {Ogilvie}}{{Barker} \&
  {Ogilvie}}{2010}]{2010MNRAS.404.1849B}
{Barker} A.~J.,  {Ogilvie} G.~I.,  2010, \mn@doi [\mnras]
  {10.1111/j.1365-2966.2010.16400.x}, \href
  {https://ui.adsabs.harvard.edu/abs/2010MNRAS.404.1849B} {404, 1849}

\bibitem[\protect\citeauthoryear{{Barnes}}{{Barnes}}{2007}]{barnes2007}
{Barnes} S.~A.,  2007, \mn@doi [\apj] {10.1086/519295}, \href
  {https://ui.adsabs.harvard.edu/abs/2007ApJ...669.1167B} {669, 1167}

\bibitem[\protect\citeauthoryear{{Benz} et~al.,}{{Benz}
  et~al.}{2021}]{benz2021}
{Benz} W.,  et~al., 2021, \mn@doi [Experimental Astronomy]
  {10.1007/s10686-020-09679-4}, \href
  {https://ui.adsabs.harvard.edu/abs/2021ExA....51..109B} {51, 109}

\bibitem[\protect\citeauthoryear{{Bonfanti}, {Ortolani}, {Piotto}  \&
  {Nascimbeni}}{{Bonfanti} et~al.}{2015}]{bonfanti2015}
{Bonfanti} A.,  {Ortolani} S.,  {Piotto} G.,   {Nascimbeni} V.,  2015, \mn@doi
  [\aap] {10.1051/0004-6361/201424951}, \href
  {https://ui.adsabs.harvard.edu/abs/2015A&A...575A..18B} {575, A18}

\bibitem[\protect\citeauthoryear{{Bonomo} et~al.,}{{Bonomo}
  et~al.}{2017}]{bonomo2017}
{Bonomo} A.~S.,  et~al., 2017, \mn@doi [\aap] {10.1051/0004-6361/201629882},
  \href {https://ui.adsabs.harvard.edu/abs/2017A&A...602A.107B} {602, A107}

\bibitem[\protect\citeauthoryear{{Borsato} et~al.,}{{Borsato}
  et~al.}{2021}]{borsato2021}
{Borsato} L.,  et~al., 2021, \mn@doi [\mnras] {10.1093/mnras/stab1782}, \href
  {https://ui.adsabs.harvard.edu/abs/2021MNRAS.506.3810B} {506, 3810}

\bibitem[\protect\citeauthoryear{{Bourrier}, {Cegla}, {Lovis}  \&
  {Wyttenbach}}{{Bourrier} et~al.}{2017}]{bourrier2017}
{Bourrier} V.,  {Cegla} H.~M.,  {Lovis} C.,   {Wyttenbach} A.,  2017, \mn@doi
  [\aap] {10.1051/0004-6361/201629973}, \href
  {https://ui.adsabs.harvard.edu/abs/2017A&A...599A..33B} {599, A33}

\bibitem[\protect\citeauthoryear{{Brown}}{{Brown}}{2014}]{2014MNRAS.442.1844B}
{Brown} D.~J.~A.,  2014, \mn@doi [\mnras] {10.1093/mnras/stu950}, \href
  {https://ui.adsabs.harvard.edu/abs/2014MNRAS.442.1844B} {442, 1844}

\bibitem[\protect\citeauthoryear{{Bryan} et~al.,}{{Bryan}
  et~al.}{2012}]{2012ApJ...750...84B}
{Bryan} M.~L.,  et~al., 2012, \mn@doi [\apj] {10.1088/0004-637X/750/1/84},
  \href {https://ui.adsabs.harvard.edu/abs/2012ApJ...750...84B} {750, 84}

\bibitem[\protect\citeauthoryear{{Buchhave} et~al.,}{{Buchhave}
  et~al.}{2010}]{bucchave2010}
{Buchhave} L.~A.,  et~al., 2010, \mn@doi [\apj] {10.1088/0004-637X/720/2/1118},
  \href {https://ui.adsabs.harvard.edu/abs/2010ApJ...720.1118B} {720, 1118}

\bibitem[\protect\citeauthoryear{{Burke} et~al.,}{{Burke}
  et~al.}{2007}]{burke2007}
{Burke} C.~J.,  et~al., 2007, \mn@doi [\apj] {10.1086/523087}, \href
  {https://ui.adsabs.harvard.edu/abs/2007ApJ...671.2115B} {671, 2115}

\bibitem[\protect\citeauthoryear{{C{\'a}ceres} et~al.,}{{C{\'a}ceres}
  et~al.}{2014}]{caceres2014}
{C{\'a}ceres} C.,  et~al., 2014, \mn@doi [\aap] {10.1051/0004-6361/201321087},
  \href {https://ui.adsabs.harvard.edu/abs/2014A&A...565A...7C} {565, A7}

\bibitem[\protect\citeauthoryear{{Canto Martins} et~al.,}{{Canto Martins}
  et~al.}{2020}]{martins2020}
{Canto Martins} B.~L.,  et~al., 2020, \mn@doi [\apjs]
  {10.3847/1538-4365/aba73f}, \href
  {https://ui.adsabs.harvard.edu/abs/2020ApJS..250...20C} {250, 20}

\bibitem[\protect\citeauthoryear{{Cauley}, {Redfield}  \& {Jensen}}{{Cauley}
  et~al.}{2017}]{2017AJ....153...81C}
{Cauley} P.~W.,  {Redfield} S.,   {Jensen} A.~G.,  2017, \mn@doi [\aj]
  {10.3847/1538-3881/153/2/81}, \href
  {https://ui.adsabs.harvard.edu/abs/2017AJ....153...81C} {153, 81}

\bibitem[\protect\citeauthoryear{{Charbonneau}, {Brown}, {Latham}  \&
  {Mayor}}{{Charbonneau} et~al.}{2000}]{Charbonneau2000}
{Charbonneau} D.,  {Brown} T.~M.,  {Latham} D.~W.,   {Mayor} M.,  2000, \mn@doi
  [\apjl] {10.1086/312457}, \href
  {https://ui.adsabs.harvard.edu/abs/2000ApJ...529L..45C} {529, L45}

\bibitem[\protect\citeauthoryear{{Charbonneau} et~al.,}{{Charbonneau}
  et~al.}{2009}]{2009Natur.462..891C}
{Charbonneau} D.,  et~al., 2009, \mn@doi [\nat] {10.1038/nature08679}, \href
  {https://ui.adsabs.harvard.edu/abs/2009Natur.462..891C} {462, 891}

\bibitem[\protect\citeauthoryear{{Ciceri} et~al.,}{{Ciceri}
  et~al.}{2013}]{ciceri2013}
{Ciceri} S.,  et~al., 2013, \mn@doi [\aap] {10.1051/0004-6361/201321669}, \href
  {https://ui.adsabs.harvard.edu/abs/2013A&A...557A..30C} {557, A30}

\bibitem[\protect\citeauthoryear{{Claret} \& {Bloemen}}{{Claret} \&
  {Bloemen}}{2011}]{2011A&A...529A..75C}
{Claret} A.,  {Bloemen} S.,  2011, \mn@doi [\aap]
  {10.1051/0004-6361/201116451}, \href
  {https://ui.adsabs.harvard.edu/abs/2011A&A...529A..75C} {529, A75}

\bibitem[\protect\citeauthoryear{{Cloutier}, {Charbonneau}, {Deming}, {Bonfils}
   \& {Astudillo-Defru}}{{Cloutier} et~al.}{2021}]{clotier2012}
{Cloutier} R.,  {Charbonneau} D.,  {Deming} D.,  {Bonfils} X.,
  {Astudillo-Defru} N.,  2021, \mn@doi [\aj] {10.3847/1538-3881/ac1584}, \href
  {https://ui.adsabs.harvard.edu/abs/2021AJ....162..174C} {162, 174}

\bibitem[\protect\citeauthoryear{{Collins}, {Kielkopf}, {Stassun}  \&
  {Hessman}}{{Collins} et~al.}{2017}]{collins2017}
{Collins} K.~A.,  {Kielkopf} J.~F.,  {Stassun} K.~G.,   {Hessman} F.~V.,  2017,
  \mn@doi [\aj] {10.3847/1538-3881/153/2/77}, \href
  {https://ui.adsabs.harvard.edu/abs/2017AJ....153...77C} {153, 77}

\bibitem[\protect\citeauthoryear{{Corrales}, {Ravi}, {King}, {May}, {Rauscher}
  \& {Reynolds}}{{Corrales} et~al.}{2021}]{2021AJ....162..287C}
{Corrales} L.,  {Ravi} S.,  {King} G.~W.,  {May} E.,  {Rauscher} E.,
  {Reynolds} M.,  2021, \mn@doi [\aj] {10.3847/1538-3881/ac2c67}, \href
  {https://ui.adsabs.harvard.edu/abs/2021AJ....162..287C} {162, 287}

\bibitem[\protect\citeauthoryear{{Cort{\'e}s-Zuleta}, {Rojo}, {Wang}, {Hinse},
  {Hoyer}, {Sanhueza}, {Correa-Amaro}  \& {Albornoz}}{{Cort{\'e}s-Zuleta}
  et~al.}{2020}]{2020A&A...636A..98C}
{Cort{\'e}s-Zuleta} P.,  {Rojo} P.,  {Wang} S.,  {Hinse} T.~C.,  {Hoyer} S.,
  {Sanhueza} B.,  {Correa-Amaro} P.,   {Albornoz} J.,  2020, \mn@doi [\aap]
  {10.1051/0004-6361/201936279}, \href
  {https://ui.adsabs.harvard.edu/abs/2020A&A...636A..98C} {636, A98}

\bibitem[\protect\citeauthoryear{{Counselman}}{{Counselman}}{1973}]{1973ApJ...180..307C}
{Counselman} Charles~C. I.,  1973, \mn@doi [\apj] {10.1086/151964}, \href
  {https://ui.adsabs.harvard.edu/abs/1973ApJ...180..307C} {180, 307}

\bibitem[\protect\citeauthoryear{{Crouzet}, {McCullough}, {Burke}  \&
  {Long}}{{Crouzet} et~al.}{2012}]{2012ApJ...761....7C}
{Crouzet} N.,  {McCullough} P.~R.,  {Burke} C.,   {Long} D.,  2012, \mn@doi
  [\apj] {10.1088/0004-637X/761/1/7}, \href
  {https://ui.adsabs.harvard.edu/abs/2012ApJ...761....7C} {761, 7}

\bibitem[\protect\citeauthoryear{{Dai}, {Winn}, {Yu}  \& {Albrecht}}{{Dai}
  et~al.}{2017}]{dai2017}
{Dai} F.,  {Winn} J.~N.,  {Yu} L.,   {Albrecht} S.,  2017, \mn@doi [\aj]
  {10.3847/1538-3881/153/1/40}, \href
  {https://ui.adsabs.harvard.edu/abs/2017AJ....153...40D} {153, 40}

\bibitem[\protect\citeauthoryear{{Damasso} et~al.,}{{Damasso}
  et~al.}{2015}]{2015A&A...575A.111D}
{Damasso} M.,  et~al., 2015, \mn@doi [\aap] {10.1051/0004-6361/201425332},
  \href {https://ui.adsabs.harvard.edu/abs/2015A&A...575A.111D} {575, A111}

\bibitem[\protect\citeauthoryear{{Desidera} et~al.,}{{Desidera}
  et~al.}{2014}]{desidera2014}
{Desidera} S.,  et~al., 2014, \mn@doi [\aap] {10.1051/0004-6361/201424339},
  \href {https://ui.adsabs.harvard.edu/abs/2014A&A...567L...6D} {567, L6}

\bibitem[\protect\citeauthoryear{{Eastman}, {Gaudi}  \& {Agol}}{{Eastman}
  et~al.}{2013}]{2013PASP..125...83E}
{Eastman} J.,  {Gaudi} B.~S.,   {Agol} E.,  2013, \mn@doi [\pasp]
  {10.1086/669497}, \href
  {https://ui.adsabs.harvard.edu/abs/2013PASP..125...83E} {125, 83}

\bibitem[\protect\citeauthoryear{{Enoch} et~al.,}{{Enoch}
  et~al.}{2011}]{enoch2011}
{Enoch} B.,  et~al., 2011, \mn@doi [\aj] {10.1088/0004-6256/142/3/86}, \href
  {https://ui.adsabs.harvard.edu/abs/2011AJ....142...86E} {142, 86}

\bibitem[\protect\citeauthoryear{{Esposito} et~al.,}{{Esposito}
  et~al.}{2017}]{esposito2017}
{Esposito} M.,  et~al., 2017, \mn@doi [\aap] {10.1051/0004-6361/201629720},
  \href {https://ui.adsabs.harvard.edu/abs/2017A&A...601A..53E} {601, A53}

\bibitem[\protect\citeauthoryear{{Essick} \& {Weinberg}}{{Essick} \&
  {Weinberg}}{2016}]{2016ApJ...816...18E}
{Essick} R.,  {Weinberg} N.~N.,  2016, \mn@doi [\apj]
  {10.3847/0004-637X/816/1/18}, \href
  {https://ui.adsabs.harvard.edu/abs/2016ApJ...816...18E} {816, 18}

\bibitem[\protect\citeauthoryear{{Fernandez}, {Holman}, {Winn}, {Torres},
  {Shporer}, {Mazeh}, {Esquerdo}  \& {Everett}}{{Fernandez}
  et~al.}{2009}]{2009AJ....137.4911F}
{Fernandez} J.~M.,  {Holman} M.~J.,  {Winn} J.~N.,  {Torres} G.,  {Shporer} A.,
   {Mazeh} T.,  {Esquerdo} G.~A.,   {Everett} M.~E.,  2009, \mn@doi [\aj]
  {10.1088/0004-6256/137/6/4911}, \href
  {https://ui.adsabs.harvard.edu/abs/2009AJ....137.4911F} {137, 4911}

\bibitem[\protect\citeauthoryear{{Foreman-Mackey}, {Hogg}, {Lang}  \&
  {Goodman}}{{Foreman-Mackey} et~al.}{2013}]{2013PASP..125..306F}
{Foreman-Mackey} D.,  {Hogg} D.~W.,  {Lang} D.,   {Goodman} J.,  2013, \mn@doi
  [\pasp] {10.1086/670067}, \href
  {https://ui.adsabs.harvard.edu/abs/2013PASP..125..306F} {125, 306}

\bibitem[\protect\citeauthoryear{{Fulton}, {Shporer}, {Winn}, {Holman},
  {P{\'a}l}  \& {Gazak}}{{Fulton} et~al.}{2011a}]{fulton2011}
{Fulton} B.~J.,  {Shporer} A.,  {Winn} J.~N.,  {Holman} M.~J.,  {P{\'a}l} A.,
  {Gazak} J.~Z.,  2011a, \mn@doi [\aj] {10.1088/0004-6256/142/3/84}, \href
  {https://ui.adsabs.harvard.edu/abs/2011AJ....142...84F} {142, 84}

\bibitem[\protect\citeauthoryear{{Fulton}, {Shporer}, {Winn}, {Holman},
  {P{\'a}l}  \& {Gazak}}{{Fulton} et~al.}{2011b}]{fulton2011h13}
{Fulton} B.~J.,  {Shporer} A.,  {Winn} J.~N.,  {Holman} M.~J.,  {P{\'a}l} A.,
  {Gazak} J.~Z.,  2011b, \mn@doi [\aj] {10.1088/0004-6256/142/3/84}, \href
  {https://ui.adsabs.harvard.edu/abs/2011AJ....142...84F} {142, 84}

\bibitem[\protect\citeauthoryear{{Gajdo{\v{s}}} et~al.,}{{Gajdo{\v{s}}}
  et~al.}{2019}]{2019MNRAS.485.3580G}
{Gajdo{\v{s}}} P.,  et~al., 2019, \mn@doi [\mnras] {10.1093/mnras/stz676},
  \href {https://ui.adsabs.harvard.edu/abs/2019MNRAS.485.3580G} {485, 3580}

\bibitem[\protect\citeauthoryear{{Gallet}}{{Gallet}}{2020}]{gallet2020}
{Gallet} F.,  2020, \mn@doi [\aap] {10.1051/0004-6361/202038058}, \href
  {https://ui.adsabs.harvard.edu/abs/2020A&A...641A..38G} {641, A38}

\bibitem[\protect\citeauthoryear{{Gao} et~al.,}{{Gao}
  et~al.}{2023}]{2023ApJ...951...96G}
{Gao} P.,  et~al., 2023, \mn@doi [\apj] {10.3847/1538-4357/acd16f}, \href
  {https://ui.adsabs.harvard.edu/abs/2023ApJ...951...96G} {951, 96}

\bibitem[\protect\citeauthoryear{{Gillon}, {Pont}, {Moutou}, {Bouchy},
  {Courbin}, {Sohy}  \& {Magain}}{{Gillon} et~al.}{2006}]{gillon2006}
{Gillon} M.,  {Pont} F.,  {Moutou} C.,  {Bouchy} F.,  {Courbin} F.,  {Sohy} S.,
    {Magain} P.,  2006, \mn@doi [\aap] {10.1051/0004-6361:20065844}, \href
  {https://ui.adsabs.harvard.edu/abs/2006A&A...459..249G} {459, 249}

\bibitem[\protect\citeauthoryear{{Gillon} et~al.,}{{Gillon}
  et~al.}{2011}]{2011A&A...533A..88G}
{Gillon} M.,  et~al., 2011, \mn@doi [\aap] {10.1051/0004-6361/201117198}, \href
  {https://ui.adsabs.harvard.edu/abs/2011A&A...533A..88G} {533, A88}

\bibitem[\protect\citeauthoryear{{Gillon} et~al.,}{{Gillon}
  et~al.}{2014}]{gillon2014}
{Gillon} M.,  et~al., 2014, \mn@doi [\aap] {10.1051/0004-6361/201322362}, \href
  {https://ui.adsabs.harvard.edu/abs/2014A&A...563A..21G} {563, A21}

\bibitem[\protect\citeauthoryear{{Gillon} et~al.,}{{Gillon}
  et~al.}{2017}]{2017Natur.542..456G}
{Gillon} M.,  et~al., 2017, \mn@doi [\nat] {10.1038/nature21360}, \href
  {https://ui.adsabs.harvard.edu/abs/2017Natur.542..456G} {542, 456}

\bibitem[\protect\citeauthoryear{{Gim{\'e}nez} \& {Bastero}}{{Gim{\'e}nez} \&
  {Bastero}}{1995}]{1995Ap&SS.226...99G}
{Gim{\'e}nez} A.,  {Bastero} M.,  1995, \mn@doi [\apss] {10.1007/BF00626903},
  \href {https://ui.adsabs.harvard.edu/abs/1995Ap&SS.226...99G} {226, 99}

\bibitem[\protect\citeauthoryear{{Goldreich} \& {Soter}}{{Goldreich} \&
  {Soter}}{1966}]{1966Icar....5..375G}
{Goldreich} P.,  {Soter} S.,  1966, \mn@doi [\icarus]
  {10.1016/0019-1035(66)90051-0}, \href
  {https://ui.adsabs.harvard.edu/abs/1966Icar....5..375G} {5, 375}

\bibitem[\protect\citeauthoryear{{Harps{\o}e} et~al.,}{{Harps{\o}e}
  et~al.}{2013}]{harpsoe2013}
{Harps{\o}e} K.~B.~W.,  et~al., 2013, \mn@doi [\aap]
  {10.1051/0004-6361/201219996}, \href
  {https://ui.adsabs.harvard.edu/abs/2013A&A...549A..10H} {549, A10}

\bibitem[\protect\citeauthoryear{{Harre} et~al.,}{{Harre}
  et~al.}{2023}]{2023A&A...669A.124H}
{Harre} J.~V.,  et~al., 2023, \mn@doi [\aap] {10.1051/0004-6361/202244529},
  \href {https://ui.adsabs.harvard.edu/abs/2023A&A...669A.124H} {669, A124}

\bibitem[\protect\citeauthoryear{{Hartman} et~al.,}{{Hartman}
  et~al.}{2015}]{hartman2015}
{Hartman} J.~D.,  et~al., 2015, \mn@doi [\aj] {10.1088/0004-6256/150/6/168},
  \href {https://ui.adsabs.harvard.edu/abs/2015AJ....150..168H} {150, 168}

\bibitem[\protect\citeauthoryear{{Hay} et~al.,}{{Hay} et~al.}{2016}]{hay2016}
{Hay} K.~L.,  et~al., 2016, \mn@doi [\mnras] {10.1093/mnras/stw2090}, \href
  {https://ui.adsabs.harvard.edu/abs/2016MNRAS.463.3276H} {463, 3276}

\bibitem[\protect\citeauthoryear{{Hinse}, {Han}, {Yoon}, {Lee}, {Kim}  \&
  {Kim}}{{Hinse} et~al.}{2015}]{hinse2015}
{Hinse} T.~C.,  {Han} W.,  {Yoon} J.-N.,  {Lee} C.-U.,  {Kim} Y.-G.,   {Kim}
  C.-H.,  2015, \mn@doi [Journal of Astronomy and Space Sciences]
  {10.5140/JASS.2015.32.1.21}, \href
  {https://ui.adsabs.harvard.edu/abs/2015JASS...32...21H} {32, 21}

\bibitem[\protect\citeauthoryear{{Howell} et~al.,}{{Howell}
  et~al.}{2014}]{keplerk2}
{Howell} S.~B.,  et~al., 2014, \mn@doi [\pasp] {10.1086/676406}, \href
  {https://ui.adsabs.harvard.edu/abs/2014PASP..126..398H} {126, 398}

\bibitem[\protect\citeauthoryear{{Ivanov}, {Papaloizou}  \& {Chernov}}{{Ivanov}
  et~al.}{2013}]{2013MNRAS.432.2339I}
{Ivanov} P.~B.,  {Papaloizou} J.~C.~B.,   {Chernov} S.~V.,  2013, \mn@doi
  [\mnras] {10.1093/mnras/stt595}, \href
  {https://ui.adsabs.harvard.edu/abs/2013MNRAS.432.2339I} {432, 2339}

\bibitem[\protect\citeauthoryear{{Jenkins} et~al.,}{{Jenkins}
  et~al.}{2016}]{jenkins2016}
{Jenkins} J.~M.,  et~al., 2016, in {Chiozzi} G.,  {Guzman} J.~C.,  eds,
  Society of Photo-Optical Instrumentation Engineers (SPIE) Conference Series
  Vol. 9913, Software and Cyberinfrastructure for Astronomy IV. p. 99133E,
  \mn@doi{10.1117/12.2233418}

\bibitem[\protect\citeauthoryear{{Johnson} et~al.,}{{Johnson}
  et~al.}{2008}]{johnson2008}
{Johnson} J.~A.,  et~al., 2008, \mn@doi [\apj] {10.1086/591078}, \href
  {https://ui.adsabs.harvard.edu/abs/2008ApJ...686..649J} {686, 649}

\bibitem[\protect\citeauthoryear{{Johnson} et~al.,}{{Johnson}
  et~al.}{2011}]{2011ApJ...735...24J}
{Johnson} J.~A.,  et~al., 2011, \mn@doi [\apj] {10.1088/0004-637X/735/1/24},
  \href {https://ui.adsabs.harvard.edu/abs/2011ApJ...735...24J} {735, 24}

\bibitem[\protect\citeauthoryear{{Kempton} et~al.,}{{Kempton}
  et~al.}{2018}]{kempton2018}
{Kempton} E. M.~R.,  et~al., 2018, \mn@doi [\pasp] {10.1088/1538-3873/aadf6f},
  \href {https://ui.adsabs.harvard.edu/abs/2018PASP..130k4401K} {130, 114401}

\bibitem[\protect\citeauthoryear{{Kjurkchieva}, {Petrov}, {Ibryamov}, {Nikolov}
   \& {Popov}}{{Kjurkchieva} et~al.}{2018}]{kjur2018}
{Kjurkchieva} D.,  {Petrov} N.,  {Ibryamov} S.,  {Nikolov} G.,   {Popov} V.,
  2018, \mn@doi [Serbian Astronomical Journal] {10.2298/SAJ1896015K}, \href
  {https://ui.adsabs.harvard.edu/abs/2018SerAJ.196...15K} {196, 15}

\bibitem[\protect\citeauthoryear{{Knutson}, {Howard}  \& {Isaacson}}{{Knutson}
  et~al.}{2010}]{2010ApJ...720.1569K}
{Knutson} H.~A.,  {Howard} A.~W.,   {Isaacson} H.,  2010, \mn@doi [\apj]
  {10.1088/0004-637X/720/2/1569}, \href
  {https://ui.adsabs.harvard.edu/abs/2010ApJ...720.1569K} {720, 1569}

\bibitem[\protect\citeauthoryear{{Knutson} et~al.,}{{Knutson}
  et~al.}{2014}]{knutson2014}
{Knutson} H.~A.,  et~al., 2014, \mn@doi [\apj] {10.1088/0004-637X/785/2/126},
  \href {https://ui.adsabs.harvard.edu/abs/2014ApJ...785..126K} {785, 126}

\bibitem[\protect\citeauthoryear{{Kundurthy}, {Agol}, {Becker}, {Barnes},
  {Williams}  \& {Mukadam}}{{Kundurthy} et~al.}{2011}]{kundurthy2011}
{Kundurthy} P.,  {Agol} E.,  {Becker} A.~C.,  {Barnes} R.,  {Williams} B.,
  {Mukadam} A.,  2011, \mn@doi [\apj] {10.1088/0004-637X/731/2/123}, \href
  {https://ui.adsabs.harvard.edu/abs/2011ApJ...731..123K} {731, 123}

\bibitem[\protect\citeauthoryear{{Lamp{\'o}n} et~al.,}{{Lamp{\'o}n}
  et~al.}{2023}]{2023A&A...673A.140L}
{Lamp{\'o}n} M.,  et~al., 2023, \mn@doi [\aap] {10.1051/0004-6361/202245649},
  \href {https://ui.adsabs.harvard.edu/abs/2023A&A...673A.140L} {673, A140}

\bibitem[\protect\citeauthoryear{{Lightkurve Collaboration}
  et~al.,}{{Lightkurve Collaboration} et~al.}{2018}]{lightkurve2018}
{Lightkurve Collaboration} et~al., 2018, {Lightkurve: Kepler and TESS time
  series analysis in Python}, Astrophysics Source Code Library (\mn@eprint
  {ascl} {1812.013})

\bibitem[\protect\citeauthoryear{{Line} et~al.,}{{Line}
  et~al.}{2021}]{2021Natur.598..580L}
{Line} M.~R.,  et~al., 2021, \mn@doi [\nat] {10.1038/s41586-021-03912-6}, \href
  {https://ui.adsabs.harvard.edu/abs/2021Natur.598..580L} {598, 580}

\bibitem[\protect\citeauthoryear{{Maciejewski} et~al.,}{{Maciejewski}
  et~al.}{2016a}]{maciejewski2016}
{Maciejewski} G.,  et~al., 2016a, \mn@doi [\actaa] {10.48550/arXiv.1603.03268},
  \href {https://ui.adsabs.harvard.edu/abs/2016AcA....66...55M} {66, 55}

\bibitem[\protect\citeauthoryear{{Maciejewski} et~al.,}{{Maciejewski}
  et~al.}{2016b}]{maci2016}
{Maciejewski} G.,  et~al., 2016b, \mn@doi [\aap] {10.1051/0004-6361/201628312},
  \href {https://ui.adsabs.harvard.edu/abs/2016A&A...588L...6M} {588, L6}

\bibitem[\protect\citeauthoryear{{Maciejewski}, {Stangret}, {Ohlert},
  {Basaran}, {Maciejczak}, {Puciata-Mroczynska}  \& {Boulanger}}{{Maciejewski}
  et~al.}{2018}]{2018IBVS.6243....1M}
{Maciejewski} G.,  {Stangret} M.,  {Ohlert} J.,  {Basaran} C.~S.,  {Maciejczak}
  J.,  {Puciata-Mroczynska} M.,   {Boulanger} E.,  2018, \mn@doi [Information
  Bulletin on Variable Stars] {10.22444/IBVS.6243}, \href
  {https://ui.adsabs.harvard.edu/abs/2018IBVS.6243....1M} {6243, 1}

\bibitem[\protect\citeauthoryear{{Maguire}, {Gibson}, {Nugroho}, {Ramkumar},
  {Fortune}, {Merritt}  \& {de Mooij}}{{Maguire} et~al.}{2022}]{maguire2022}
{Maguire} C.,  {Gibson} N.~P.,  {Nugroho} S.~K.,  {Ramkumar} S.,  {Fortune} M.,
   {Merritt} S.~R.,   {de Mooij} E.,  2022, arXiv e-prints, \href
  {https://ui.adsabs.harvard.edu/abs/2022arXiv221109621M} {p. arXiv:2211.09621}

\bibitem[\protect\citeauthoryear{{Mallonn} et~al.,}{{Mallonn}
  et~al.}{2018}]{2018A&A...614A..35M}
{Mallonn} M.,  et~al., 2018, \mn@doi [\aap] {10.1051/0004-6361/201732300},
  \href {https://ui.adsabs.harvard.edu/abs/2018A&A...614A..35M} {614, A35}

\bibitem[\protect\citeauthoryear{{Mallonn} et~al.,}{{Mallonn}
  et~al.}{2019}]{mallonn2019}
{Mallonn} M.,  et~al., 2019, \mn@doi [\aap] {10.1051/0004-6361/201834194},
  \href {https://ui.adsabs.harvard.edu/abs/2019A&A...622A..81M} {622, A81}

\bibitem[\protect\citeauthoryear{{Mancini} et~al.,}{{Mancini}
  et~al.}{2013}]{2013MNRAS.430.2932M}
{Mancini} L.,  et~al., 2013, \mn@doi [\mnras] {10.1093/mnras/stt095}, \href
  {https://ui.adsabs.harvard.edu/abs/2013MNRAS.430.2932M} {430, 2932}

\bibitem[\protect\citeauthoryear{{Mancini} et~al.,}{{Mancini}
  et~al.}{2014}]{mancini2014}
{Mancini} L.,  et~al., 2014, \mn@doi [\mnras] {10.1093/mnras/stu1286}, \href
  {https://ui.adsabs.harvard.edu/abs/2014MNRAS.443.2391M} {443, 2391}

\bibitem[\protect\citeauthoryear{{Mancini} et~al.,}{{Mancini}
  et~al.}{2015}]{mancini2015}
{Mancini} L.,  et~al., 2015, \mn@doi [\aap] {10.1051/0004-6361/201526030},
  \href {https://ui.adsabs.harvard.edu/abs/2015A&A...579A.136M} {579, A136}

\bibitem[\protect\citeauthoryear{{Mancini} et~al.,}{{Mancini}
  et~al.}{2017}]{mancini2017}
{Mancini} L.,  et~al., 2017, \mn@doi [\mnras] {10.1093/mnras/stw1987}, \href
  {https://ui.adsabs.harvard.edu/abs/2017MNRAS.465..843M} {465, 843}

\bibitem[\protect\citeauthoryear{{Mancini} et~al.,}{{Mancini}
  et~al.}{2018}]{mancini2018}
{Mancini} L.,  et~al., 2018, \mn@doi [\aap] {10.1051/0004-6361/201732234},
  \href {https://ui.adsabs.harvard.edu/abs/2018A&A...613A..41M} {613, A41}

\bibitem[\protect\citeauthoryear{{Mancini} et~al.,}{{Mancini}
  et~al.}{2022}]{mancini2022}
{Mancini} L.,  et~al., 2022, \mn@doi [\aap] {10.1051/0004-6361/202243742},
  \href {https://ui.adsabs.harvard.edu/abs/2022A&A...664A.162M} {664, A162}

\bibitem[\protect\citeauthoryear{{Mansfield} et~al.,}{{Mansfield}
  et~al.}{2022}]{2022AJ....163..261M}
{Mansfield} M.,  et~al., 2022, \mn@doi [\aj] {10.3847/1538-3881/ac658f}, \href
  {https://ui.adsabs.harvard.edu/abs/2022AJ....163..261M} {163, 261}

\bibitem[\protect\citeauthoryear{{Maxted} et~al.,}{{Maxted}
  et~al.}{2013}]{2013PASP..125...48M}
{Maxted} P.~F.~L.,  et~al., 2013, \mn@doi [\pasp] {10.1086/669231}, \href
  {https://ui.adsabs.harvard.edu/abs/2013PASP..125...48M} {125, 48}

\bibitem[\protect\citeauthoryear{{Ment}, {Fischer}, {Bakos}, {Howard}  \&
  {Isaacson}}{{Ment} et~al.}{2018}]{kristo2018}
{Ment} K.,  {Fischer} D.~A.,  {Bakos} G.,  {Howard} A.~W.,   {Isaacson} H.,
  2018, \mn@doi [\aj] {10.3847/1538-3881/aae1f5}, \href
  {https://ui.adsabs.harvard.edu/abs/2018AJ....156..213M} {156, 213}

\bibitem[\protect\citeauthoryear{{Montalto}, {Iro}, {Santos}, {Desidera},
  {Martins}, {Figueira}  \& {Alonso}}{{Montalto} et~al.}{2015}]{montalto2015}
{Montalto} M.,  {Iro} N.,  {Santos} N.~C.,  {Desidera} S.,  {Martins} J.~H.~C.,
   {Figueira} P.,   {Alonso} R.,  2015, \mn@doi [\apj]
  {10.1088/0004-637X/811/1/55}, \href
  {https://ui.adsabs.harvard.edu/abs/2015ApJ...811...55M} {811, 55}

\bibitem[\protect\citeauthoryear{{Moutou} et~al.,}{{Moutou}
  et~al.}{2011}]{moutou2011}
{Moutou} C.,  et~al., 2011, \mn@doi [\aap] {10.1051/0004-6361/201116760}, \href
  {https://ui.adsabs.harvard.edu/abs/2011A&A...533A.113M} {533, A113}

\bibitem[\protect\citeauthoryear{{Mo{\v{c}}nik}, {Southworth}  \&
  {Hellier}}{{Mo{\v{c}}nik} et~al.}{2017}]{mocnik2017}
{Mo{\v{c}}nik} T.,  {Southworth} J.,   {Hellier} C.,  2017, \mn@doi [\mnras]
  {10.1093/mnras/stx1557}, \href
  {https://ui.adsabs.harvard.edu/abs/2017MNRAS.471..394M} {471, 394}

\bibitem[\protect\citeauthoryear{{Narita}, {Hirano}, {Sato}, {Harakawa},
  {Fukui}, {Aoki}  \& {Tamura}}{{Narita} et~al.}{2011}]{narita2011}
{Narita} N.,  {Hirano} T.,  {Sato} B.,  {Harakawa} H.,  {Fukui} A.,  {Aoki} W.,
    {Tamura} M.,  2011, \mn@doi [\pasj] {10.1093/pasj/63.6.L67}, \href
  {https://ui.adsabs.harvard.edu/abs/2011PASJ...63L..67N} {63, L67}

\bibitem[\protect\citeauthoryear{{Narita} et~al.,}{{Narita}
  et~al.}{2013}]{narita2013}
{Narita} N.,  et~al., 2013, \mn@doi [\apj] {10.1088/0004-637X/773/2/144}, \href
  {https://ui.adsabs.harvard.edu/abs/2013ApJ...773..144N} {773, 144}

\bibitem[\protect\citeauthoryear{{Nascimbeni}, {Piotto}, {Bedin}, {Damasso},
  {Malavolta}  \& {Borsato}}{{Nascimbeni} et~al.}{2011}]{nascimbeni2011}
{Nascimbeni} V.,  {Piotto} G.,  {Bedin} L.~R.,  {Damasso} M.,  {Malavolta} L.,
   {Borsato} L.,  2011, \mn@doi [\aap] {10.1051/0004-6361/201116830}, \href
  {https://ui.adsabs.harvard.edu/abs/2011A&A...532A..24N} {532, A24}

\bibitem[\protect\citeauthoryear{{Nascimbeni} et~al.,}{{Nascimbeni}
  et~al.}{2015}]{nascimbeni2015}
{Nascimbeni} V.,  et~al., 2015, \mn@doi [\aap] {10.1051/0004-6361/201425350},
  \href {https://ui.adsabs.harvard.edu/abs/2015A&A...579A.113N} {579, A113}

\bibitem[\protect\citeauthoryear{Nelder \& Mead}{Nelder \&
  Mead}{1965}]{10.1093/comjnl/7.4.308}
Nelder J.~A.,  Mead R.,  1965, \mn@doi [The Computer Journal]
  {10.1093/comjnl/7.4.308}, 7, 308

\bibitem[\protect\citeauthoryear{{Ngo} et~al.,}{{Ngo} et~al.}{2015}]{ngo2_2015}
{Ngo} H.,  et~al., 2015, \mn@doi [\apj] {10.1088/0004-637X/800/2/138}, \href
  {https://ui.adsabs.harvard.edu/abs/2015ApJ...800..138N} {800, 138}

\bibitem[\protect\citeauthoryear{{Ngo} et~al.,}{{Ngo} et~al.}{2016}]{ngo2016}
{Ngo} H.,  et~al., 2016, \mn@doi [\apj] {10.3847/0004-637X/827/1/8}, \href
  {https://ui.adsabs.harvard.edu/abs/2016ApJ...827....8N} {827, 8}

\bibitem[\protect\citeauthoryear{{Noyes}, {Hartmann}, {Baliunas}, {Duncan}  \&
  {Vaughan}}{{Noyes} et~al.}{1984}]{1984ApJ...279..763N}
{Noyes} R.~W.,  {Hartmann} L.~W.,  {Baliunas} S.~L.,  {Duncan} D.~K.,
  {Vaughan} A.~H.,  1984, \mn@doi [\apj] {10.1086/161945}, \href
  {https://ui.adsabs.harvard.edu/abs/1984ApJ...279..763N} {279, 763}

\bibitem[\protect\citeauthoryear{{Orell-Miquel} et~al.,}{{Orell-Miquel}
  et~al.}{2022}]{2022A&A...659A..55O}
{Orell-Miquel} J.,  et~al., 2022, \mn@doi [\aap] {10.1051/0004-6361/202142455},
  \href {https://ui.adsabs.harvard.edu/abs/2022A&A...659A..55O} {659, A55}

\bibitem[\protect\citeauthoryear{{P{\'a}l}, {S{\'a}rneczky}, {Szab{\'o}},
  {Szing}, {Kiss}, {Mez{\H{o}}}  \& {Reg{\'a}ly}}{{P{\'a}l}
  et~al.}{2011}]{pal2011}
{P{\'a}l} A.,  {S{\'a}rneczky} K.,  {Szab{\'o}} G.~M.,  {Szing} A.,  {Kiss}
  L.~L.,  {Mez{\H{o}}} G.,   {Reg{\'a}ly} Z.,  2011, \mn@doi [\mnras]
  {10.1111/j.1745-3933.2011.01029.x}, \href
  {https://ui.adsabs.harvard.edu/abs/2011MNRAS.413L..43P} {413, L43}

\bibitem[\protect\citeauthoryear{{Parviainen}, {Aigrain}, {Thatte}, {Barstow},
  {Evans}  \& {Gibson}}{{Parviainen} et~al.}{2015}]{2015MNRAS.453.3875P}
{Parviainen} H.,  {Aigrain} S.,  {Thatte} N.,  {Barstow} J.~K.,  {Evans} T.~M.,
    {Gibson} N.,  2015, \mn@doi [\mnras] {10.1093/mnras/stv1839}, \href
  {https://ui.adsabs.harvard.edu/abs/2015MNRAS.453.3875P} {453, 3875}

\bibitem[\protect\citeauthoryear{{Patra}, {Winn}, {Holman}, {Yu}, {Deming}  \&
  {Dai}}{{Patra} et~al.}{2017}]{2017AJ....154....4P}
{Patra} K.~C.,  {Winn} J.~N.,  {Holman} M.~J.,  {Yu} L.,  {Deming} D.,   {Dai}
  F.,  2017, \mn@doi [\aj] {10.3847/1538-3881/aa6d75}, \href
  {https://ui.adsabs.harvard.edu/abs/2017AJ....154....4P} {154, 4}

\bibitem[\protect\citeauthoryear{{Pearson}, {Turner}  \& {Sagan}}{{Pearson}
  et~al.}{2014}]{2014NewA...27..102P}
{Pearson} K.~A.,  {Turner} J.~D.,   {Sagan} T.~G.,  2014, \mn@doi [\na]
  {10.1016/j.newast.2013.08.002}, \href
  {https://ui.adsabs.harvard.edu/abs/2014NewA...27..102P} {27, 102}

\bibitem[\protect\citeauthoryear{{Pepper} et~al.,}{{Pepper}
  et~al.}{2013}]{2013ApJ...773...64P}
{Pepper} J.,  et~al., 2013, \mn@doi [\apj] {10.1088/0004-637X/773/1/64}, \href
  {https://ui.adsabs.harvard.edu/abs/2013ApJ...773...64P} {773, 64}

\bibitem[\protect\citeauthoryear{{Pinsonneault}, {DePoy}  \&
  {Coffee}}{{Pinsonneault} et~al.}{2001}]{2001ApJ...556L..59P}
{Pinsonneault} M.~H.,  {DePoy} D.~L.,   {Coffee} M.,  2001, \mn@doi [\apjl]
  {10.1086/323531}, \href
  {https://ui.adsabs.harvard.edu/abs/2001ApJ...556L..59P} {556, L59}

\bibitem[\protect\citeauthoryear{{Piskorz}, {Knutson}, {Ngo}, {Muirhead},
  {Batygin}, {Crepp}, {Hinkley}  \& {Morton}}{{Piskorz}
  et~al.}{2015}]{friends3}
{Piskorz} D.,  {Knutson} H.~A.,  {Ngo} H.,  {Muirhead} P.~S.,  {Batygin} K.,
  {Crepp} J.~R.,  {Hinkley} S.,   {Morton} T.~D.,  2015, \mn@doi [\apj]
  {10.1088/0004-637X/814/2/148}, \href
  {https://ui.adsabs.harvard.edu/abs/2015ApJ...814..148P} {814, 148}

\bibitem[\protect\citeauthoryear{{Pont}, {Zucker}  \& {Queloz}}{{Pont}
  et~al.}{2006}]{pont2006}
{Pont} F.,  {Zucker} S.,   {Queloz} D.,  2006, \mn@doi [\mnras]
  {10.1111/j.1365-2966.2006.11012.x}, \href
  {https://ui.adsabs.harvard.edu/abs/2006MNRAS.373..231P} {373, 231}

\bibitem[\protect\citeauthoryear{{Queloz} et~al.,}{{Queloz}
  et~al.}{2010}]{2010A&A...517L...1Q}
{Queloz} D.,  et~al., 2010, \mn@doi [\aap] {10.1051/0004-6361/201014768}, \href
  {https://ui.adsabs.harvard.edu/abs/2010A&A...517L...1Q} {517, L1}

\bibitem[\protect\citeauthoryear{{Rackham} et~al.,}{{Rackham}
  et~al.}{2017}]{2017ApJ...834..151R}
{Rackham} B.,  et~al., 2017, \mn@doi [\apj] {10.3847/1538-4357/aa4f6c}, \href
  {https://ui.adsabs.harvard.edu/abs/2017ApJ...834..151R} {834, 151}

\bibitem[\protect\citeauthoryear{Raftery}{Raftery}{1995}]{e0988035-67ff-3d00-8d22-b00bd6518fb4}
Raftery A.~E.,  1995, Sociological Methodology, 25, 111

\bibitem[\protect\citeauthoryear{{Rasio}, {Tout}, {Lubow}  \& {Livio}}{{Rasio}
  et~al.}{1996}]{1996ApJ...470.1187R}
{Rasio} F.~A.,  {Tout} C.~A.,  {Lubow} S.~H.,   {Livio} M.,  1996, \mn@doi
  [\apj] {10.1086/177941}, \href
  {https://ui.adsabs.harvard.edu/abs/1996ApJ...470.1187R} {470, 1187}

\bibitem[\protect\citeauthoryear{{Reggiani}, {Schlaufman}, {Healy},
  {Lothringer}  \& {Sing}}{{Reggiani} et~al.}{2022}]{2022AJ....163..159R}
{Reggiani} H.,  {Schlaufman} K.~C.,  {Healy} B.~F.,  {Lothringer} J.~D.,
  {Sing} D.~K.,  2022, \mn@doi [\aj] {10.3847/1538-3881/ac4d9f}, \href
  {https://ui.adsabs.harvard.edu/abs/2022AJ....163..159R} {163, 159}

\bibitem[\protect\citeauthoryear{{Rein} \& {Liu}}{{Rein} \&
  {Liu}}{2012}]{rebound}
{Rein} H.,  {Liu} S.~F.,  2012, \mn@doi [\aap] {10.1051/0004-6361/201118085},
  \href {https://ui.adsabs.harvard.edu/abs/2012A&A...537A.128R} {537, A128}

\bibitem[\protect\citeauthoryear{{Rein} \& {Spiegel}}{{Rein} \&
  {Spiegel}}{2015}]{ias15}
{Rein} H.,  {Spiegel} D.~S.,  2015, \mn@doi [\mnras] {10.1093/mnras/stu2164},
  \href {https://ui.adsabs.harvard.edu/abs/2015MNRAS.446.1424R} {446, 1424}

\bibitem[\protect\citeauthoryear{{Ricker} et~al.,}{{Ricker}
  et~al.}{2015}]{ricker2015}
{Ricker} G.~R.,  et~al., 2015, \mn@doi [Journal of Astronomical Telescopes,
  Instruments, and Systems] {10.1117/1.JATIS.1.1.014003}, \href
  {https://ui.adsabs.harvard.edu/abs/2015JATIS...1a4003R} {1, 014003}

\bibitem[\protect\citeauthoryear{{Ridden-Harper}, {Turner}  \&
  {Jayawardhana}}{{Ridden-Harper} et~al.}{2020}]{2020AJ....160..249R}
{Ridden-Harper} A.,  {Turner} J.~D.,   {Jayawardhana} R.,  2020, \mn@doi [\aj]
  {10.3847/1538-3881/abba1e}, \href
  {https://ui.adsabs.harvard.edu/abs/2020AJ....160..249R} {160, 249}

\bibitem[\protect\citeauthoryear{{Sada}}{{Sada}}{2018}]{sada2018}
{Sada} P.~V.,  2018, in AAS/Division for Planetary Sciences Meeting Abstracts
  \#50. p. 413.01

\bibitem[\protect\citeauthoryear{{Sada} \& {Ram{\'o}n-Fox}}{{Sada} \&
  {Ram{\'o}n-Fox}}{2016}]{sada2016}
{Sada} P.~V.,  {Ram{\'o}n-Fox} F.~G.,  2016, \mn@doi [\pasp]
  {10.1088/1538-3873/128/960/024402}, \href
  {https://ui.adsabs.harvard.edu/abs/2016PASP..128b4402S} {128, 024402}

\bibitem[\protect\citeauthoryear{{Sada} et~al.,}{{Sada}
  et~al.}{2012}]{sada2012}
{Sada} P.~V.,  et~al., 2012, \mn@doi [\pasp] {10.1086/665043}, \href
  {https://ui.adsabs.harvard.edu/abs/2012PASP..124..212S} {124, 212}

\bibitem[\protect\citeauthoryear{{Sing} et~al.,}{{Sing}
  et~al.}{2012}]{sing2012}
{Sing} D.~K.,  et~al., 2012, \mn@doi [\mnras]
  {10.1111/j.1365-2966.2012.21938.x}, \href
  {https://ui.adsabs.harvard.edu/abs/2012MNRAS.426.1663S} {426, 1663}

\bibitem[\protect\citeauthoryear{{Southworth} et~al.,}{{Southworth}
  et~al.}{2009}]{southworth2009}
{Southworth} J.,  et~al., 2009, \mn@doi [\mnras]
  {10.1111/j.1365-2966.2009.14767.x}, \href
  {https://ui.adsabs.harvard.edu/abs/2009MNRAS.396.1023S} {396, 1023}

\bibitem[\protect\citeauthoryear{{Southworth}, {Bruni}, {Mancini}  \&
  {Gregorio}}{{Southworth} et~al.}{2012}]{southworth2012}
{Southworth} J.,  {Bruni} I.,  {Mancini} L.,   {Gregorio} J.,  2012, \mn@doi
  [\mnras] {10.1111/j.1365-2966.2011.20230.x}, \href
  {https://ui.adsabs.harvard.edu/abs/2012MNRAS.420.2580S} {420, 2580}

\bibitem[\protect\citeauthoryear{{Spake} et~al.,}{{Spake}
  et~al.}{2022}]{2022ApJ...939L..11S}
{Spake} J.~J.,  et~al., 2022, \mn@doi [\apjl] {10.3847/2041-8213/ac88c9}, \href
  {https://ui.adsabs.harvard.edu/abs/2022ApJ...939L..11S} {939, L11}

\bibitem[\protect\citeauthoryear{{Su{\'a}rez Mascare{\~n}o}, {Rebolo},
  {Gonz{\'a}lez Hern{\'a}ndez}  \& {Esposito}}{{Su{\'a}rez Mascare{\~n}o}
  et~al.}{2015}]{2015MNRAS.452.2745S}
{Su{\'a}rez Mascare{\~n}o} A.,  {Rebolo} R.,  {Gonz{\'a}lez Hern{\'a}ndez}
  J.~I.,   {Esposito} M.,  2015, \mn@doi [\mnras] {10.1093/mnras/stv1441},
  \href {https://ui.adsabs.harvard.edu/abs/2015MNRAS.452.2745S} {452, 2745}

\bibitem[\protect\citeauthoryear{{Sun}, {Gu}, {Wang}, {Bai}, {Schmitt},
  {Perdelwitz}  \& {Ioannidis}}{{Sun} et~al.}{2023}]{sunetal2023}
{Sun} L.,  {Gu} S.,  {Wang} X.,  {Bai} L.,  {Schmitt} J.~H.~M.~M.,
  {Perdelwitz} V.,   {Ioannidis} P.,  2023, \mn@doi [\mnras]
  {10.1093/mnras/stad204}, \href
  {https://ui.adsabs.harvard.edu/abs/2023MNRAS.520.1642S} {520, 1642}

\bibitem[\protect\citeauthoryear{{Szab{\'o}} et~al.,}{{Szab{\'o}}
  et~al.}{2010}]{szabo2010}
{Szab{\'o}} G.~M.,  et~al., 2010, \mn@doi [\aap] {10.1051/0004-6361/201015172},
  \href {https://ui.adsabs.harvard.edu/abs/2010A&A...523A..84S} {523, A84}

\bibitem[\protect\citeauthoryear{{Tejada Arevalo}, {Winn}  \&
  {Anderson}}{{Tejada Arevalo} et~al.}{2021}]{2021ApJ...919..138T}
{Tejada Arevalo} R.~A.,  {Winn} J.~N.,   {Anderson} K.~R.,  2021, \mn@doi
  [\apj] {10.3847/1538-4357/ac1429}, \href
  {https://ui.adsabs.harvard.edu/abs/2021ApJ...919..138T} {919, 138}

\bibitem[\protect\citeauthoryear{{Tregloan-Reed} \&
  {Southworth}}{{Tregloan-Reed} \& {Southworth}}{2013}]{tregloan2013}
{Tregloan-Reed} J.,  {Southworth} J.,  2013, \mn@doi [\mnras]
  {10.1093/mnras/stt227}, \href
  {https://ui.adsabs.harvard.edu/abs/2013MNRAS.431..966T} {431, 966}

\bibitem[\protect\citeauthoryear{{Trifonov} et~al.,}{{Trifonov}
  et~al.}{2021}]{2021AJ....162..283T}
{Trifonov} T.,  et~al., 2021, \mn@doi [\aj] {10.3847/1538-3881/ac1bbe}, \href
  {https://ui.adsabs.harvard.edu/abs/2021AJ....162..283T} {162, 283}

\bibitem[\protect\citeauthoryear{{Turner} et~al.,}{{Turner}
  et~al.}{2016}]{turner2016}
{Turner} J.~D.,  et~al., 2016, \mn@doi [\mnras] {10.1093/mnras/stw574}, \href
  {https://ui.adsabs.harvard.edu/abs/2016MNRAS.459..789T} {459, 789}

\bibitem[\protect\citeauthoryear{{VanderPlas}}{{VanderPlas}}{2018}]{2018ApJS..236...16V}
{VanderPlas} J.~T.,  2018, \mn@doi [\apjs] {10.3847/1538-4365/aab766}, \href
  {https://ui.adsabs.harvard.edu/abs/2018ApJS..236...16V} {236, 16}

\bibitem[\protect\citeauthoryear{{Vissapragada} et~al.,}{{Vissapragada}
  et~al.}{2022}]{Vissapragada2022}
{Vissapragada} S.,  et~al., 2022, \mn@doi [\apjl] {10.3847/2041-8213/aca47e},
  \href {https://ui.adsabs.harvard.edu/abs/2022ApJ...941L..31V} {941, L31}

\bibitem[\protect\citeauthoryear{{Wakeford} et~al.,}{{Wakeford}
  et~al.}{2013}]{wakeford2013}
{Wakeford} H.~R.,  et~al., 2013, \mn@doi [\mnras] {10.1093/mnras/stt1536},
  \href {https://ui.adsabs.harvard.edu/abs/2013MNRAS.435.3481W} {435, 3481}

\bibitem[\protect\citeauthoryear{{Wang}, {Gu}, {Collier Cameron}, {Wang},
  {Hui}, {Kwok}, {Yeung}  \& {Leung}}{{Wang} et~al.}{2014}]{wang2014}
{Wang} X.-b.,  {Gu} S.-h.,  {Collier Cameron} A.,  {Wang} Y.-b.,  {Hui} H.-K.,
  {Kwok} C.-T.,  {Yeung} B.,   {Leung} K.-C.,  2014, \mn@doi [\aj]
  {10.1088/0004-6256/147/4/92}, \href
  {https://ui.adsabs.harvard.edu/abs/2014AJ....147...92W} {147, 92}

\bibitem[\protect\citeauthoryear{{Wang} et~al.,}{{Wang}
  et~al.}{2021}]{wang2021}
{Wang} X.-Y.,  et~al., 2021, \mn@doi [\apjs] {10.3847/1538-4365/ac0835}, \href
  {https://ui.adsabs.harvard.edu/abs/2021ApJS..255...15W} {255, 15}

\bibitem[\protect\citeauthoryear{{Weinberg}, {Davachi}, {Essick}, {Yu}, {Arras}
   \& {Belland}}{{Weinberg} et~al.}{2023}]{2023arXiv230511974W}
{Weinberg} N.~N.,  {Davachi} N.,  {Essick} R.,  {Yu} H.,  {Arras} P.,
  {Belland} B.,  2023, \mn@doi [arXiv e-prints] {10.48550/arXiv.2305.11974},
  \href {https://ui.adsabs.harvard.edu/abs/2023arXiv230511974W} {p.
  arXiv:2305.11974}

\bibitem[\protect\citeauthoryear{{West} et~al.,}{{West}
  et~al.}{2009}]{west2009}
{West} R.~G.,  et~al., 2009, \mn@doi [\aap] {10.1051/0004-6361/200810973},
  \href {https://ui.adsabs.harvard.edu/abs/2009A&A...502..395W} {502, 395}

\bibitem[\protect\citeauthoryear{{Winn} et~al.,}{{Winn}
  et~al.}{2007}]{winn2007}
{Winn} J.~N.,  et~al., 2007, \mn@doi [\aj] {10.1086/521599}, \href
  {https://ui.adsabs.harvard.edu/abs/2007AJ....134.1707W} {134, 1707}

\bibitem[\protect\citeauthoryear{{Winn} et~al.,}{{Winn}
  et~al.}{2008}]{2008ApJ...683.1076W}
{Winn} J.~N.,  et~al., 2008, \mn@doi [\apj] {10.1086/589737}, \href
  {https://ui.adsabs.harvard.edu/abs/2008ApJ...683.1076W} {683, 1076}

\bibitem[\protect\citeauthoryear{{Winn} et~al.,}{{Winn}
  et~al.}{2010a}]{winn2010}
{Winn} J.~N.,  et~al., 2010a, \mn@doi [\apj] {10.1088/0004-637X/718/1/575},
  \href {https://ui.adsabs.harvard.edu/abs/2010ApJ...718..575W} {718, 575}

\bibitem[\protect\citeauthoryear{{Winn}, {Fabrycky}, {Albrecht}  \&
  {Johnson}}{{Winn} et~al.}{2010b}]{2010ApJ...718L.145W}
{Winn} J.~N.,  {Fabrycky} D.,  {Albrecht} S.,   {Johnson} J.~A.,  2010b,
  \mn@doi [\apjl] {10.1088/2041-8205/718/2/L145}, \href
  {https://ui.adsabs.harvard.edu/abs/2010ApJ...718L.145W} {718, L145}

\bibitem[\protect\citeauthoryear{{W{\"o}llert} \& {Brandner}}{{W{\"o}llert} \&
  {Brandner}}{2015}]{2015A&A...579A.129W}
{W{\"o}llert} M.,  {Brandner} W.,  2015, \mn@doi [\aap]
  {10.1051/0004-6361/201526525}, \href
  {https://ui.adsabs.harvard.edu/abs/2015A&A...579A.129W} {579, A129}

\bibitem[\protect\citeauthoryear{{Wong} et~al.,}{{Wong}
  et~al.}{2021}]{2021AJ....162..127W}
{Wong} I.,  et~al., 2021, \mn@doi [\aj] {10.3847/1538-3881/ac0c7d}, \href
  {https://ui.adsabs.harvard.edu/abs/2021AJ....162..127W} {162, 127}

\bibitem[\protect\citeauthoryear{{Yal{\c{c}}{\i}nkaya}
  et~al.,}{{Yal{\c{c}}{\i}nkaya} et~al.}{2021}]{yalcinkaya2021}
{Yal{\c{c}}{\i}nkaya} S.,  et~al., 2021, \mn@doi [\actaa]
  {10.32023/0001-5237/71.3.3}, \href
  {https://ui.adsabs.harvard.edu/abs/2021AcA....71..223Y} {71, 223}

\bibitem[\protect\citeauthoryear{{de Mooij} et~al.,}{{de Mooij}
  et~al.}{2012}]{mooij2012}
{de Mooij} E.~J.~W.,  et~al., 2012, \mn@doi [\aap]
  {10.1051/0004-6361/201117205}, \href
  {https://ui.adsabs.harvard.edu/abs/2012A&A...538A..46D} {538, A46}

\makeatother
\end{thebibliography}



\appendix
\section{Light Curves}
\label{sec:appendixa}
\begin{figure}
  \centering
  \subfloat[T100 2021/07/19*]{\includegraphics[width=4cm]{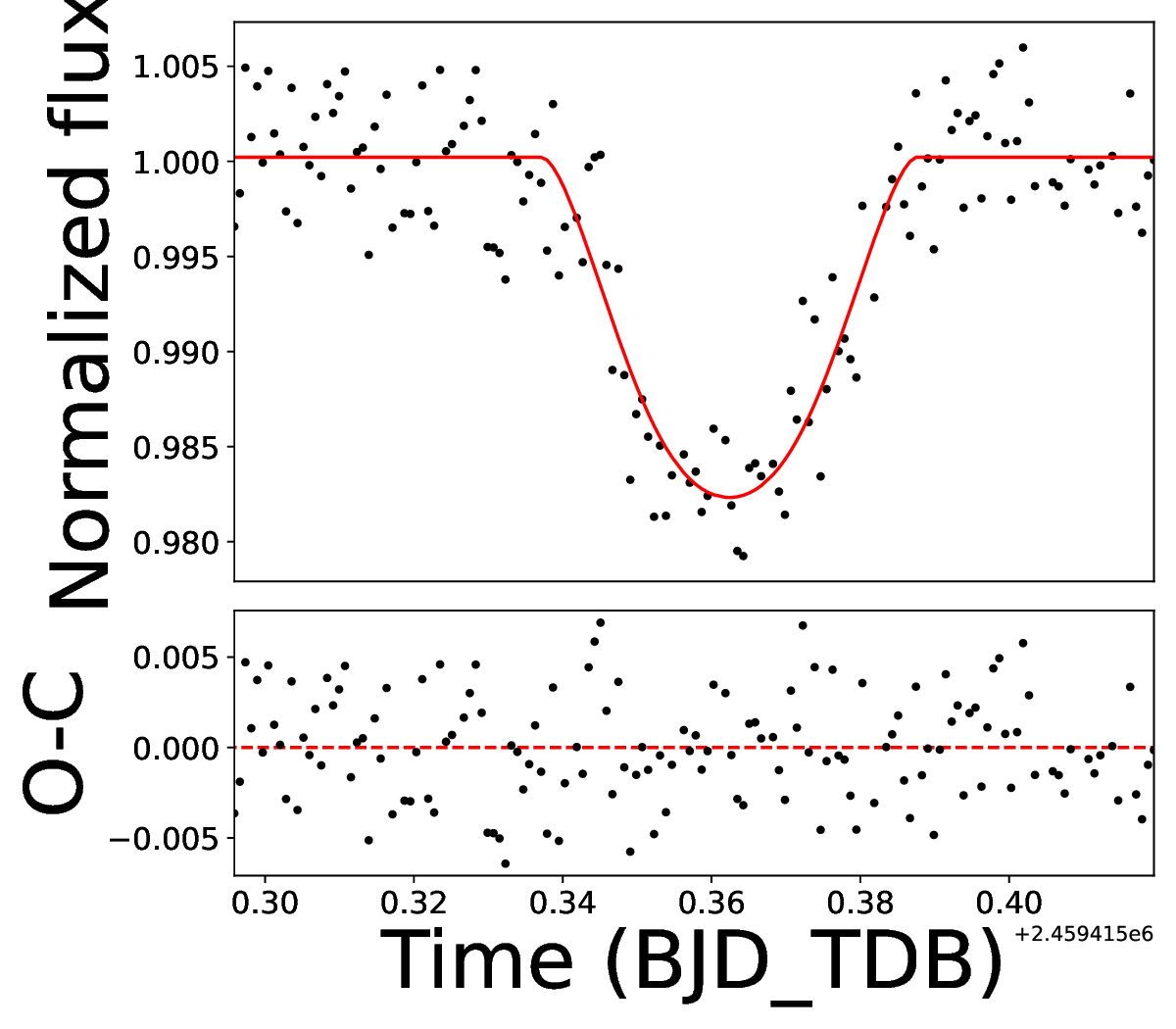}}
  \hfill
  \subfloat[T100 2021/04/23]{\includegraphics[width=4cm]{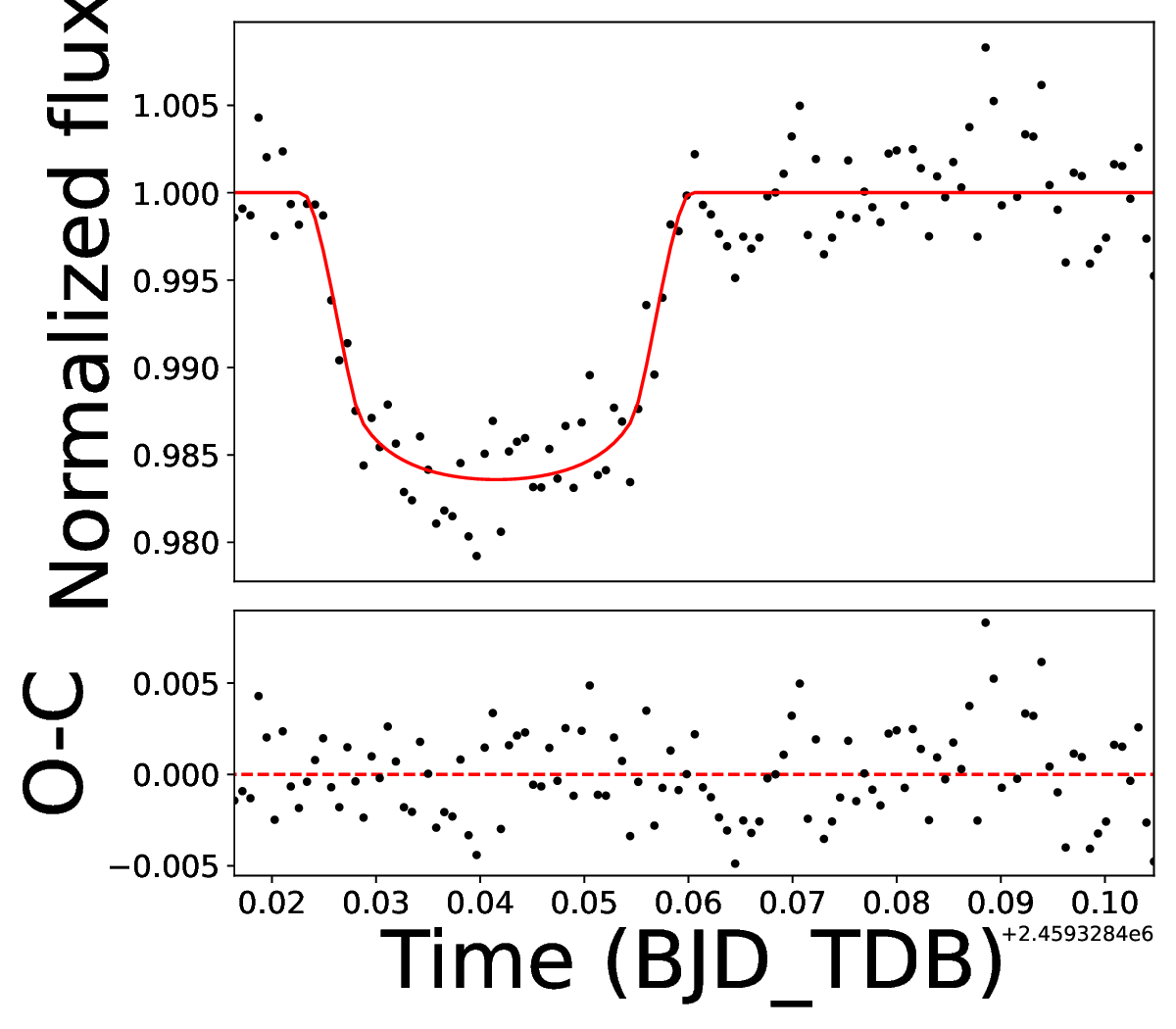}}
  \qquad
  \subfloat[T100 2020/07/03]{\includegraphics[width=4cm]{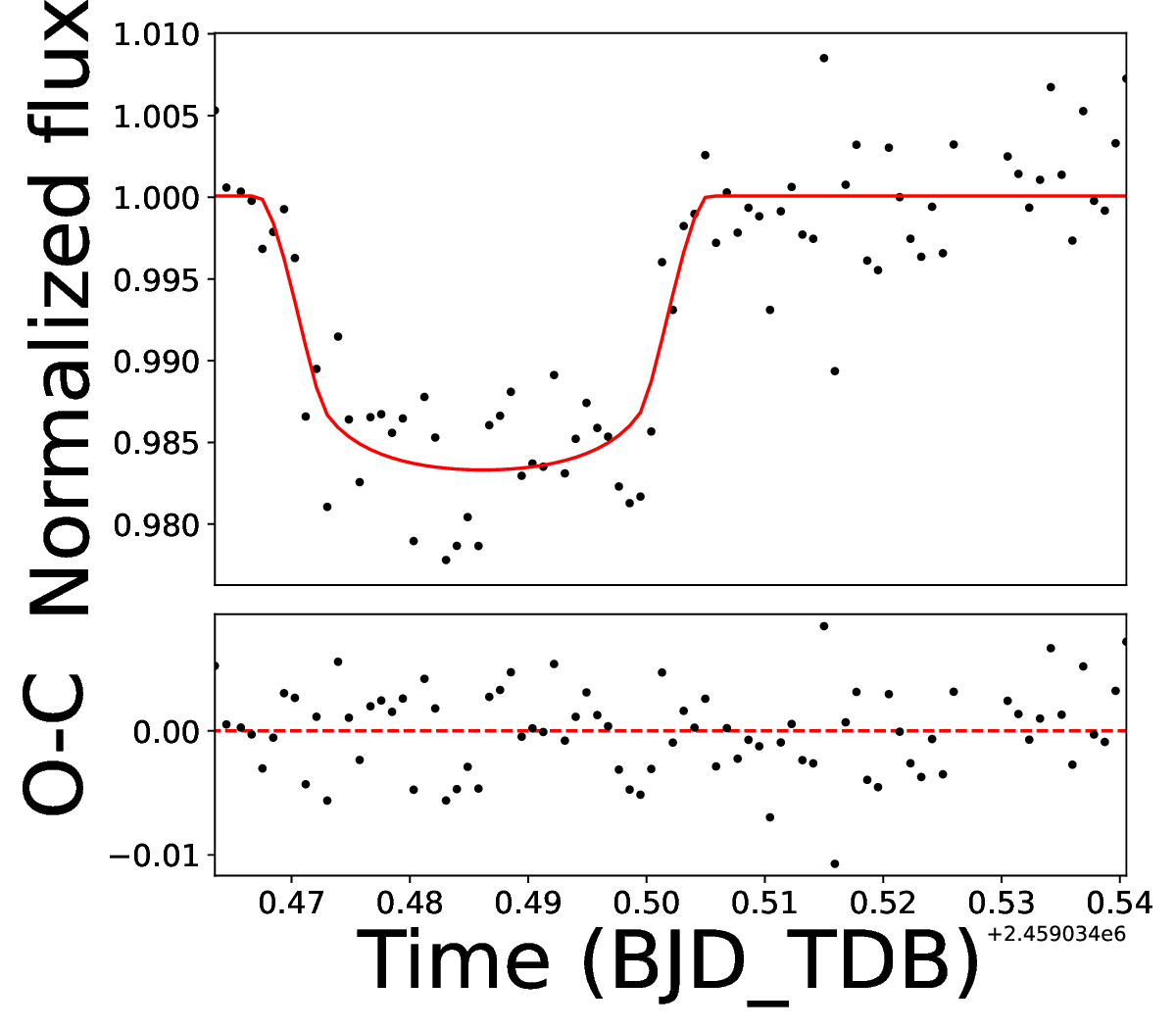}}
  \hfill
  \subfloat[T100 2020/06/11]{\includegraphics[width=4cm]{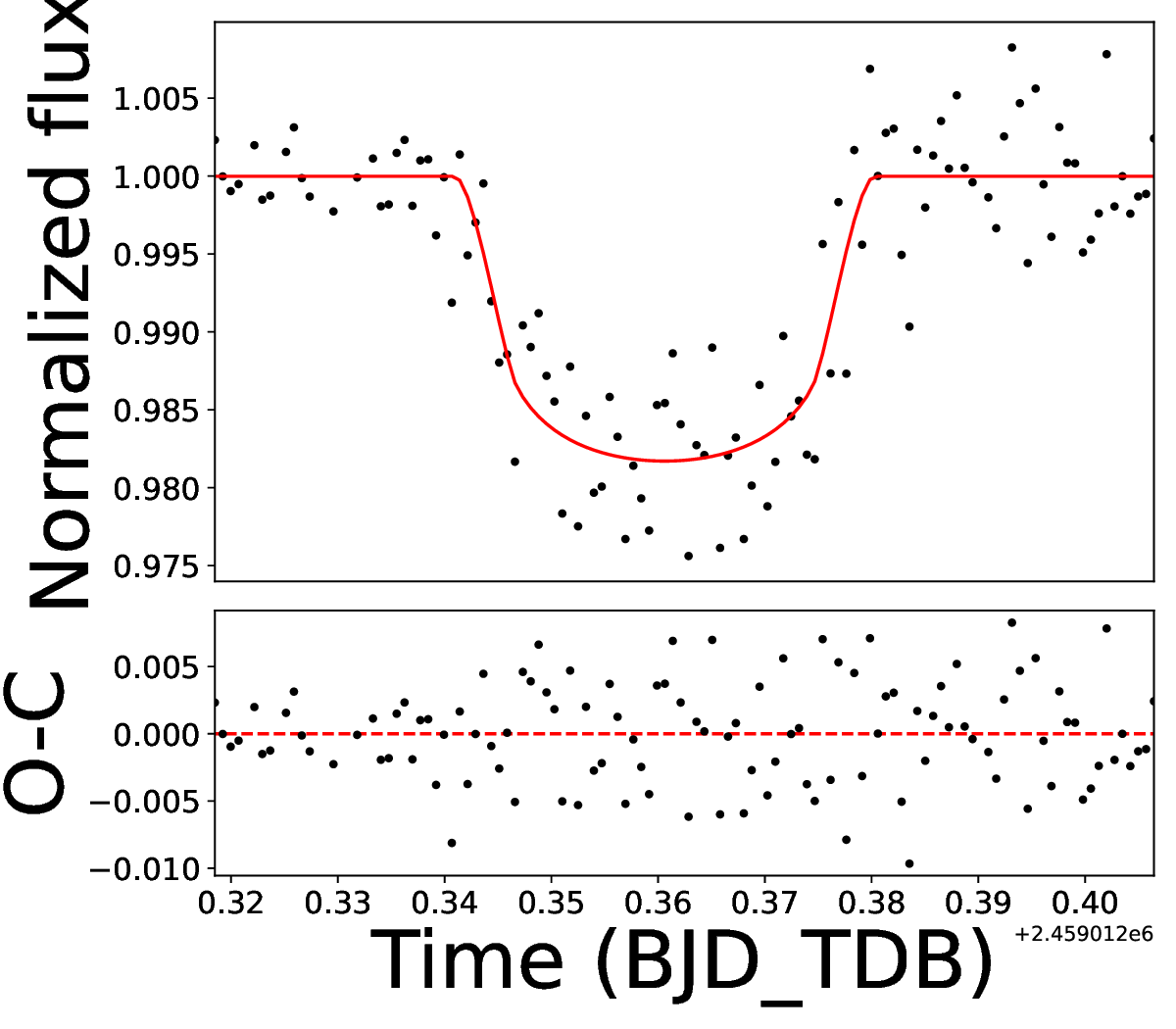}}
  \qquad
  \subfloat[T80 2020/05/12]{\includegraphics[width=4cm]{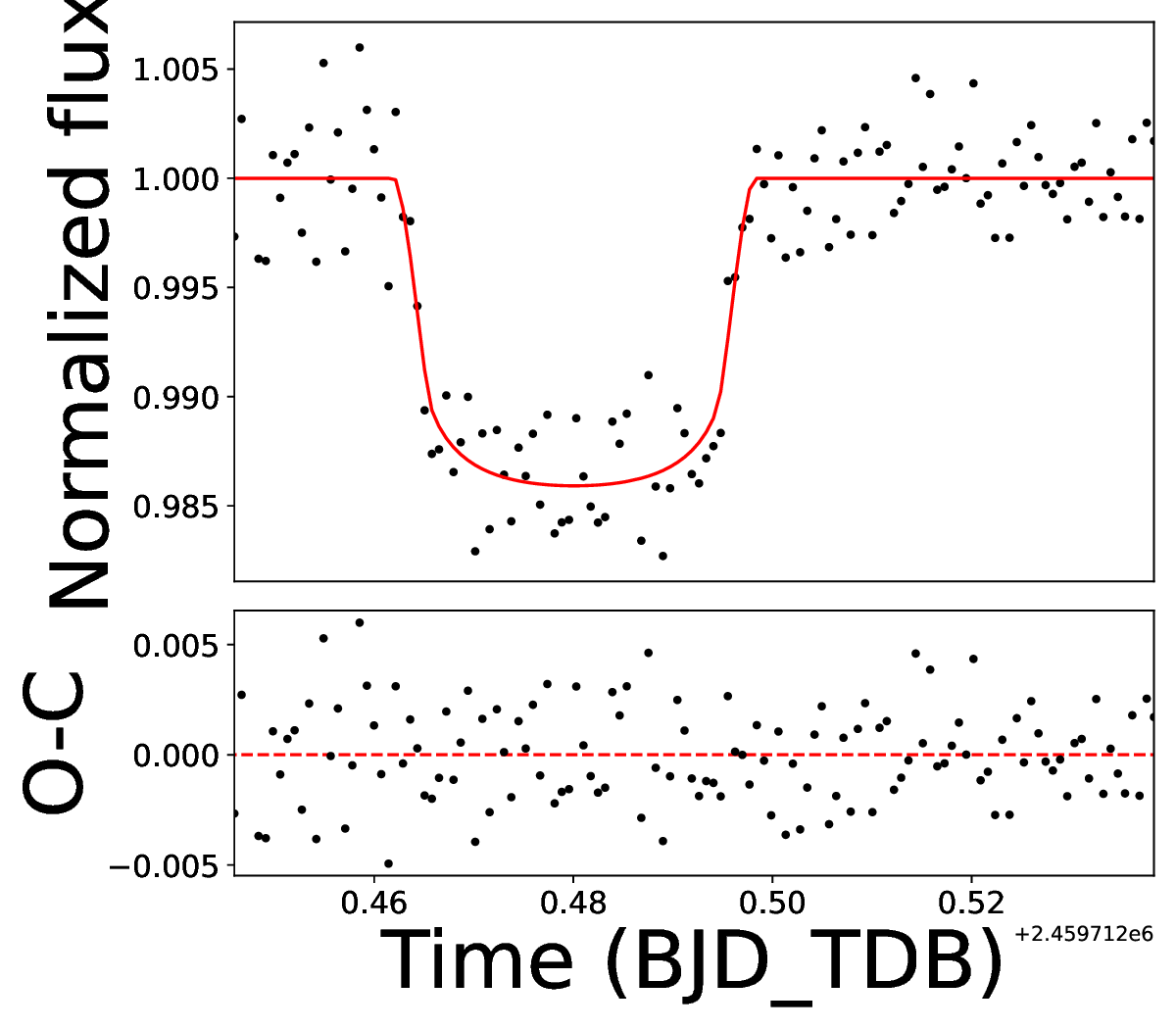}}
  \hfill
    \caption{GJ 1214\,b Light Curves. Black dots are data points while the red continuous curve is for the {\sc exofast} model in all the light curves presented in this section. The ones that were eliminated based on our quantitative light curve selection criteria, therefore not used in timing analyses are marked with asterisks.}
    \label{fig:GJ1224_lcs}
\end{figure}

\begin{figure}
  \centering
  \subfloat[CAHA 2013/09/02 b filter]{\includegraphics[width=4cm]{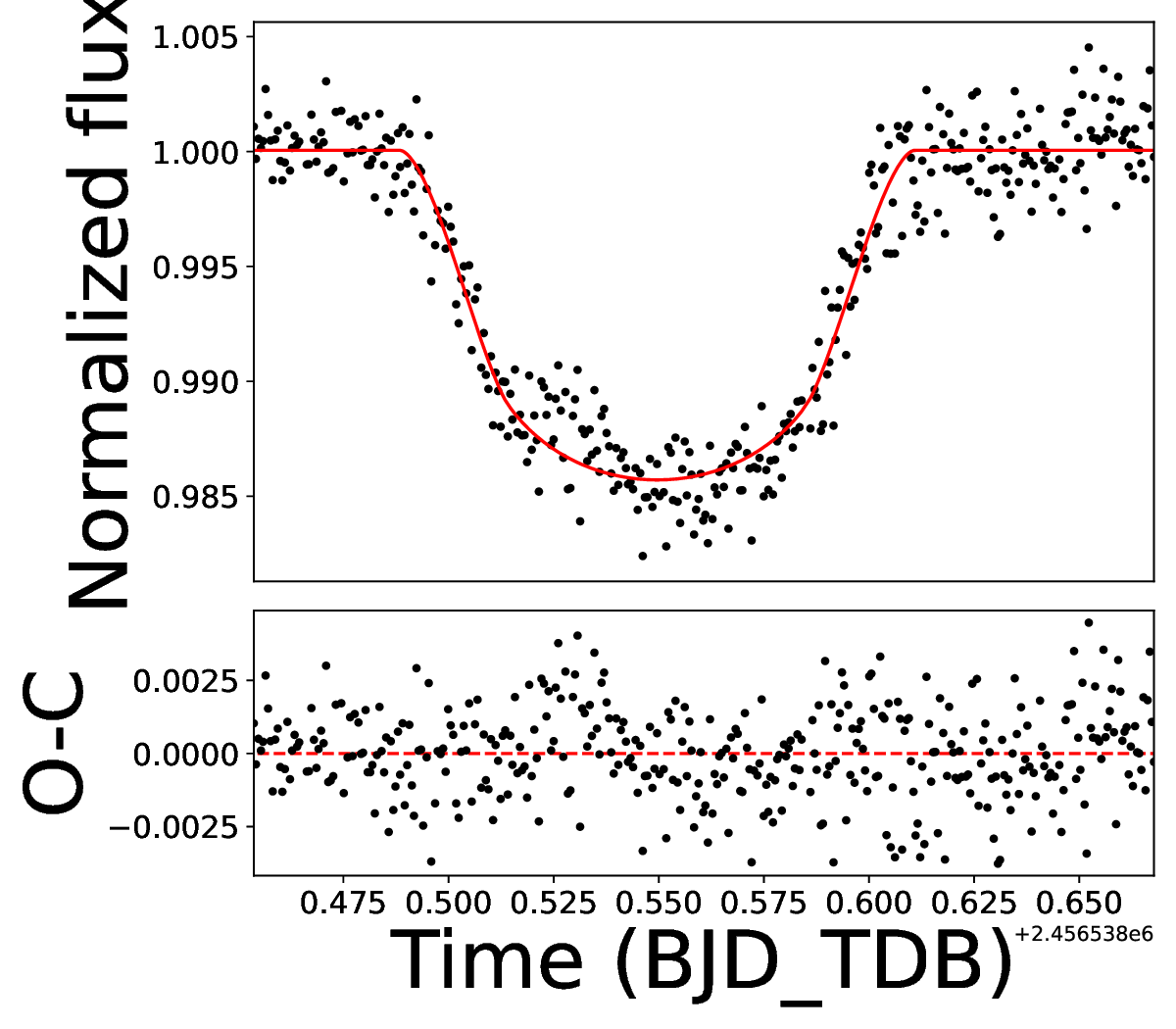}}
  \hfill
  \subfloat[CAHA 2013/09/02 u filter]{\includegraphics[width=4cm]{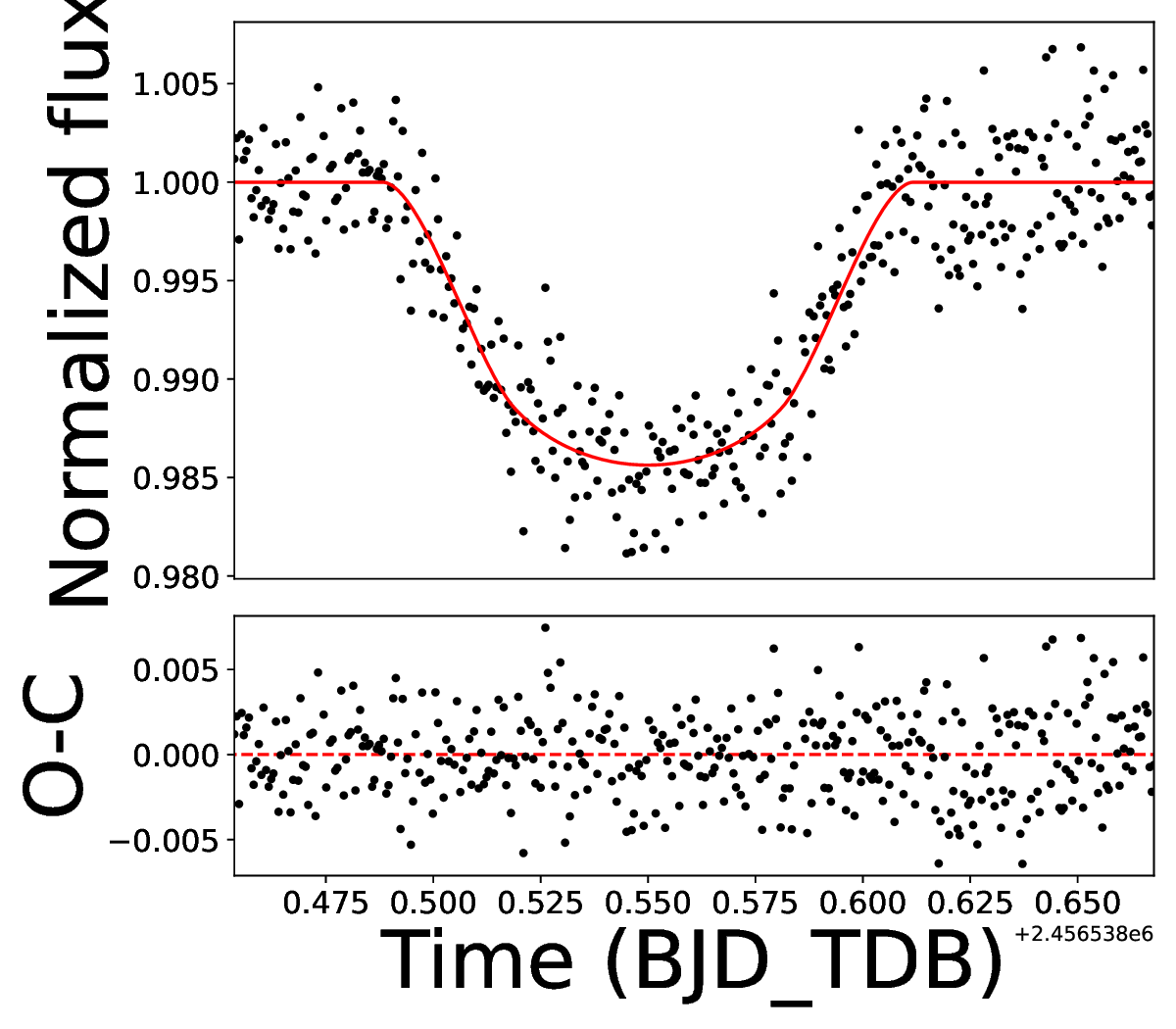}}
  \qquad
  \subfloat[CAHA 2013/09/02 y filter]{\includegraphics[width=4cm]{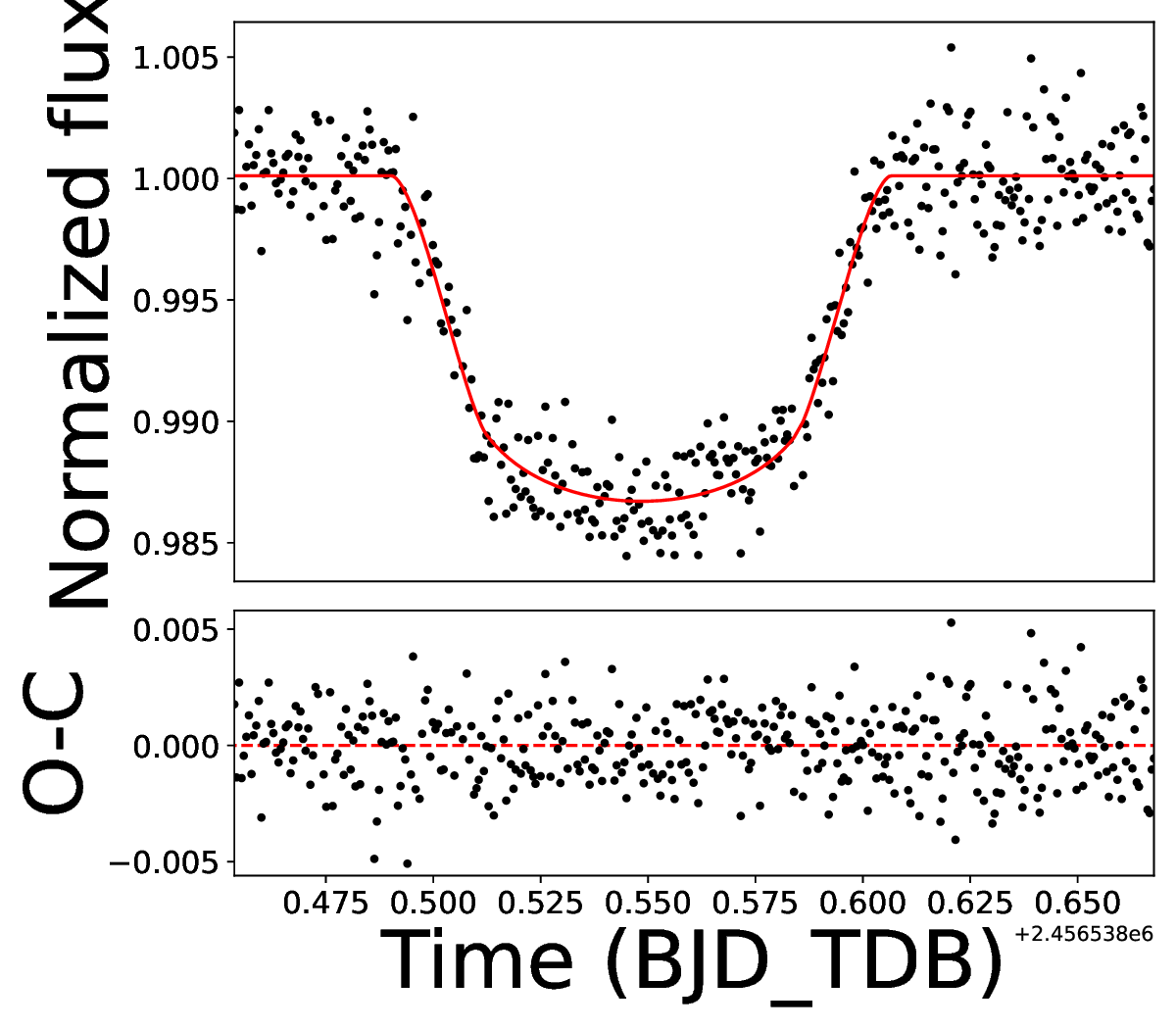}}
  \hfill
  \subfloat[CAHA 2013/09/02 z filter]{\includegraphics[width=4cm]{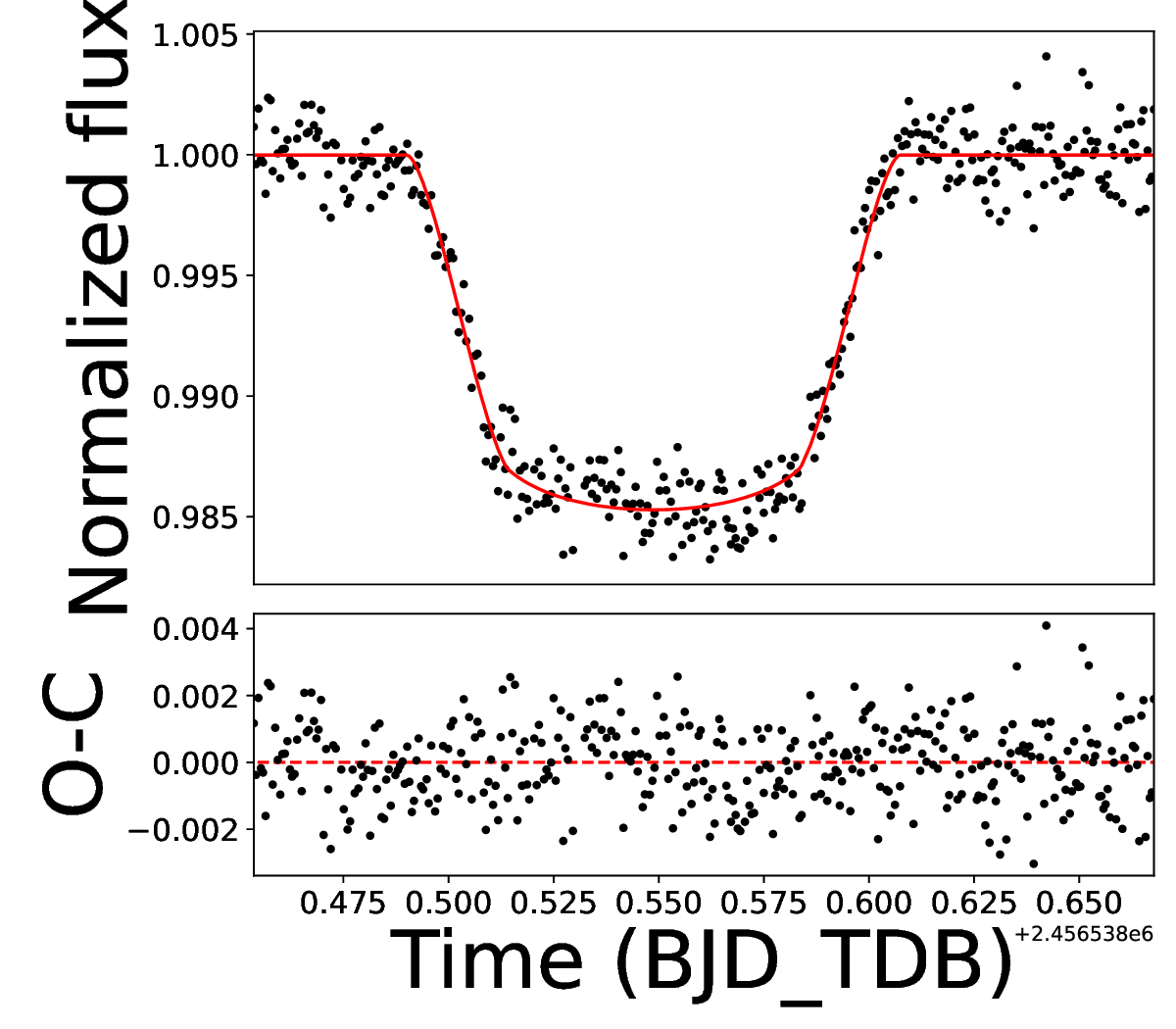}}
  \qquad
  \subfloat[ATA50 2022/09/15]{\includegraphics[width=4cm]{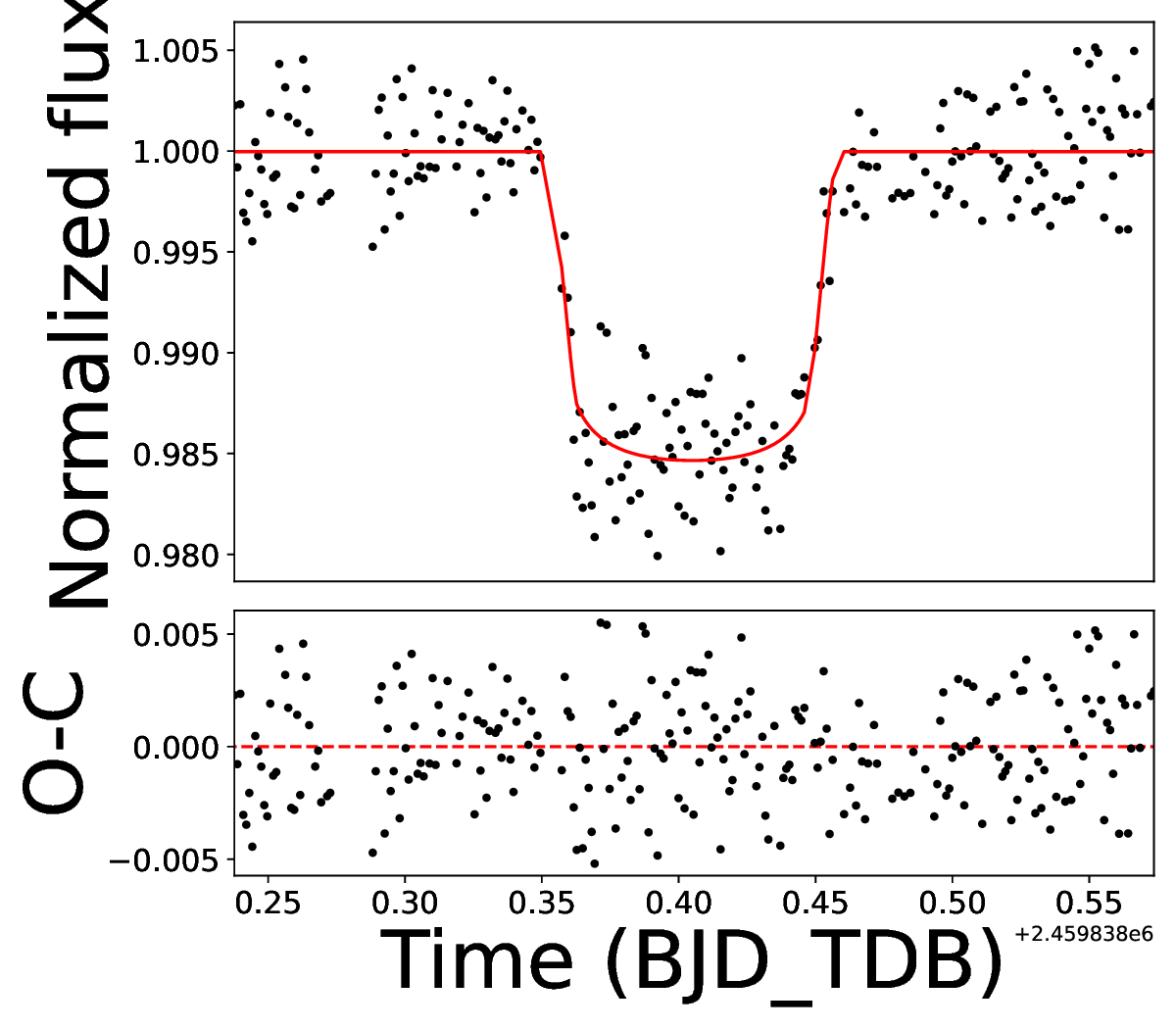}}
  \hfill
    \caption{HAT-P-1\,b Light Curves.}
    \label{fig:HAT-P-1_lcs}
\end{figure}

\begin{figure}
  \centering
  \subfloat[T100 2013/01/15]{\includegraphics[width=4cm]{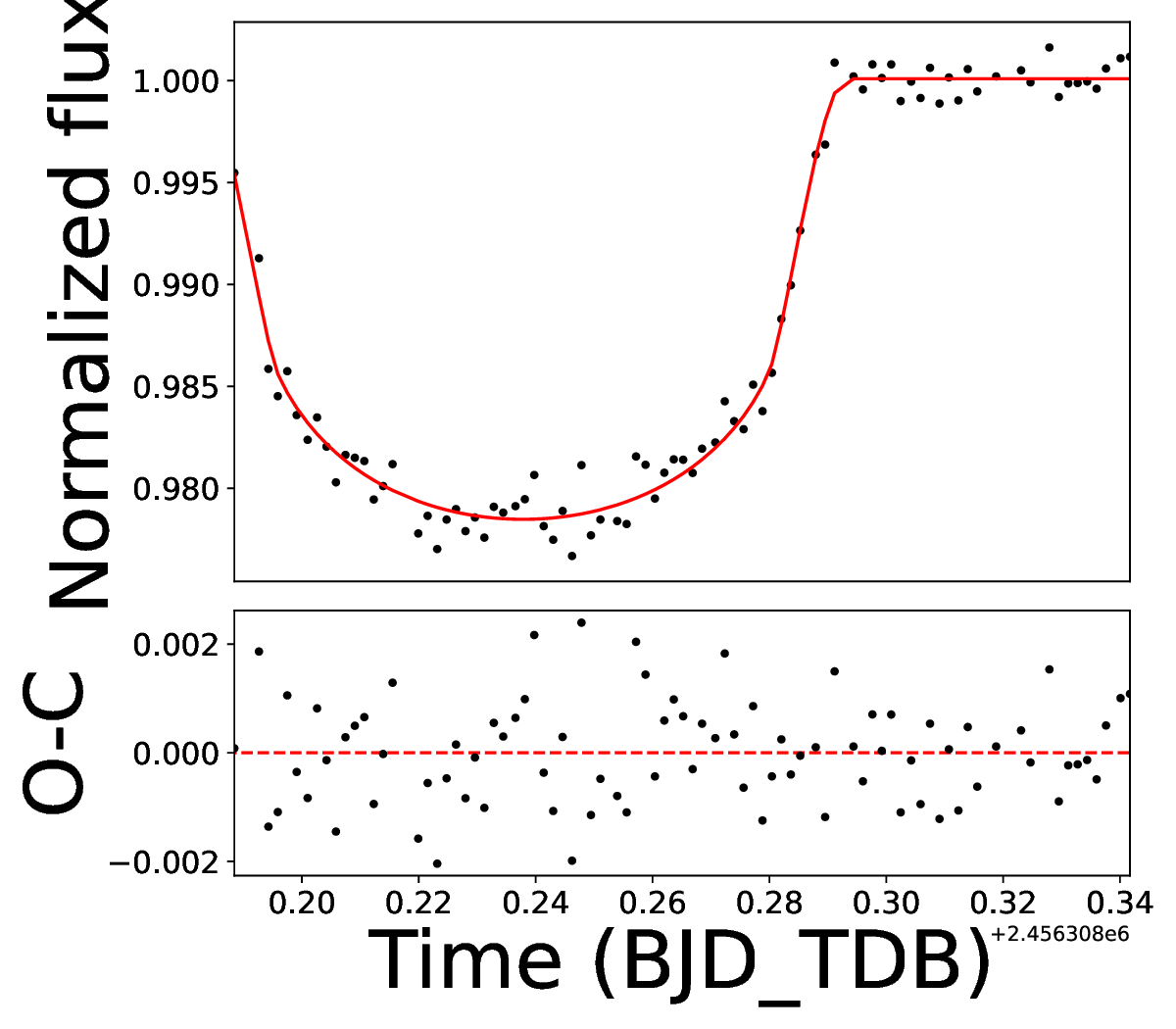}}
  \hfill
  \subfloat[T100 2020/10/25*]{\includegraphics[width=4cm]{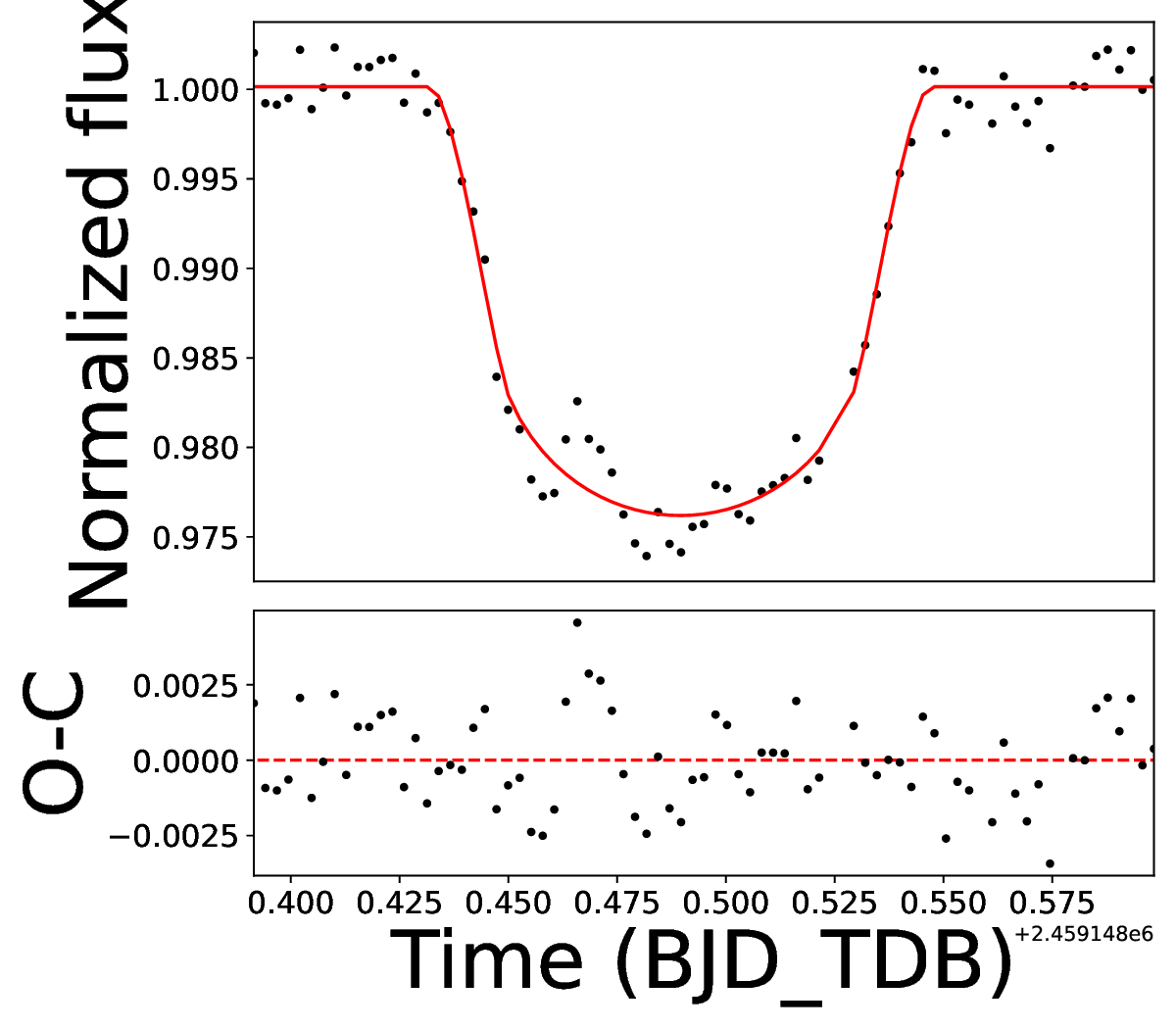}}
  \qquad
    \caption{HAT-P-10\,b Light Curves.}
    \label{fig:HAT-P-10_lcs}
\end{figure}

\begin{figure}
  \centering
  \subfloat[CAHA 2014/01/12]{\includegraphics[width=4cm]{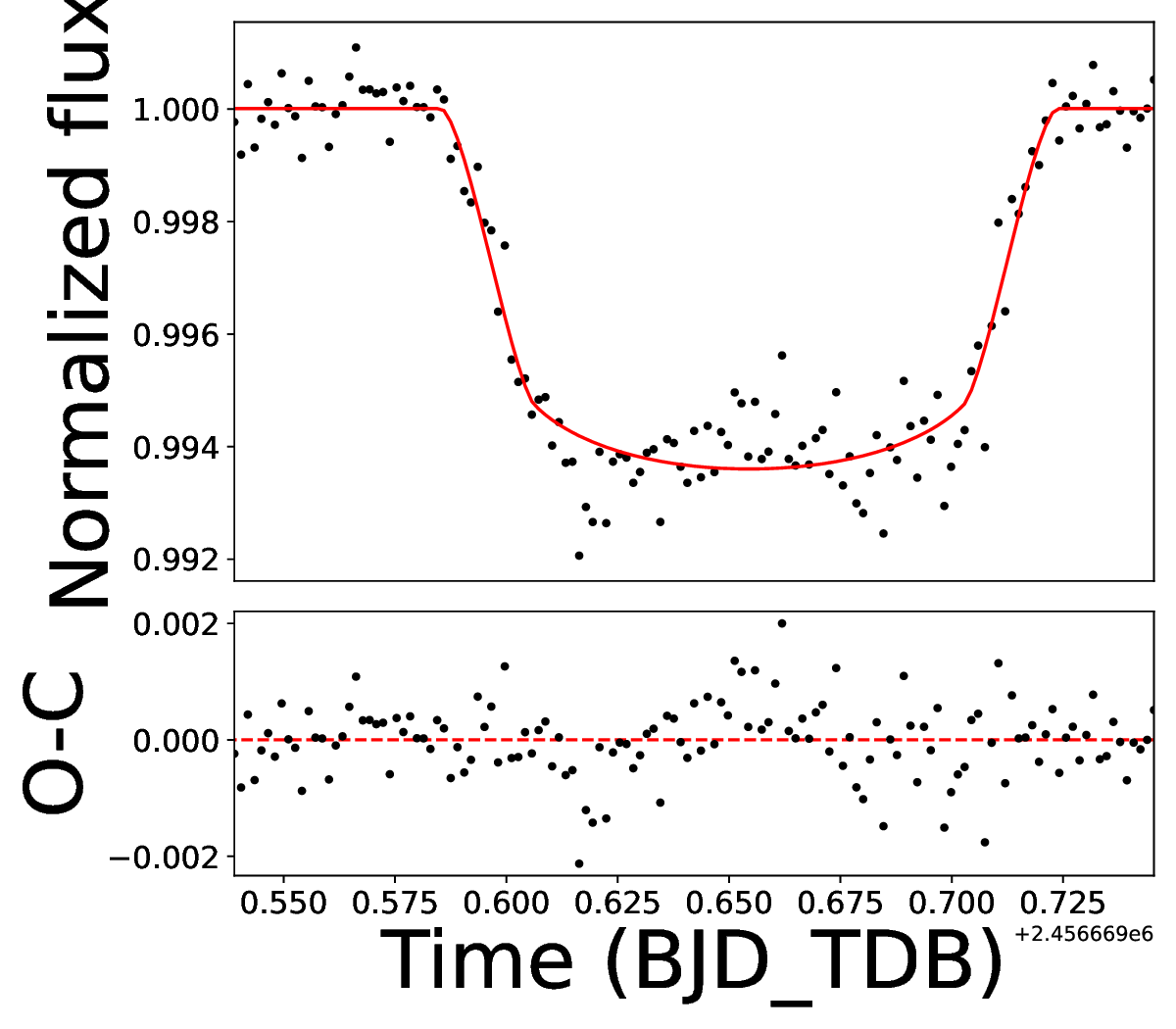}}
  \hfill
  \subfloat[CAHA 2015/03/10 ]{\includegraphics[width=4cm]{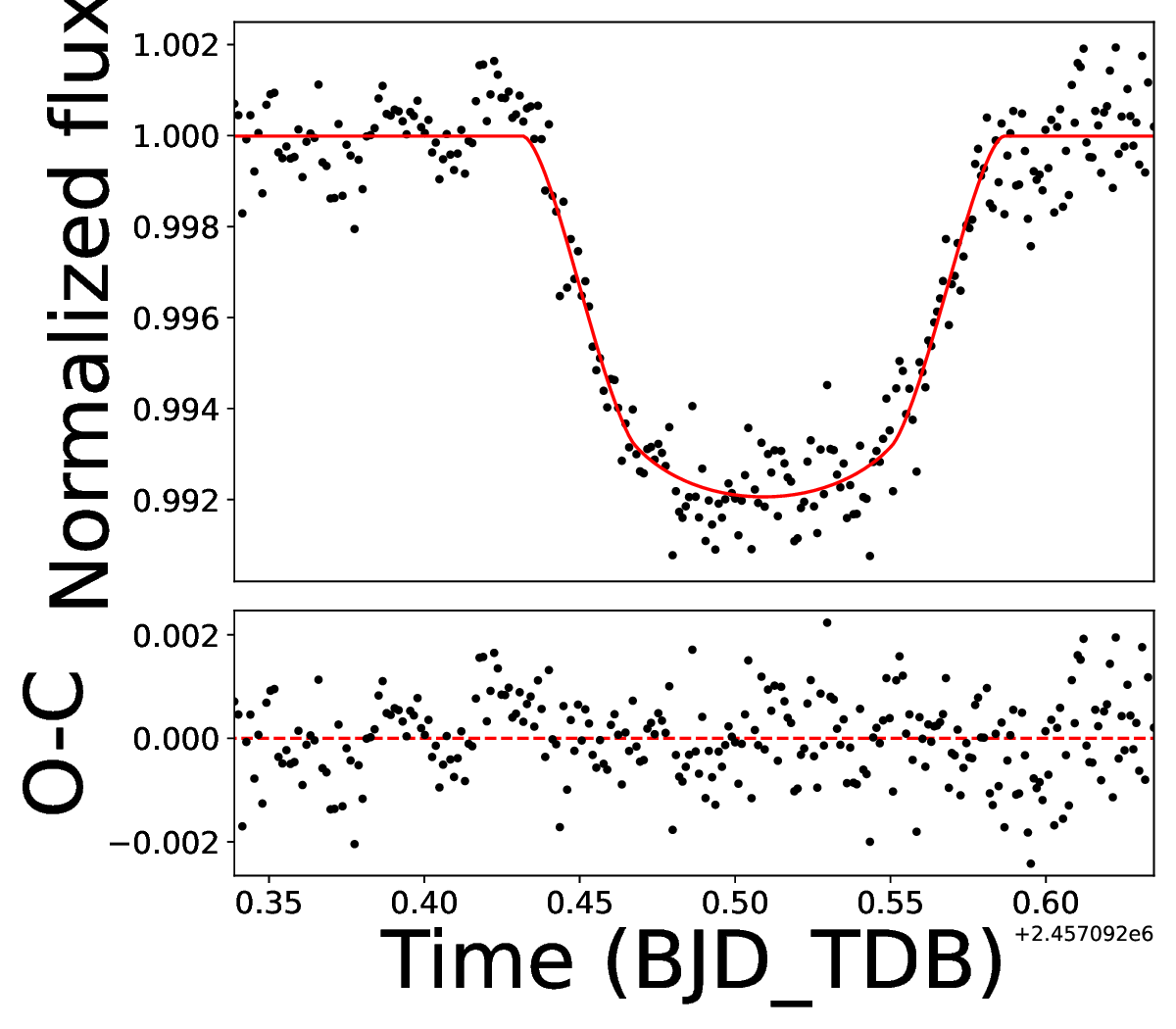}}
  \qquad
  \subfloat[T100 2021/01/03]{\includegraphics[width=4cm]{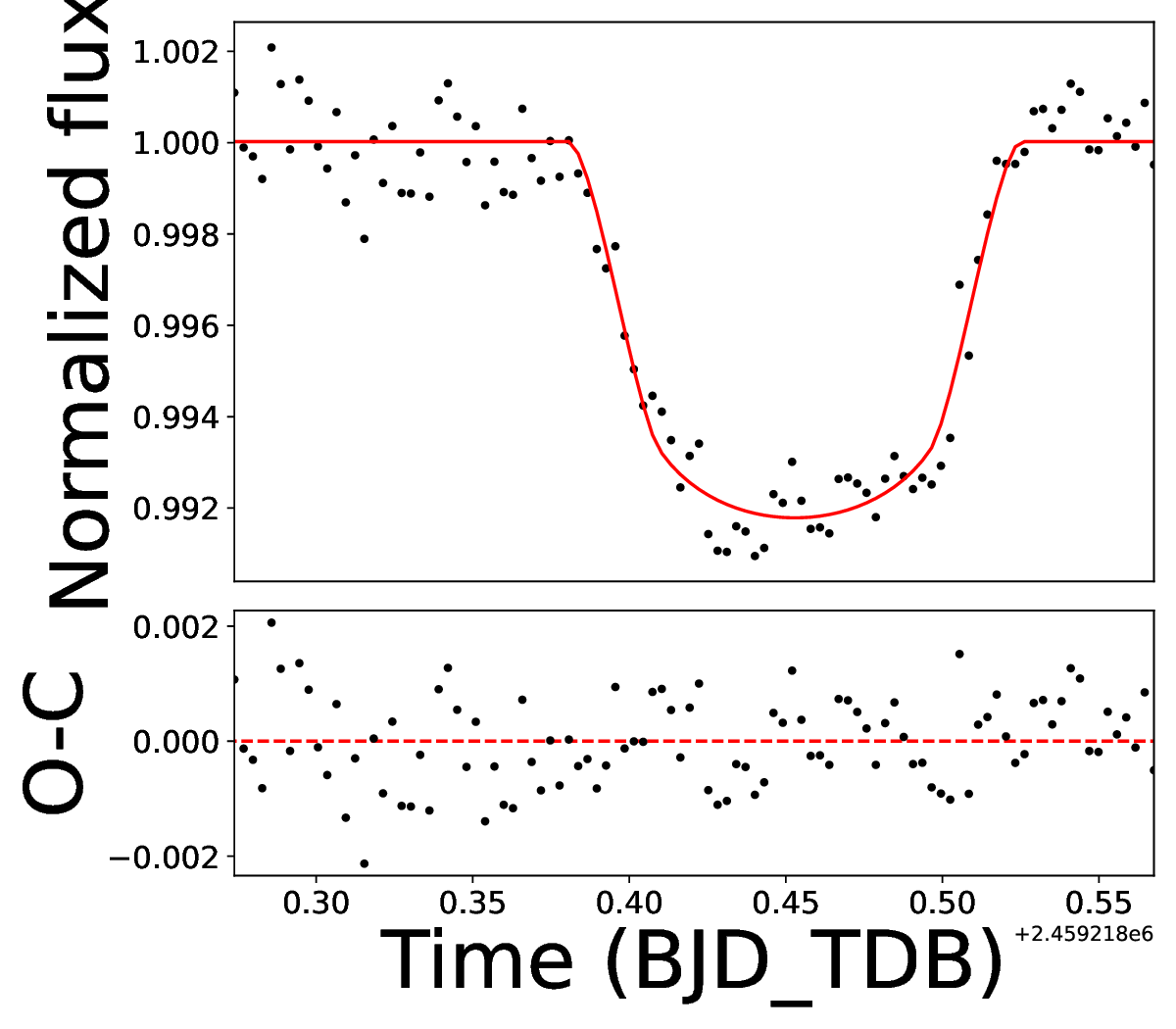}}
  \hfill
    \caption{HAT-P-13\,b Light Curves.}
    \label{fig:HAT-P-13_lcs}
\end{figure}

\begin{figure}
  \centering
  \subfloat[ATA50 2020/10/07]{\includegraphics[width=4cm]{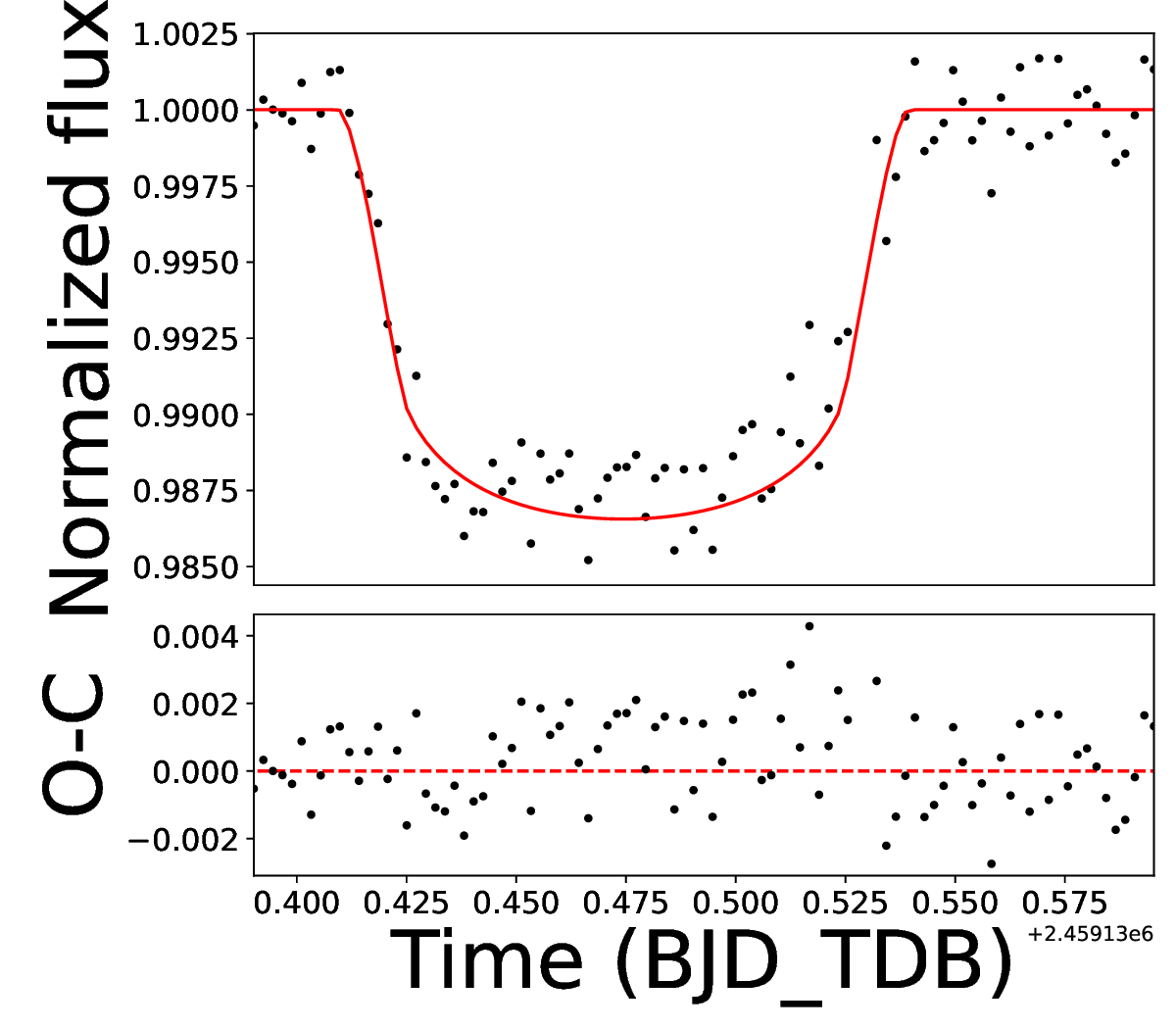}}
  \hfill
  \subfloat[UT50 2020/10/21*]{\includegraphics[width=4cm]{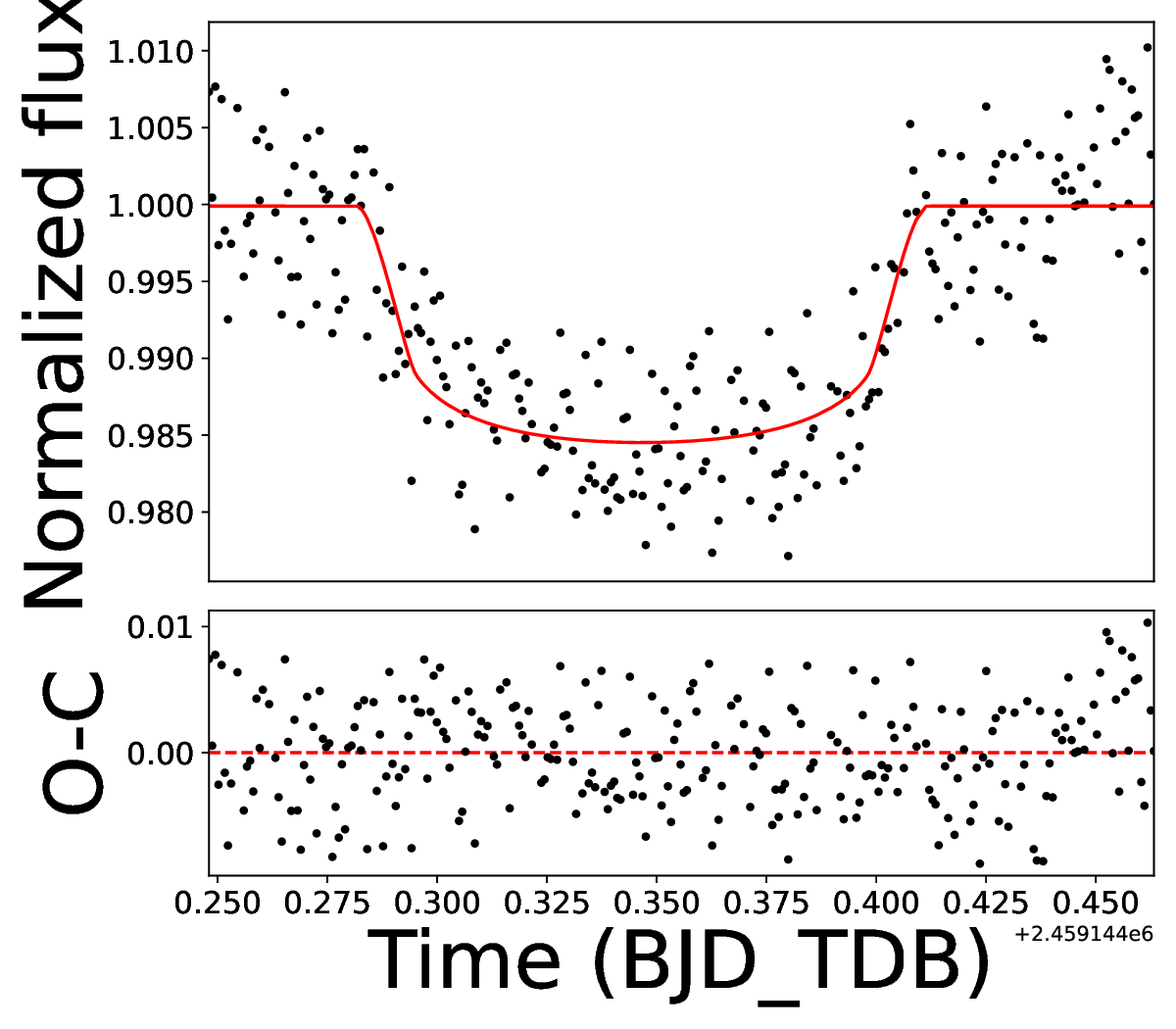}}
  \qquad
    \caption{HAT-P-16\,b Light Curves.}
    \label{fig:HAT-P-16_lcs}
\end{figure}

\begin{figure}
  \centering
  \subfloat[T100 2014/02/17]{\includegraphics[width=4cm]{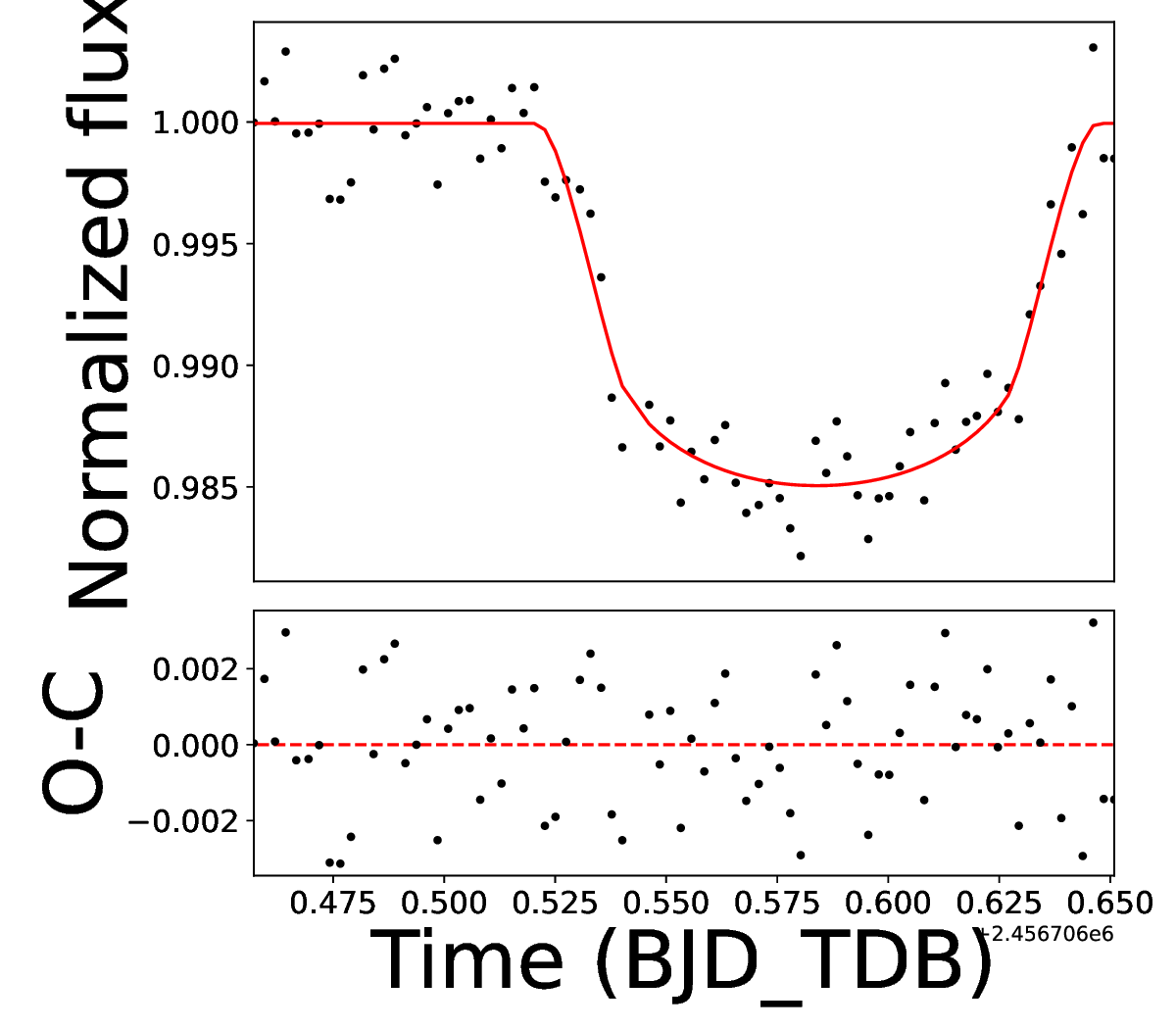}}
  \hfill
  \subfloat[T100 2021/02/14]{\includegraphics[width=4cm]{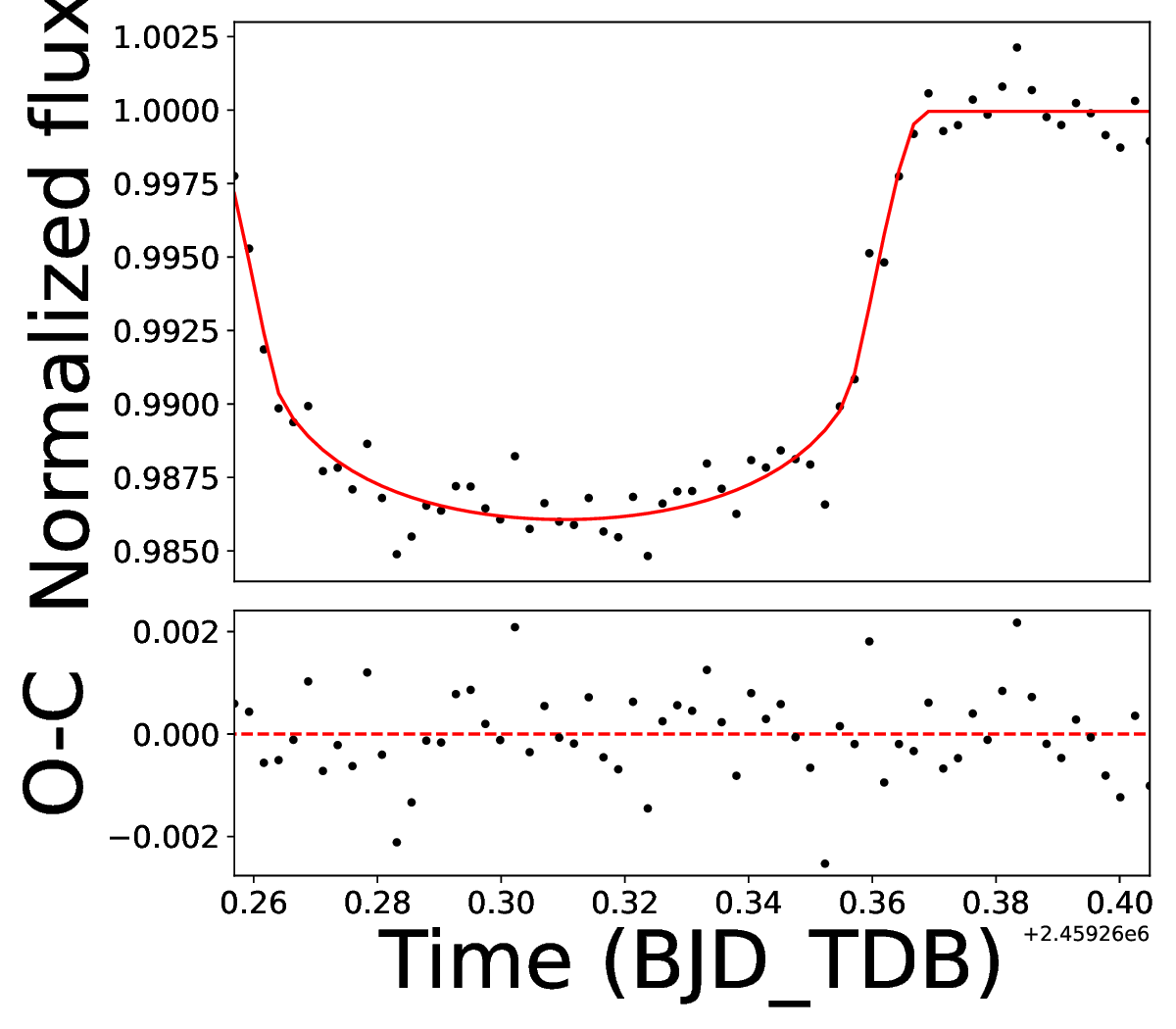}}
  \qquad
    \caption{HAT-P-22\,b Light Curves.}
    \label{fig:HAT-P-22_lcs}
\end{figure}

\begin{figure}
  \centering
  \subfloat[ATA50 2020/11/08*]{\includegraphics[width=4cm]{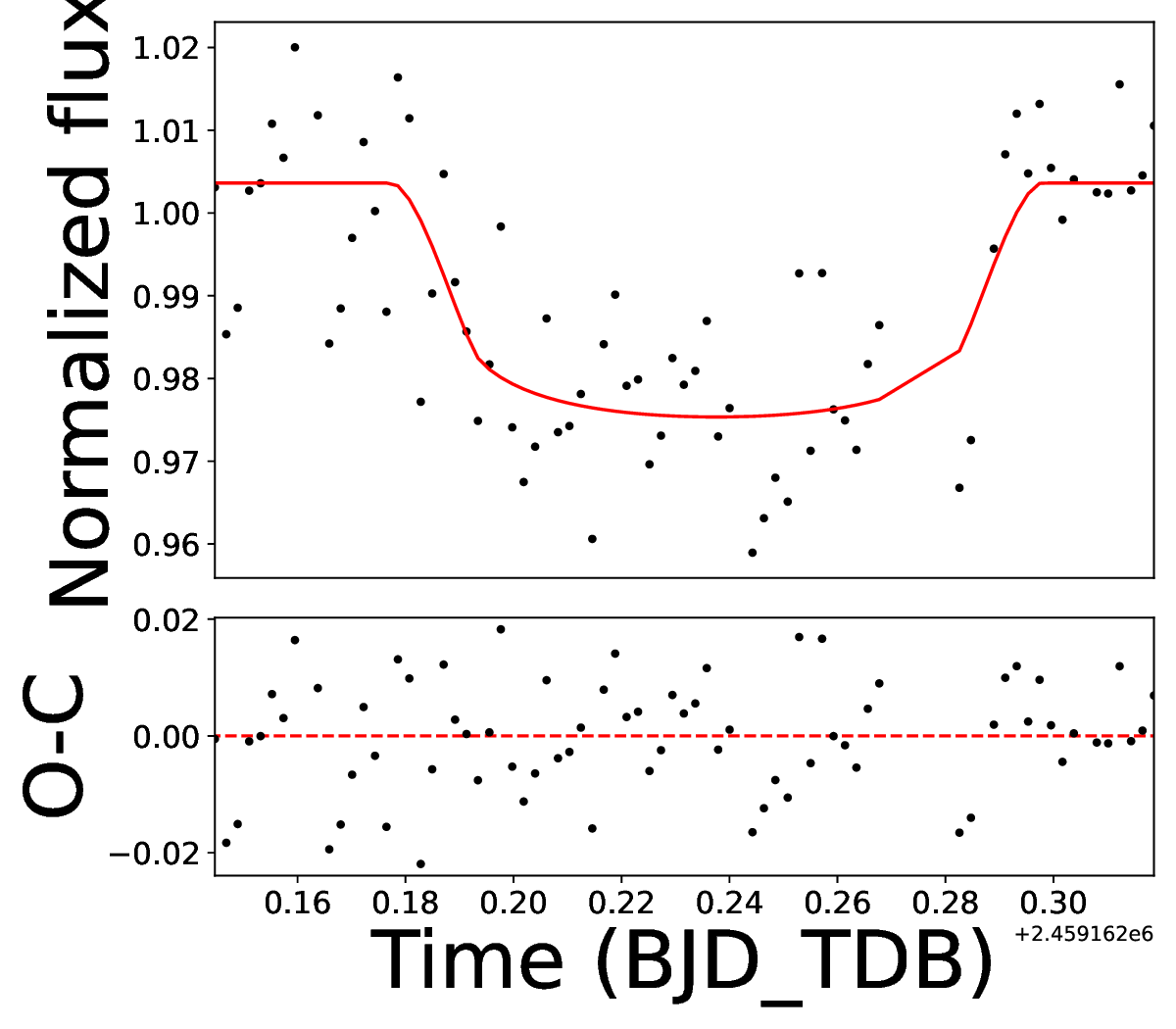}}
  \hfill
    \caption{HAT-P-53\,b Light Curves.}
    \label{fig:HAT-P-53_lcs}
\end{figure}

\begin{figure}
  \centering
  \subfloat[T100 2014/02/18]{\includegraphics[width=4cm]{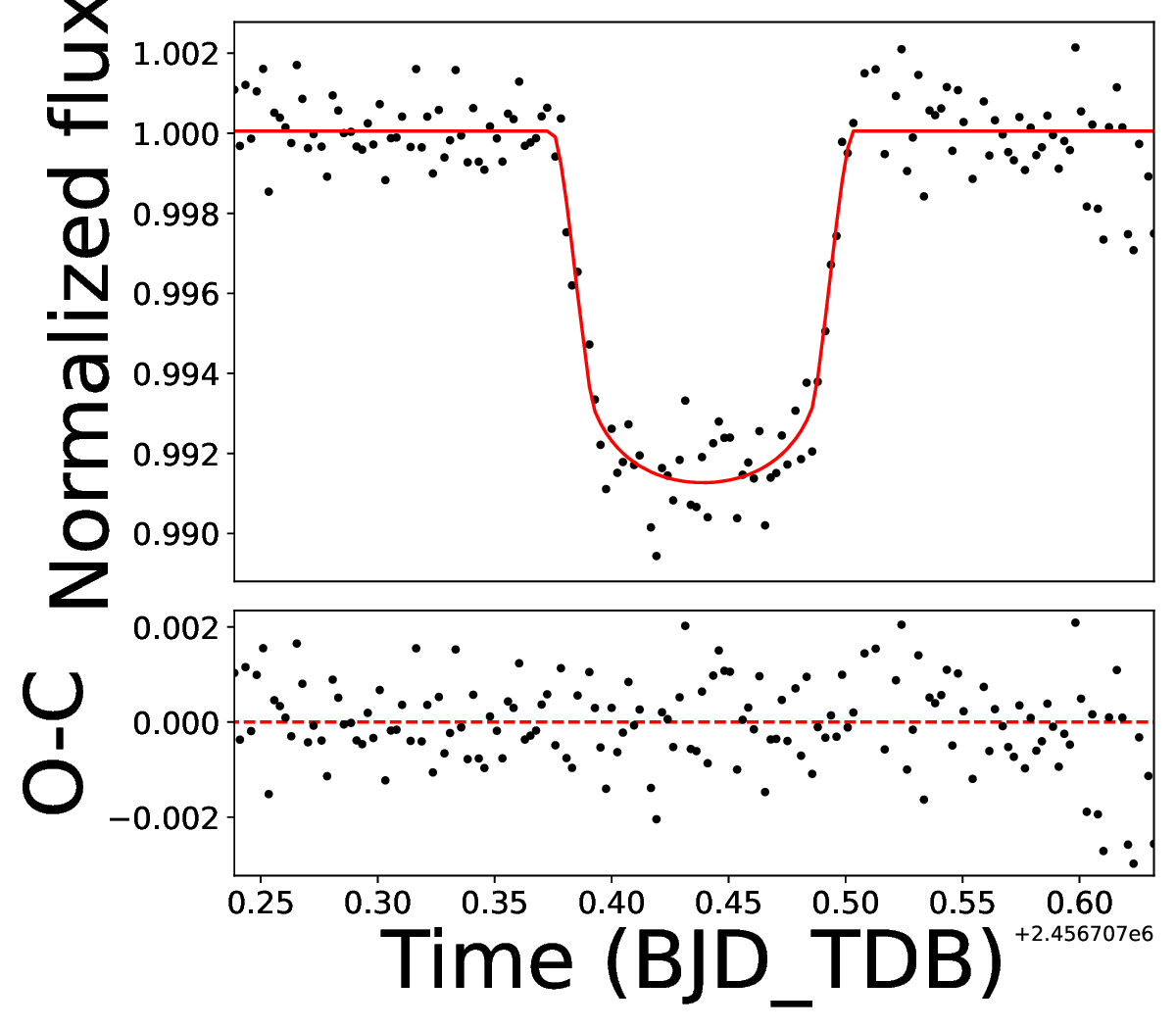}}
  \hfill
    \caption{KELT-3\,b Light Curves.}
    \label{fig:KELT-3_lcs}
\end{figure}

\begin{figure}
  \centering
  \subfloat[T100 2019/02/17]{\includegraphics[width=4cm]{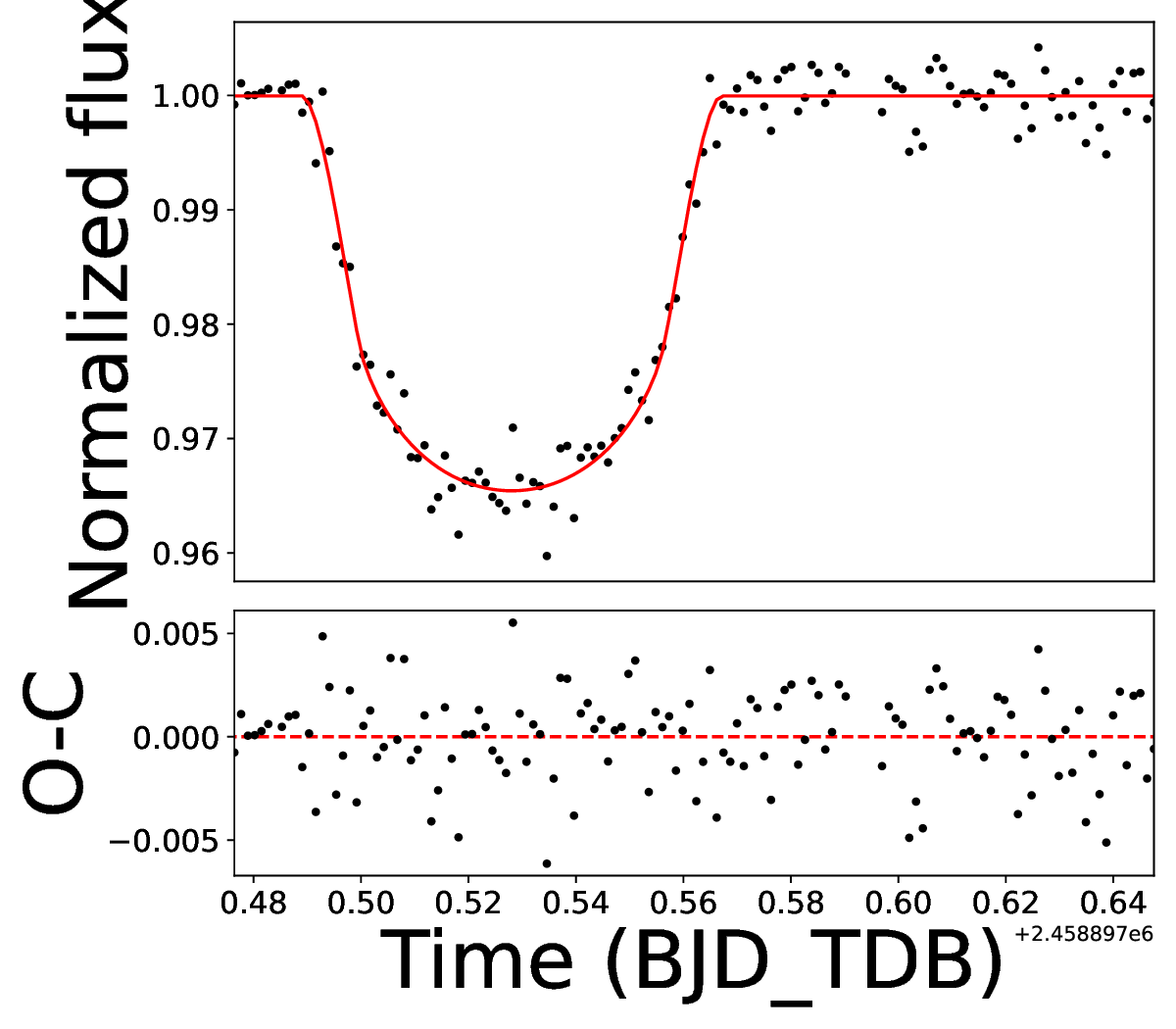}}
  \hfill
    \caption{QATAR-2\,b Light Curves.}
    \label{fig:QATAR-2_lcs}
\end{figure}

\begin{figure}
  \centering
  \subfloat[T100 2020/08/27]{\includegraphics[width=4cm]{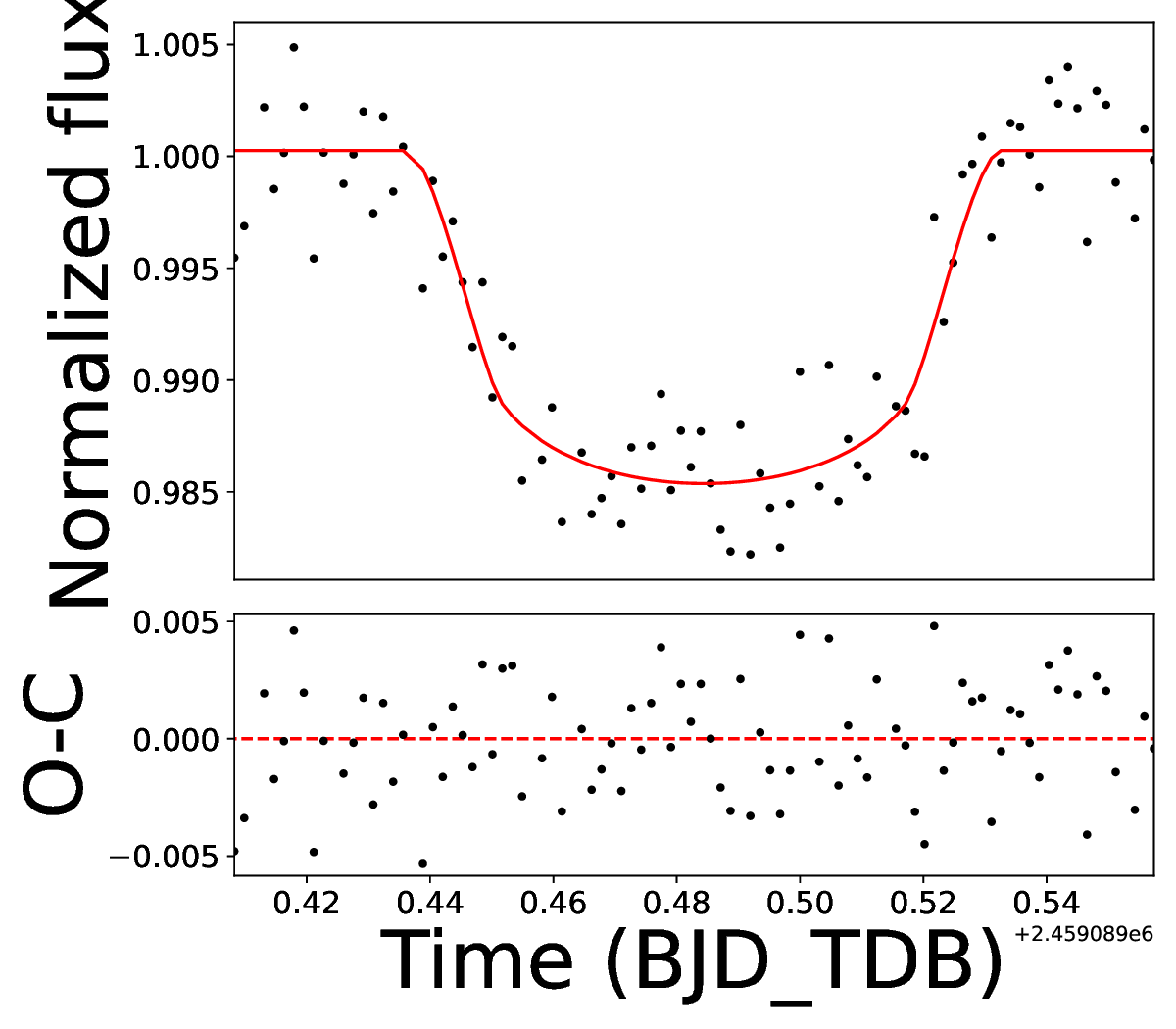}}
  \hfill
    \caption{WASP-44\,b Light Curves.}
    \label{fig:WASP-44_lcs}
\end{figure}

\begin{figure}
  \centering
  \subfloat[T100 2019/10/29]{\includegraphics[width=4cm]{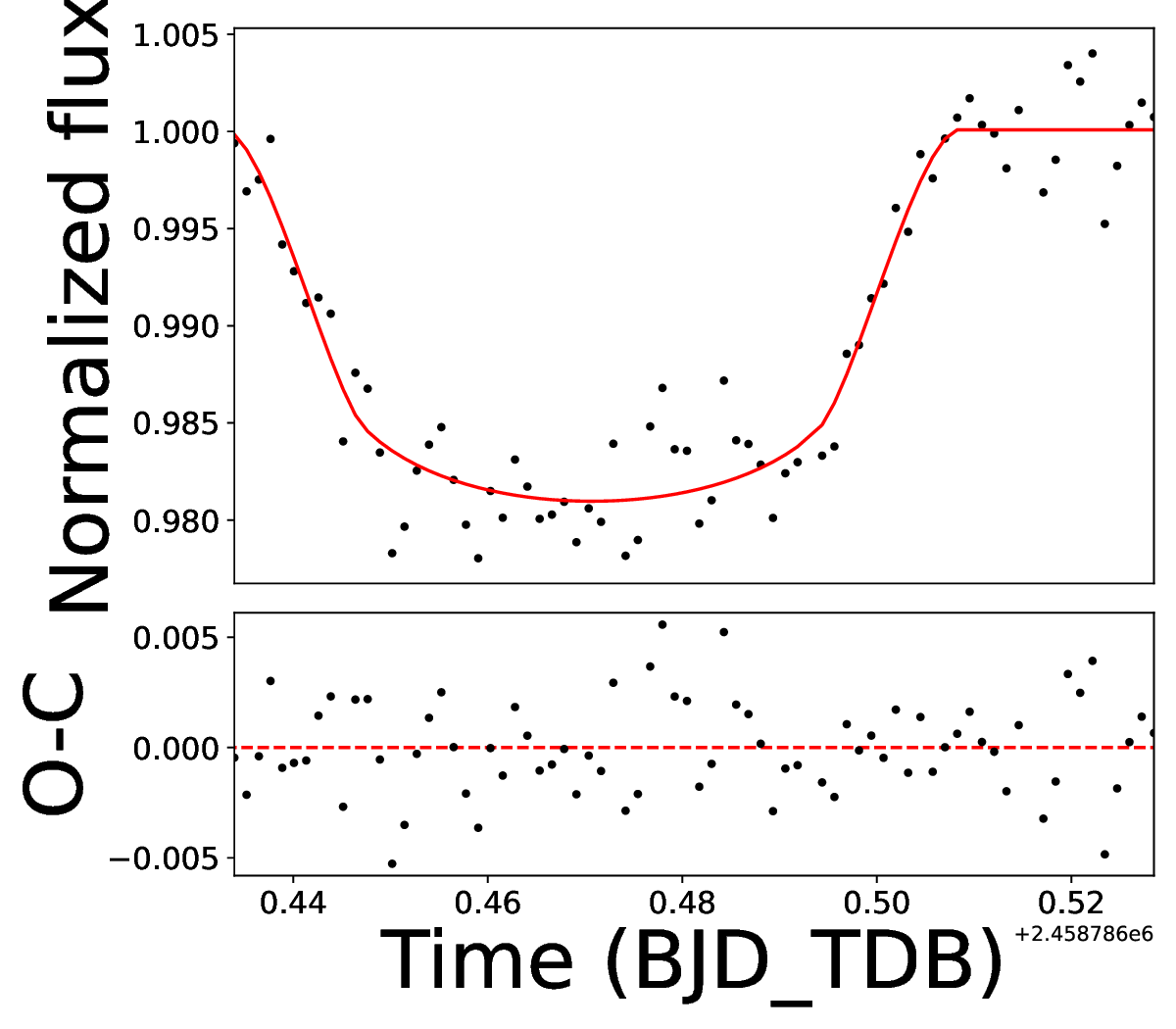}}
  \hfill
  \subfloat[T100 2020/10/09]{\includegraphics[width=4cm]{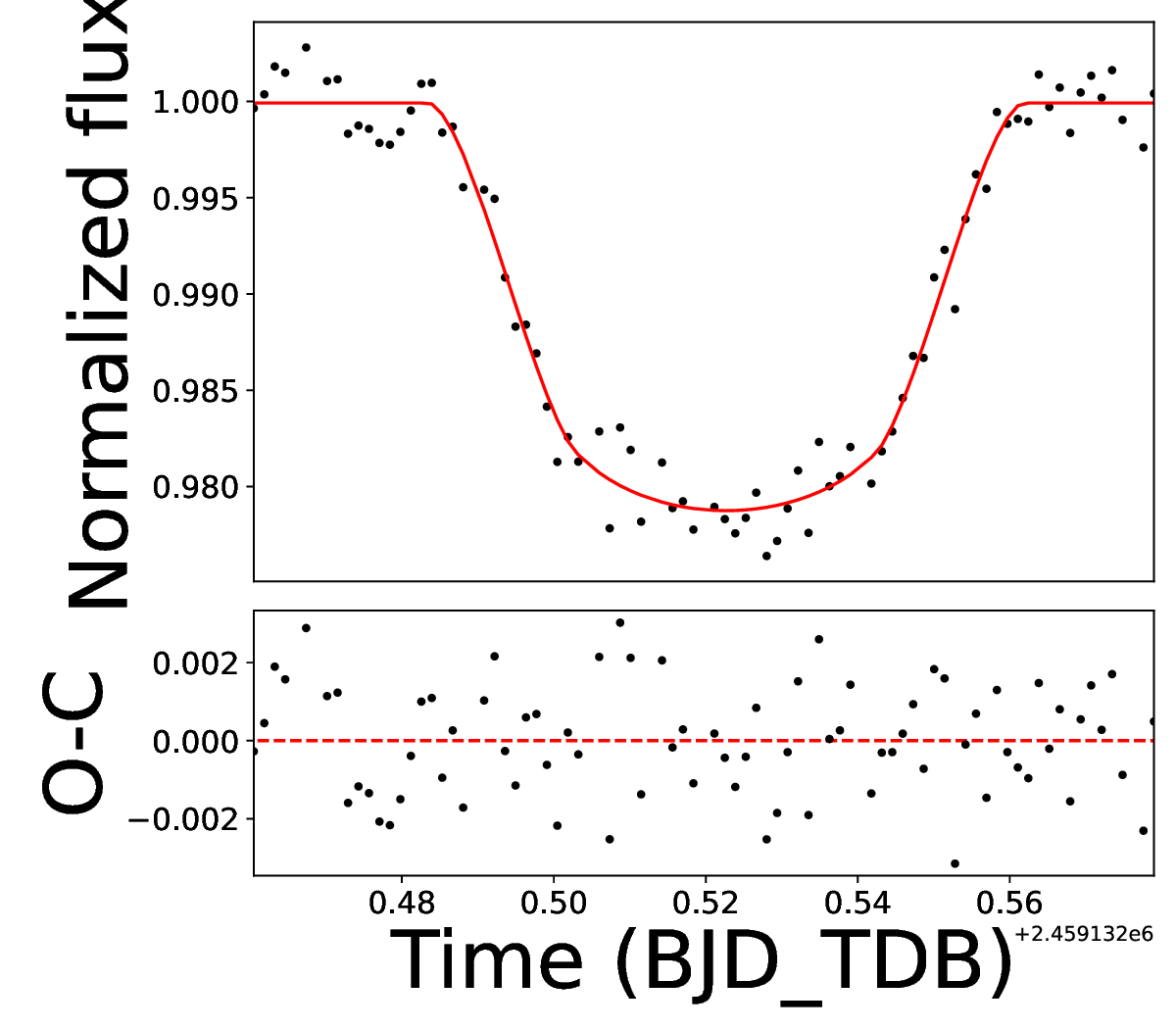}}
  \qquad
    \caption{WASP-50\,b Light Curves.}
    \label{fig:WASP-50_lcs}
\end{figure}

\begin{figure}
  \centering
  \subfloat[ATA50 2020/10/26]{\includegraphics[width=4cm]{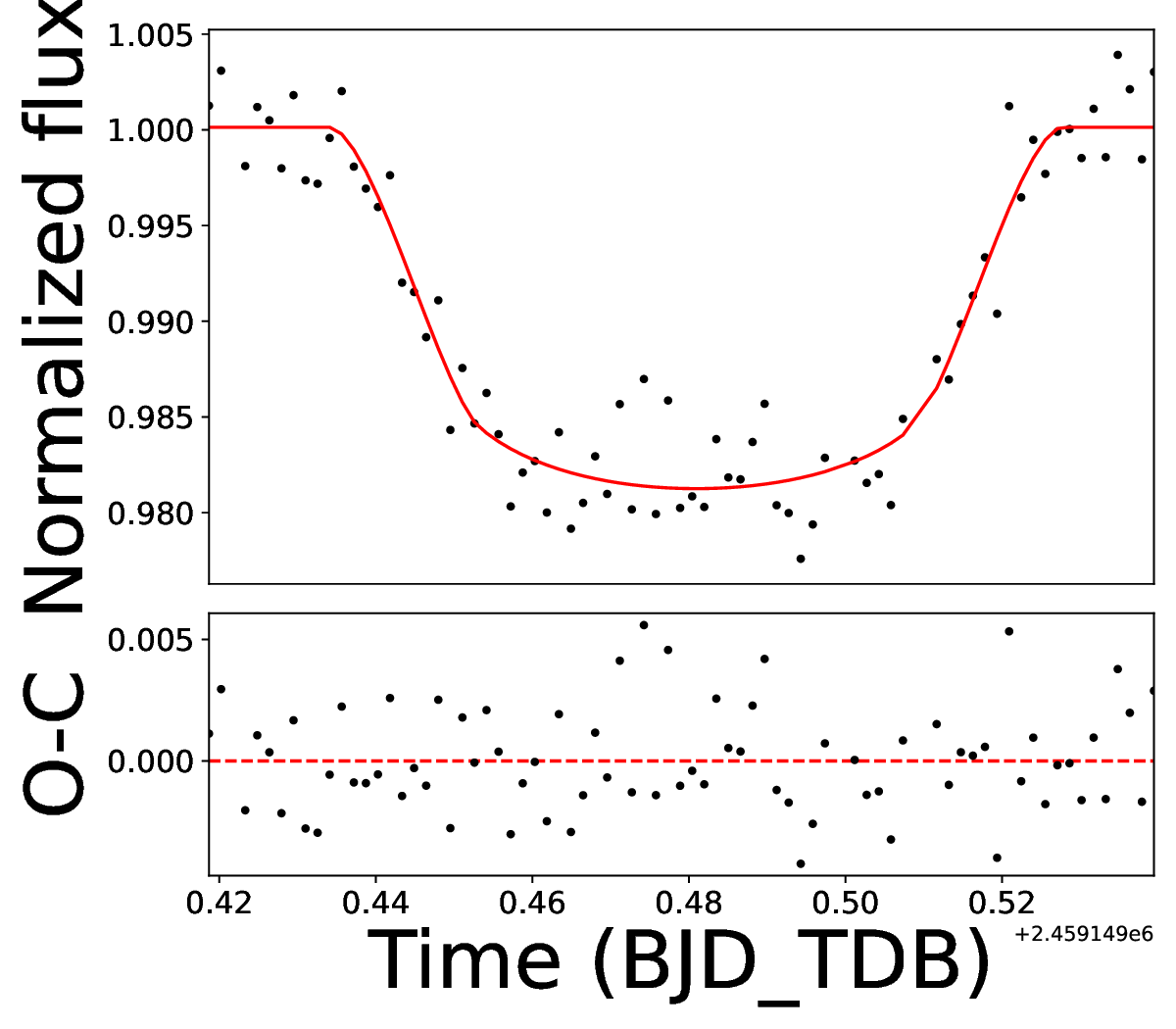}}
  \hfill
  \subfloat[ATA50 2021/10/16]{\includegraphics[width=4cm]{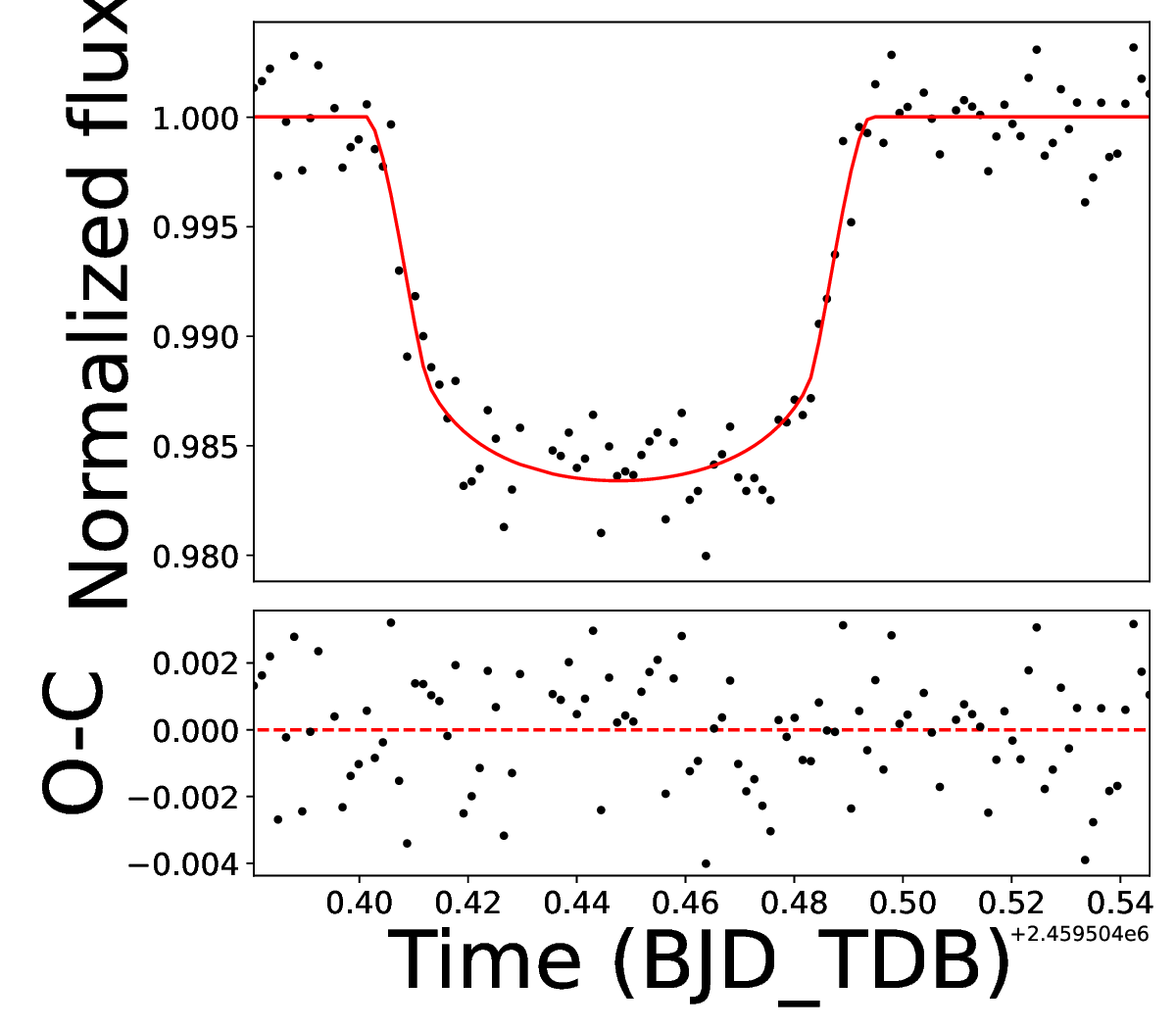}}
  \qquad
    \caption{WASP-77\,b Light Curves.}
    \label{fig:WASP-77_lcs}
\end{figure}

\begin{figure}
  \centering
  \subfloat[T100 2020/01/12]{\includegraphics[width=4cm]{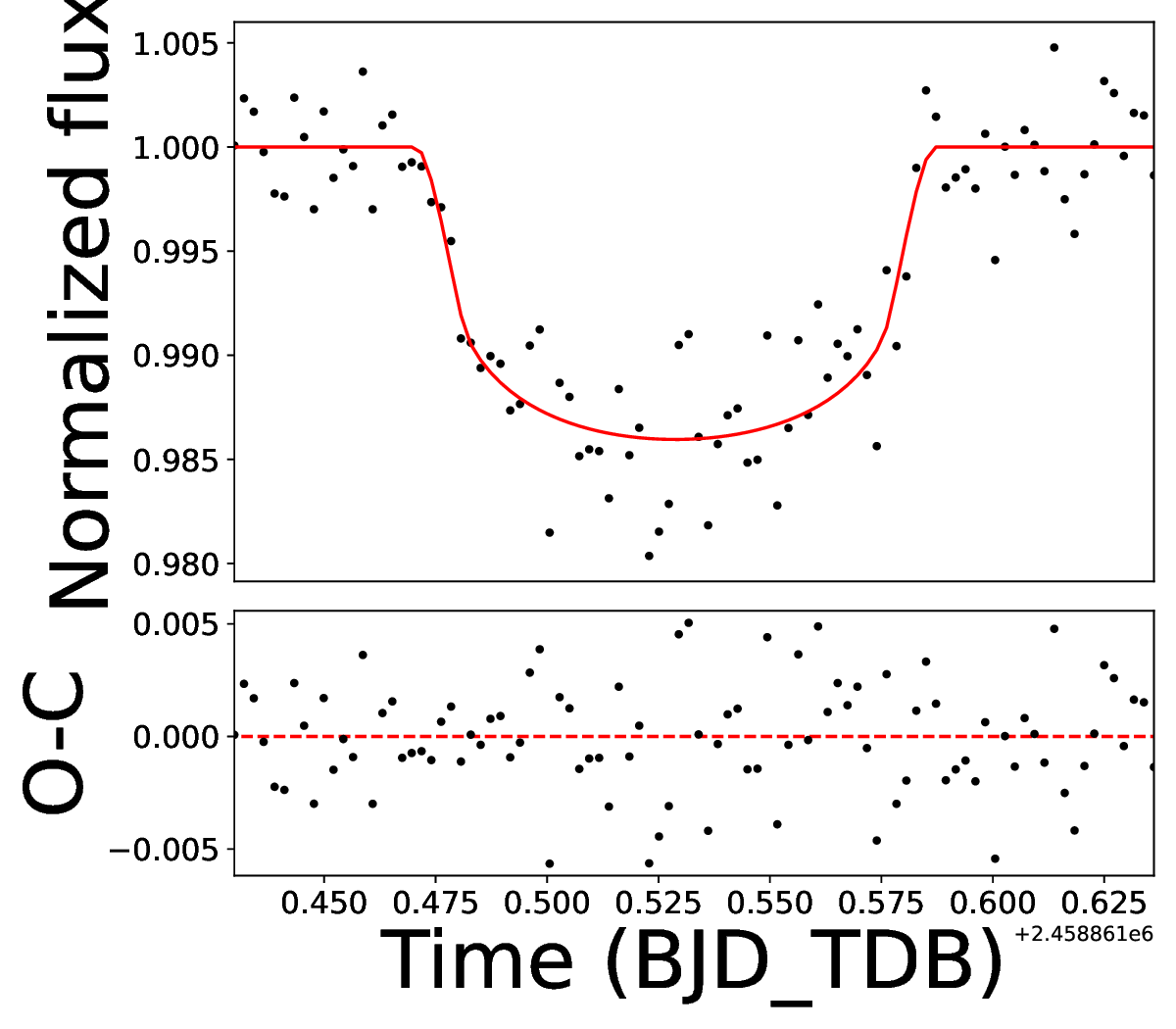}}
  \hfill
  \subfloat[T100 2020/12/25]{\includegraphics[width=4cm]{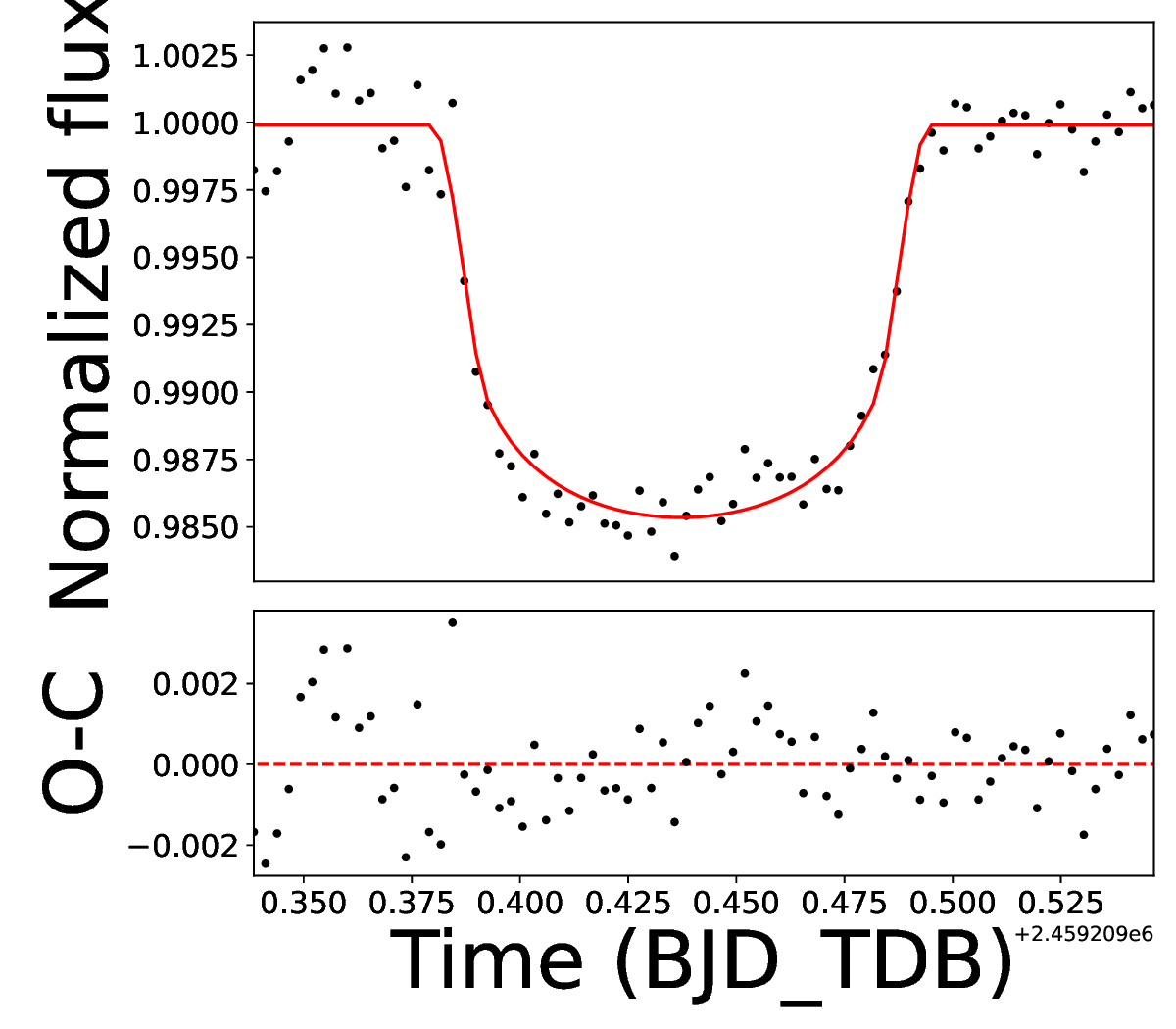}}
  \qquad
  \subfloat[T100 2020/12/25]{\includegraphics[width=4cm]{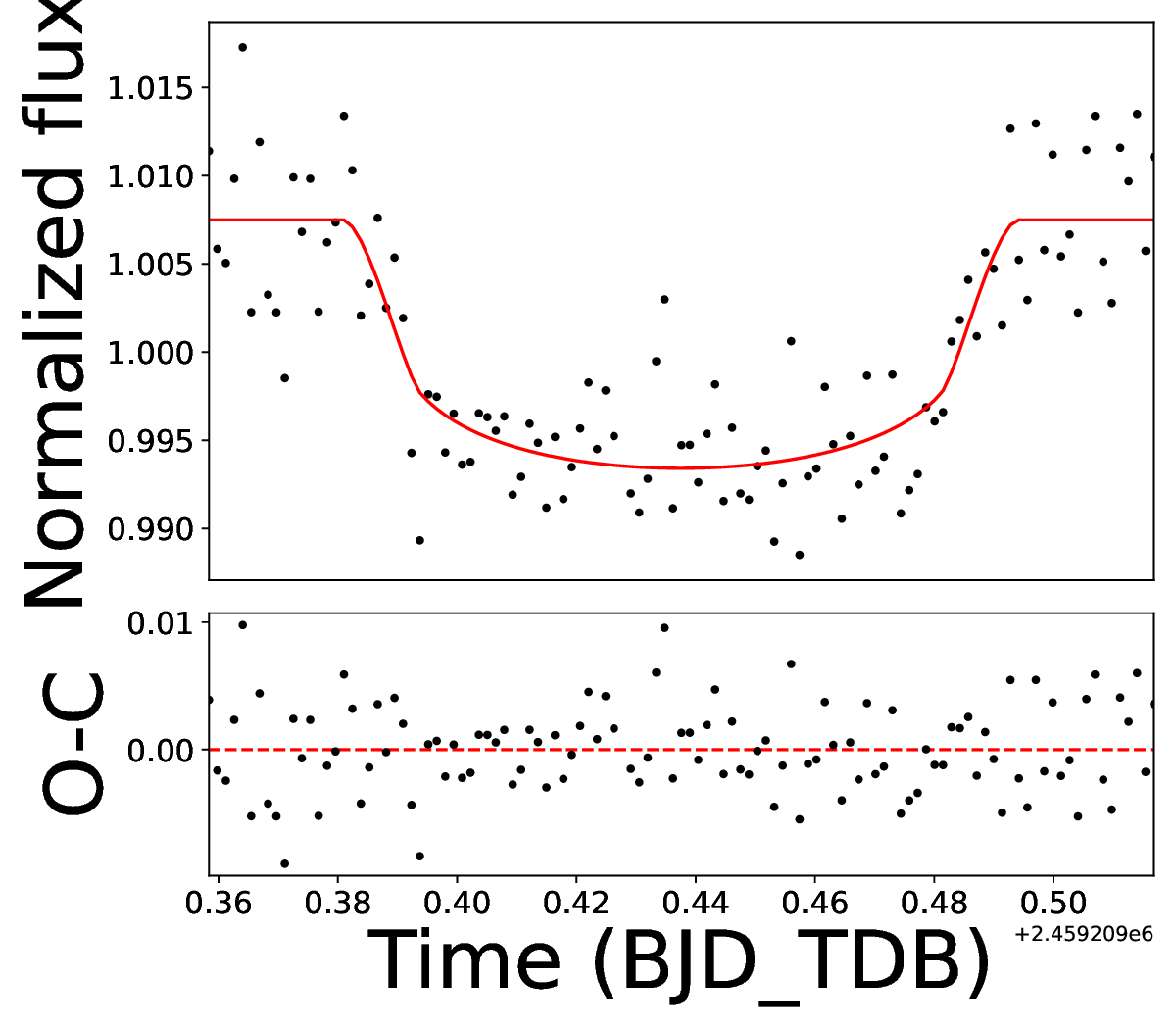}}
  \hfill
    \caption{XO-2\,b Light Curves.}
    \label{fig:XO-2_lcs}
\end{figure}

We provide our own light curves (black data points) and {\sc exofast} models (red continuous curves) in Figs.\ref{fig:GJ1224_lcs} - \ref{fig:XO-2_lcs}.


\begin{figure}
  \centering
  \subfloat[T100 2020/01/12]{\includegraphics[width=4cm]{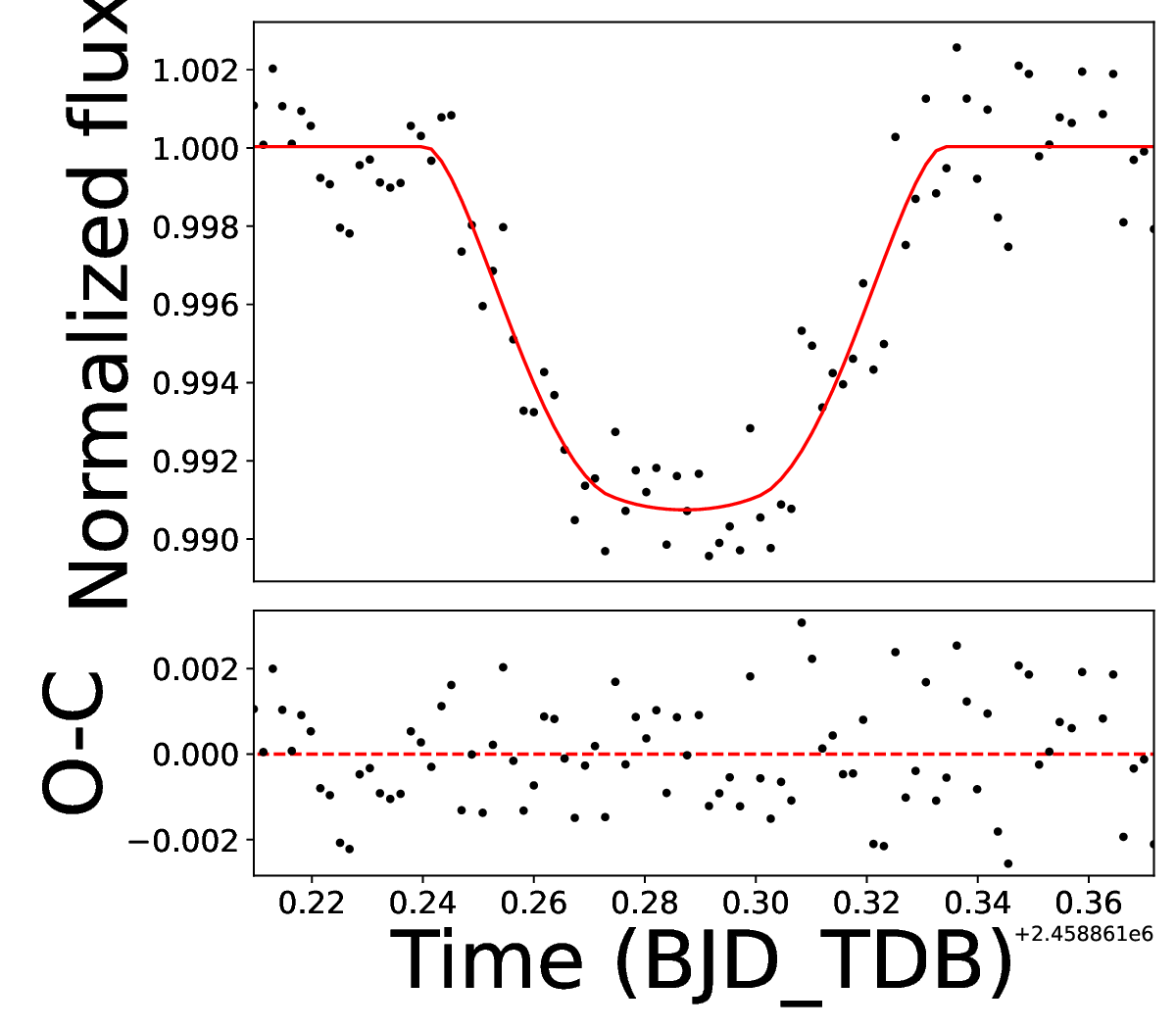}}
  \hfill
  \subfloat[T100 2020/09/22*]{\includegraphics[width=4cm]{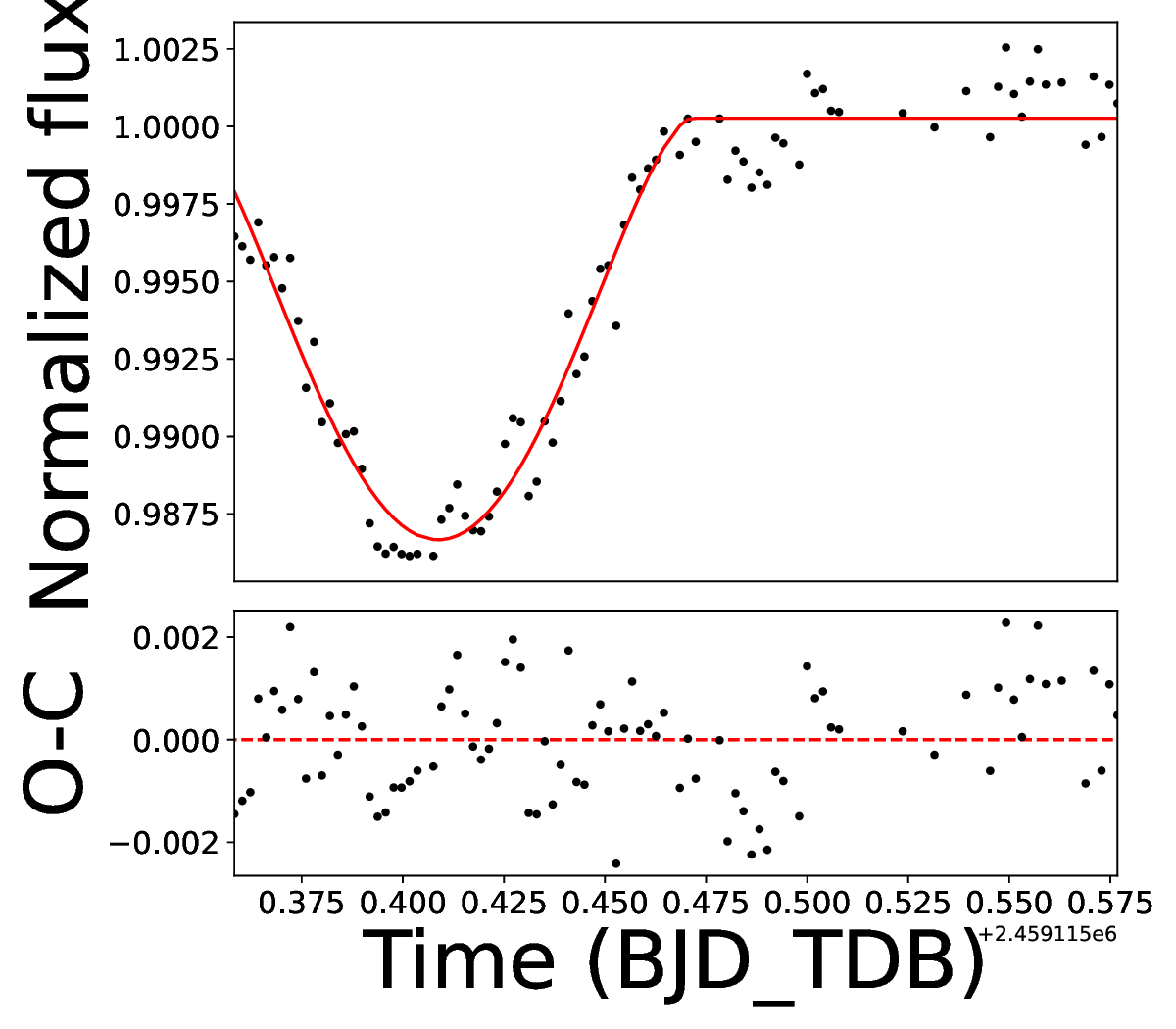}}
  \qquad
  \subfloat[T100 2019/10/30]{\includegraphics[width=4cm]{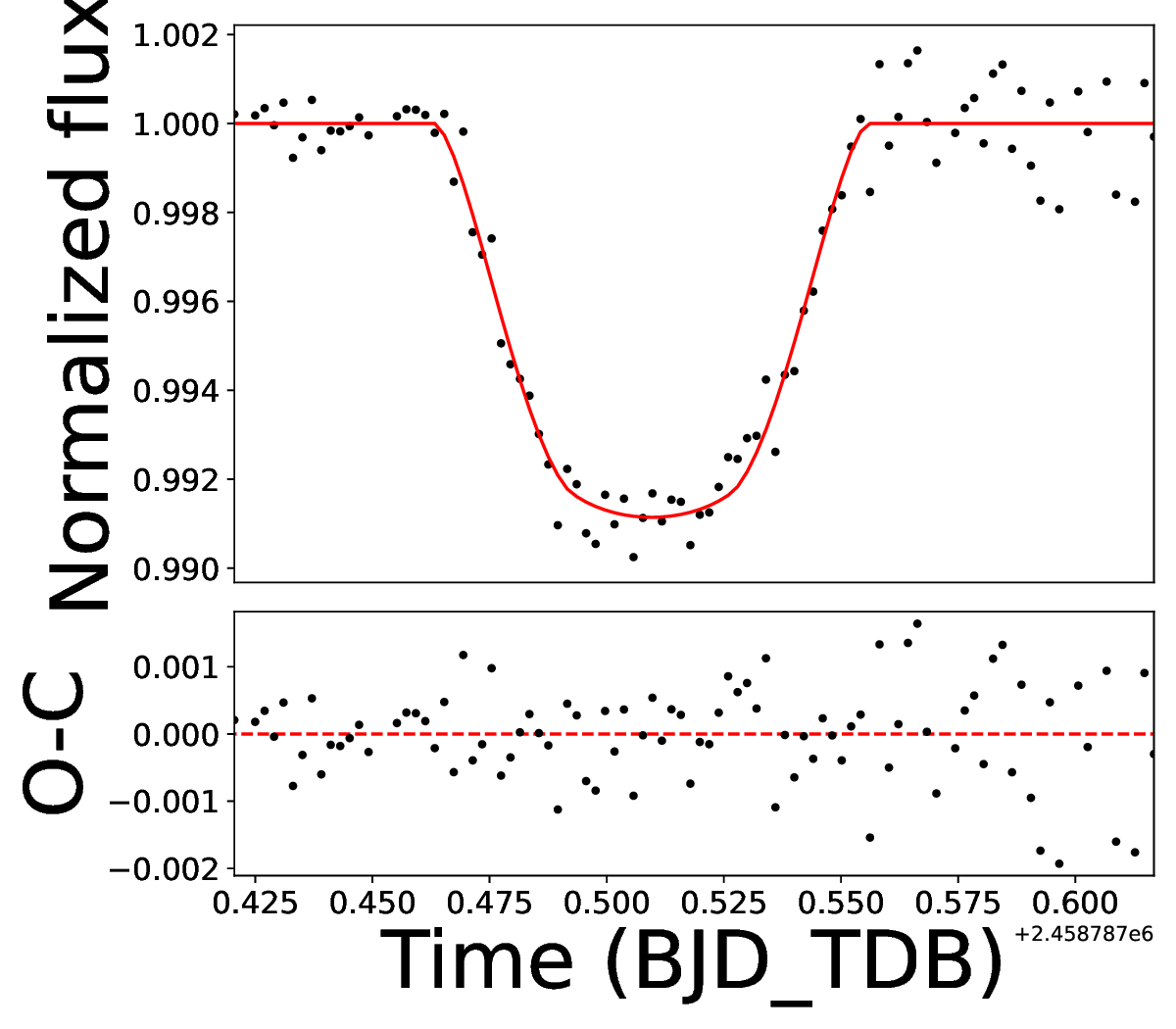}}
  \hfill
  \subfloat[T100 2020/11/24]{\includegraphics[width=4cm]{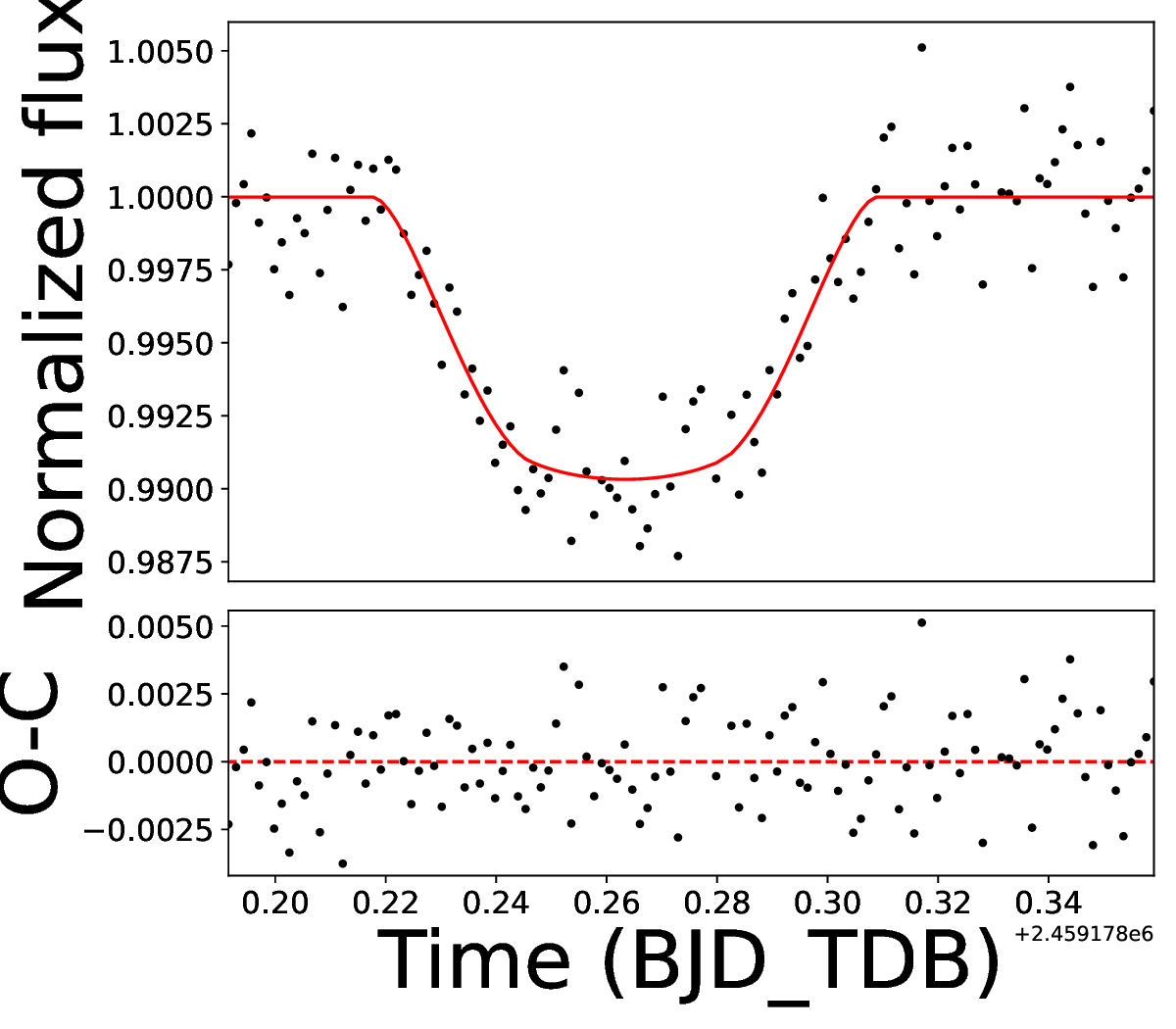}}
  \qquad
  \subfloat[UT50 2020/12/24]{\includegraphics[width=4cm]{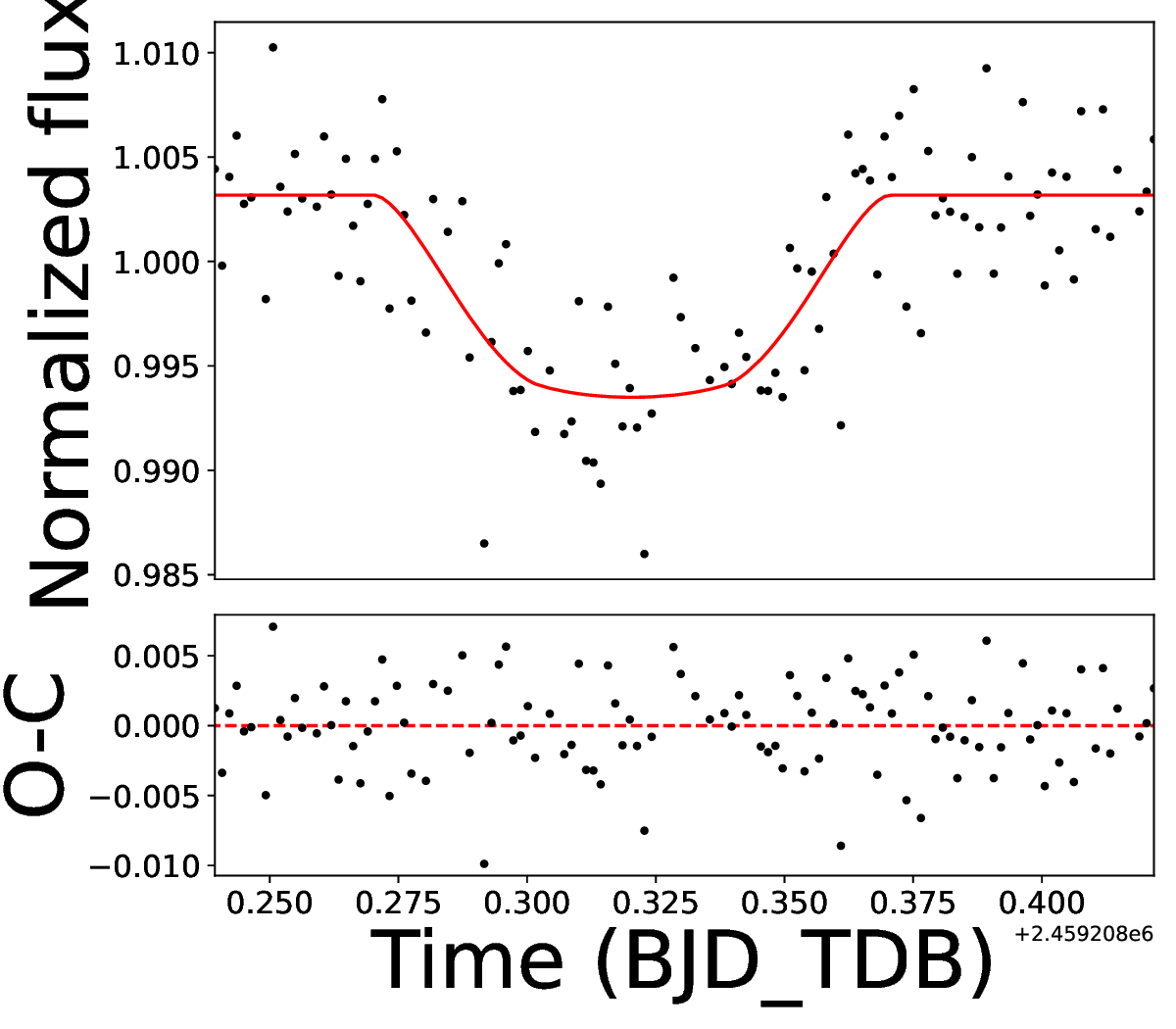}}
  \hfill
  \subfloat[T100 2021/07/25*]{\includegraphics[width=4cm]{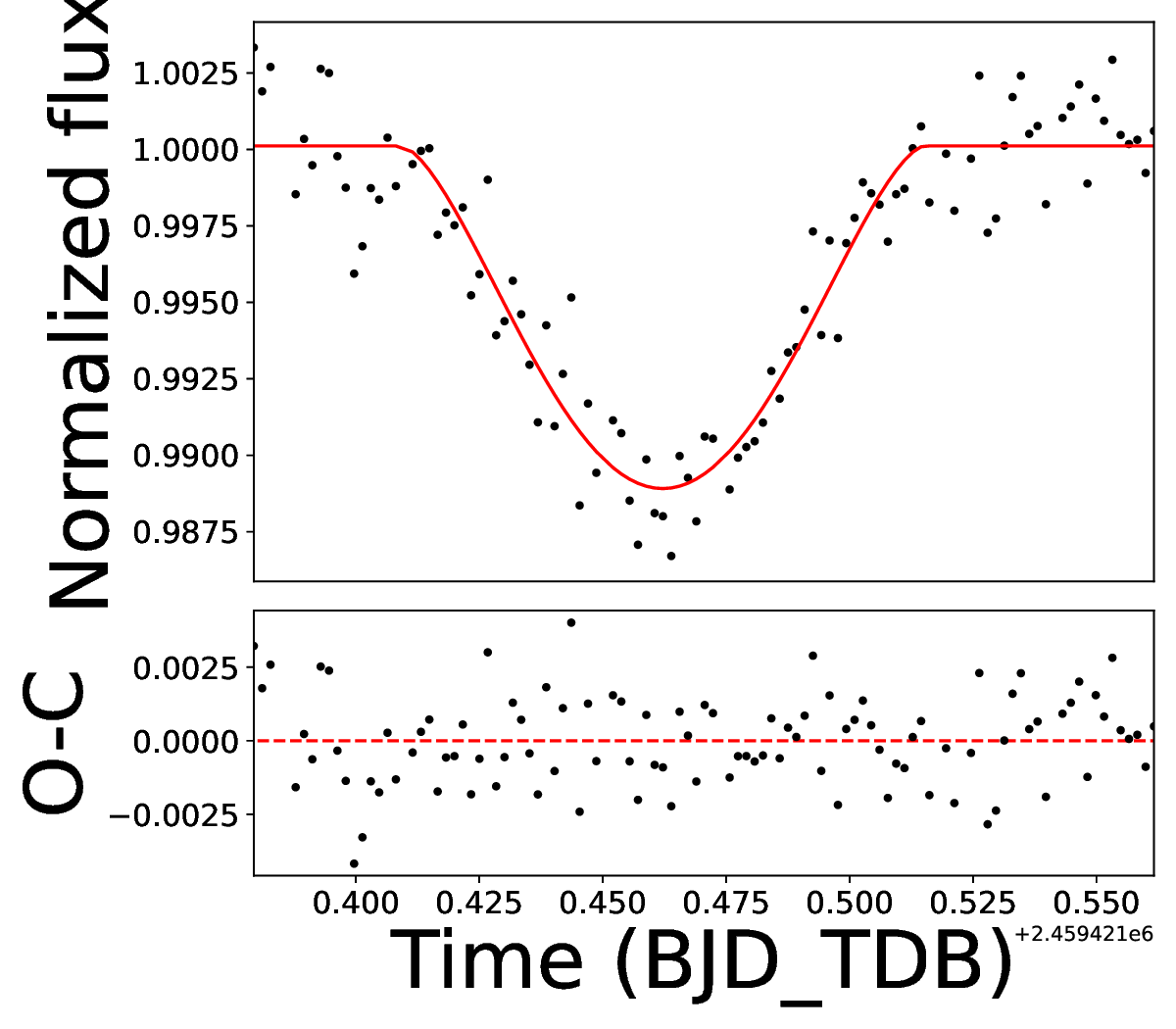}}
  \qquad
  \subfloat[ATA50 2021/09/26]{\includegraphics[width=4cm]{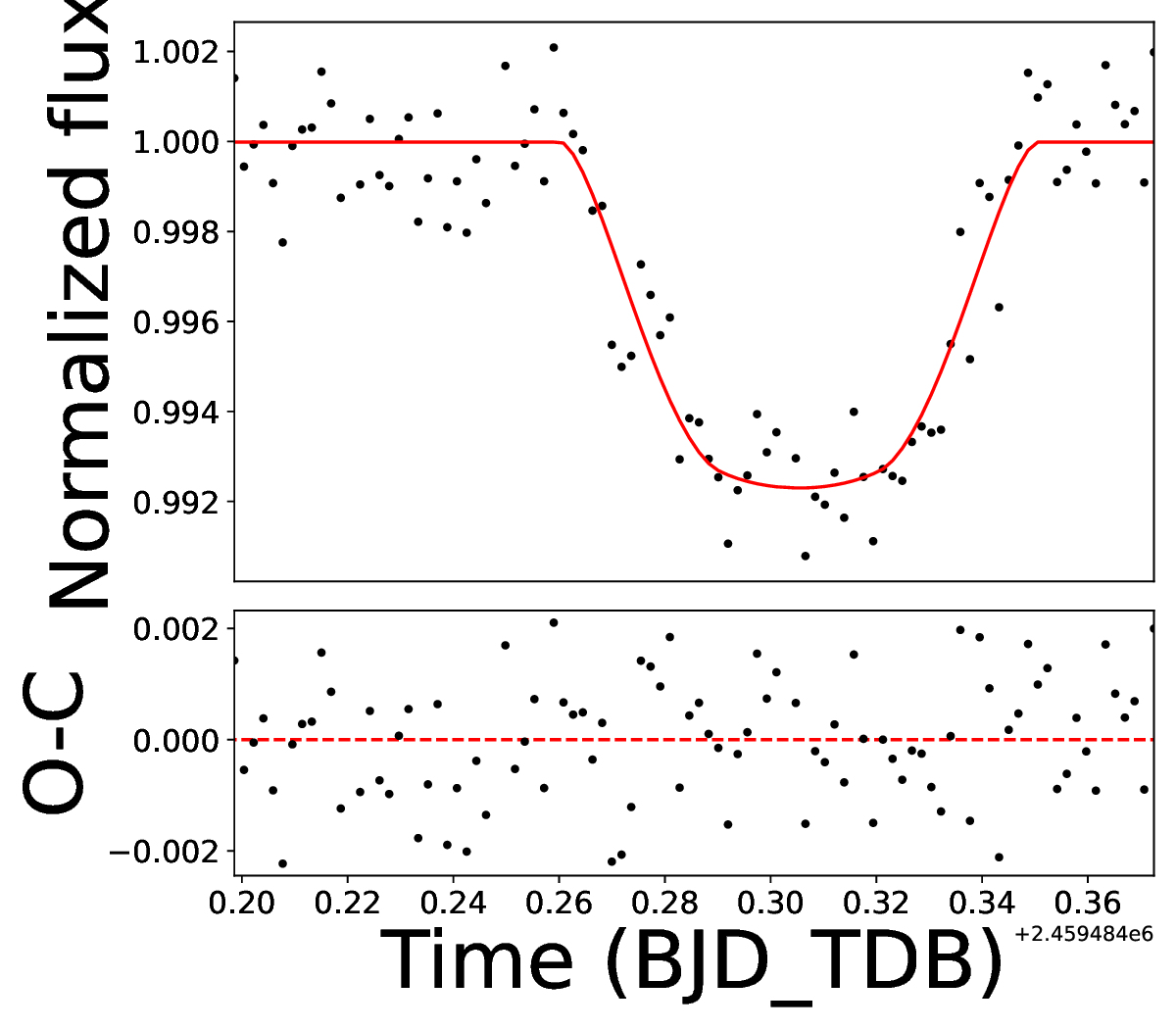}}
  \hfill
    \caption{WASP-93\,b Light Curves.}
    \label{fig:WASP-93_lcs}
\end{figure}

\onecolumn
\section{List of ETD Observers}
\label{sec:appendixb}
\begin{longtable}{|c|c|c|c|c|}
\caption{ETD observers whose light curves were used in this study.} 
\label{tab:ETDobs} \\

\hline {\textbf{System}} & {\textbf{ETD Number}} & {\textbf{Filter}} & {\textbf{Observer}} & {\textbf{TRESCA Protocol Number}} \\ \hline
\endfirsthead

\multicolumn{5}{c}%
{{\bfseries \tablename\ \thetable{} -- continued from previous page}} \\
\hline {\textbf{System}} & {\textbf{ETD Number}} & {\textbf{Filter}} & {\textbf{Observer}} & {\textbf{TRESCA Protocol Number}} \\ \hline
\endhead

\hline \multicolumn{5}{|r|}{{Continued on next page}} \\ \hline
\endfoot

\hline \hline
\endlastfoot
GJ 1214 & ETD44 & Clear & Thomas Sauer & 1278597072  \\ 
GJ 1214 & ETD48 & I & Johannes Ohlert & 1281369862  \\ 
GJ 1214 & ETD71 & Clear & Esseiva Nicolas & 1368129605  \\ 
GJ 1214 & ETD86 & Clear & Fran Campos & 1436715932  \\ 
GJ 1214 & ETD97 & I & Marc Bretton & 1531068175  \\ 
GJ 1214 & ETD105 & Clear & Paul Benni & 1616843398  \\ 
HAT-P-1 & ETD37 & R & Ramon Naves & 1378191469  \\ 
HAT-P-1 & ETD44 & V & Marc Bretton & 1506648757  \\ 
HAT-P-10 & ETD9 & Clear & Luboš Brát & 1252467945  \\ 
HAT-P-10 & ETD18 & Clear & Samuel Durrance, Stacy Irwin & 1292271351  \\ 
HAT-P-10 & ETD21 & Clear & Luboš Brát & 1317438434  \\ 
HAT-P-10 & ETD36 & R & Mark Salisbury & 1350197834  \\ 
HAT-P-10 & ETD39 & Clear & Alfonso Carreno & 1354463447  \\ 
HAT-P-10 & ETD44 & Clear & Paul Benni & 1380843090  \\ 
HAT-P-10 & ETD71 & R & Ferran Grau Horta & 1419237611  \\ 
HAT-P-10 & ETD85 & Clear & David Molina & 1476120518  \\ 
HAT-P-10 & ETD87 & R &  Wonseok Kang & 1484834873  \\ 
HAT-P-10 & ETD88 & R &  Wonseok Kang & 1484831725  \\ 
HAT-P-10 & ETD99 & V & Yves Jongen & 1544871912  \\ 
HAT-P-10 & ETD100 & Clear & Bruno Fontaine & 1546017209  \\ 
HAT-P-10 & ETD101 & Clear &  Yves Jongen & 1567668774  \\ 
HAT-P-10 & ETD102 & R &  Veli-Pekka Hentunen & 1569003995  \\ 
HAT-P-10 & ETD103 & V & Yves Jongen & 1569911364  \\ 
HAT-P-10 & ETD113 & Clear & Matthieu Bachschmidt & 1638791203  \\ 
HAT-P-13 & ETD39 & R & Ramon Naves & 1296839849  \\ 
HAT-P-13 & ETD82 & Clear & Marc Bretton & 1446972948  \\ 
HAT-P-13 & ETD100 & Clear & Manfred Raetz & 1579296693  \\ 
HAT-P-13 & ETD104 & R & Manfred Raetz & 1585809439  \\ 
HAT-P-13 & ETD107 & R & Yves Jongen & 1612792652  \\ 
HAT-P-16 & ETD10 & R & Ruth ODougherty Sanchez & 1291563452  \\ 
HAT-P-16 & ETD13 & Clear & Stan Shadick & 1318731754  \\ 
HAT-P-16 & ETD15 & Clear & Jaroslav Trnka & 1317463132  \\ 
HAT-P-16 & ETD16 & R & Thomas Sauer & 1317639008  \\ 
HAT-P-16 & ETD2 & Clear & Jaroslav Trnka & 1278846773  \\ 
HAT-P-16 & ETD23 & I & Stan Shadick & 1349233966  \\ 
HAT-P-16 & ETD27 & R & Petri Kehusmaa, Caisey Harlingten & 1351258657  \\ 
HAT-P-16 & ETD31 & R & Mark Salisbury & 1378590453  \\ 
HAT-P-16 & ETD36 & Clear & Stan Shadick & 1383606249  \\ 
HAT-P-16 & ETD37 & Clear & Paul Benni & 1383673356  \\ 
HAT-P-16 & ETD38 & Clear & Anthony Ayiomamitis & 1383875650  \\ 
HAT-P-16 & ETD39 & R & Francesco Scaggiante, Danilo Zardin & 1386236706  \\ 
HAT-P-16 & ETD45 & Clear & Luca Rizzuti & 1440519027  \\ 
HAT-P-16 & ETD46 & Clear & David Molina & 1446965044  \\ 
HAT-P-16 & ETD47 & R & Fran Campos & 1448107877  \\ 
HAT-P-16 & ETD5 & Clear & Martin Vrašťák & 1285498403  \\ 
HAT-P-16 & ETD57 & R & Ramon Naves & 1473584609  \\ 
HAT-P-16 & ETD59 & Clear & Trnka J. & 1477940875  \\ 
HAT-P-16 & ETD6 & Clear & J. Világi, Š. Gajdoš & 1288224520  \\ 
HAT-P-16 & ETD60 & Clear & David Molina & 1479143435  \\ 
HAT-P-16 & ETD61 & I & Kevin B. Alton & 1507651821  \\ 
HAT-P-16 & ETD63 & R & Veli-Pekka Hentunen & 1534770239  \\ 
HAT-P-16 & ETD64 & Clear & Anael Wunsche & 1537205125  \\ 
HAT-P-16 & ETD65 & R & Josep Gaitan & 1545555524  \\ 
HAT-P-16 & ETD69 & V & Yves Jongen & 1567852468  \\ 
HAT-P-16 & ETD70 & R & Francesco Scaggiante, Danilo Zardin & 1569339710  \\ 
HAT-P-16 & ETD70 & R & Marco Fiaschi & 1569339710  \\ 
HAT-P-16 & ETD7 & I & Stan Shadick & 1286159180  \\ 
HAT-P-16 & ETD72 & V & Mario Morales & 1571209785  \\ 
HAT-P-16 & ETD73 & V & Truman State Observer & 1583901308  \\ 
HAT-P-16 & ETD76 & Clear & Anael Wunsche & 1596031262  \\ 
HAT-P-16 & ETD78 & R & Yves Jongen & 1599988642  \\ 
HAT-P-16 & ETD79 & R & Yves Jongen & 1602144863  \\ 
HAT-P-16 & ETD80 & V & Snaevarr Gudmundsson & 1605914982  \\ 
HAT-P-16 & ETD9 & Clear & Stan Shadick & 1288134822  \\ 
HAT-P-22 & ETD53 & B & Yves Jongen & 1550908562  \\ 
HAT-P-22 & ETD55 & R & Josep Gaitan & 1552034513  \\ 
HAT-P-22 & ETD68 & R & Manfred Raetz & 1586181879  \\ 
HAT-P-22 & ETD78 & I & Manfred Raetz & 1614798234  \\ 
HAT-P-22 & ETD79 & Clear & Matthieu Bachschmidt & 1617521673  \\ 
HAT-P-30 & ETD14 & Clear & Stan Shadick & 1331321038  \\ 
HAT-P-30 & ETD15 & I & Stan Shadick & 1333050479  \\ 
HAT-P-30 & ETD16 & Clear & Juanjo Gonzalez & 1329671594  \\ 
HAT-P-30 & ETD18 & R & Stan Shadick & 1331919572  \\ 
HAT-P-30 & ETD22 & Clear & Stan Shadick & 1358100549  \\ 
HAT-P-30 & ETD26 & I & Giuseppe Marino & 1392576172  \\ 
HAT-P-30 & ETD30 & R & Andre Christophe, Clement Jacques & 1420878866  \\ 
HAT-P-30 & ETD30 & R & Nougayrede Jean-Philippe & 1420878866  \\ 
HAT-P-30 & ETD32 & R & Marc Bretton & 1425771388  \\ 
HAT-P-30 & ETD33 & Clear & Martin Zibar & 1425774283  \\ 
HAT-P-30 & ETD39 & R & Francesco Scaggiante, Danilo Zardin & 1516875191  \\ 
HAT-P-30 & ETD41 & Clear & Marc Bretton & 1519263504  \\ 
HAT-P-30 & ETD47 & V & Yves Jongen & 1580746869  \\ 
HAT-P-30 & ETD51 & Clear & Marc Bretton & 1586081207  \\ 
HAT-P-30 & ETD52 & R & Yves Jongen & 1607422166  \\ 
HAT-P-30 & ETD53 & R & Jean-Claude Mario & 1607859317  \\ 
HAT-P-30 & ETD55 & R & Yves Jongen & 1612549479  \\ 
HAT-P-30 & ETD56 & R & Yves Jongen & 1612791934  \\ 
HAT-P-30 & ETD58 & R &  Josep Gaitan & 1613042887  \\ 
HAT-P-30 & ETD59 & R & Anael Wunsche & 1613037601  \\ 
HAT-P-53 & ETD11 & R & P. Farissier, S. Combe, L. Bret-Morel & 1463150362  \\ 
HAT-P-53 & ETD11 & R & C. Gillier, R. Montaigut & 1463150362  \\ 
HAT-P-53 & ETD18 & Clear & Marc Bretton & 1480634960  \\ 
HAT-P-53 & ETD4 & Clear & Marc Bretton & 1440128332  \\ 
HAT-P-53 & ETD43 & Clear & Marc Deldem & 1600626997  \\ 
HAT-P-53 & ETD45 & Clear & Yves Jongen & 1599293169  \\ 
HAT-P-53 & ETD48 & Clear & Manfred Raetz & 1605559616  \\ 
HAT-P-53 & ETD50 & R & Jordi Lopesino & 1609010356  \\ 
KELT-3 & ETD2 & R & Ramon Naves & 1357282983  \\ 
KELT-3 & ETD20 & V & Marc Bretton & 1428978215  \\ 
KELT-3 & ETD24 & Clear & David Molina, Antoni Vives Sureda & 1461153031  \\ 
KELT-3 & ETD29 & R & Josep Gaitan & 1553504934  \\ 
KELT-3 & ETD3 & Clear & Anthony Ayiomamitis & 1362940487  \\ 
KELT-3 & ETD6 & Clear & Paul Benni & 1387025732  \\ 
KELT-3 & ETD9 & Clear & Gustavo Javier, Muler Schteinman & 1393492000  \\ 
Qatar-2 & ETD19 & Clear & Nico Montigiani, Massimiliano Mannucci & 1335686620  \\ 
Qatar-2 & ETD30 & Clear & C. Colazo, R. Melia, N. Marcionni & 1369893611  \\ 
Qatar-2 & ETD37 & Clear & Thomas Sauer & 1403609906  \\ 
Qatar-2 & ETD70 & Clear & Yves Jongen & 1581427637  \\ 
Qatar-2 & ETD83 & Clear & Yves Jongen & 1591620453  \\ 
Qatar-2 & ETD87 & Clear & Yves Jongen & 1618771546  \\ 
Qatar-2 & ETD90 & Clear & Yves Jongen & 1619795469  \\ 
WASP-8 & ETD2 & R & Yves Jongen & 1605261835  \\ 
WASP-44 & ETD12 & Clear &  Bernasconi Laurent & 1477991767  \\ 
WASP-44 & ETD18 & IR-UV & Yves Jongen & 1538296133  \\ 
WASP-44 & ETD2 & Clear & Phil Evans & 1315683470  \\ 
WASP-44 & ETD20 & Clear & Laloum Didier & 1569947071  \\ 
WASP-44 & ETD22 & IR-UV & Yves Jongen & 1594936390  \\ 
WASP-44 & ETD23 & R & Yves Jongen & 1596286136  \\ 
WASP-44 & ETD25 & Clear & Yves Jongen & 1626442442  \\ 
WASP-44 & ETD26 & V & Anael Wunsche & 1626433743  \\ 
WASP-44 & ETD27 & Clear & Yves Jongen & 1627914023  \\ 
WASP-44 & ETD4 & Clear & František Lomoz & 1316937777  \\ 
WASP-44 & ETD5 & R & Thomas Sauer & 1342738906  \\ 
WASP-44 & ETD6 & Clear & Phil Evans & 1380404329  \\ 
WASP-44 & ETD7 & Clear & Rene Roy & 1386089485  \\ 
WASP-50 & ETD11 & V & Christopher Allen & 1354306402  \\ 
WASP-50 & ETD16 & V & Vanessa Logan and Karen Lewis & 1392655969  \\ 
WASP-50 & ETD18 & Clear & Esseiva Nicolas & 1384452957  \\ 
WASP-50 & ETD29 & V & Andrew Stewart & 1494365750  \\ 
WASP-50 & ETD3 & Clear & Fernando Tifner & 1320032307  \\ 
WASP-50 & ETD33 & R & Tianyu Ma & 1479154021  \\ 
WASP-50 & ETD36 & Clear & Bernasconi Laurent & 1478164712  \\ 
WASP-50 & ETD37 & R &  Christoper Michael & 1481483736  \\ 
WASP-50 & ETD4 & Clear & Parijat Singh & 1323109321  \\ 
WASP-50 & ETD44 & Clear & C. Colazo, R. Melia, M. Starck & 1514898205  \\ 
WASP-50 & ETD45 & R & Marc Bretton & 1542158605  \\ 
WASP-50 & ETD55 & V & Yves Jongen & 1575812213  \\ 
WASP-50 & ETD56 & V & Yves Jongen & 1579635308  \\ 
WASP-50 & ETD57 & R & Yves Jongen & 1601116953  \\ 
WASP-50 & ETD58 & R & Yves Jongen & 1601412777  \\ 
WASP-50 & ETD59 & R & Yves Jongen & 1601502007  \\ 
WASP-50 & ETD6 & V & Nicole Makely, Melissa Hutcheson & 1340302725  \\ 
WASP-50 & ETD60 & R & Yves Jongen & 1605913925  \\ 
WASP-50 & ETD63 & Clear & Esseiva nicolas & 1606397868  \\ 
WASP-77 & ETD1 & Clear & Juanjo Gonzalez & 1376989484  \\ 
WASP-77 & ETD16 & Clear & David Molina & 1475258342  \\ 
WASP-77 & ETD18 & I & Phil Evans & 1478155148  \\ 
WASP-77 & ETD2 & Clear & Paul Benni & 1383268828  \\ 
WASP-77 & ETD22 & R & Josep Gaitan & 1481360368  \\ 
WASP-77 & ETD23 & Clear & Š. Gajdoš, J. Šubjak & 1488671491  \\ 
WASP-77 & ETD24 & V & Napoleao T., Silva S., Kulh D. & 1539901020  \\ 
WASP-77 & ETD27 & R & Pavel Pintr & 1578518743  \\ 
WASP-77 & ETD28 & R & Yves Jongen & 1596751821  \\ 
WASP-77 & ETD29 & R & Yves Jongen & 1602265968  \\ 
WASP-77 & ETD3 & Clear & Paul Benni & 1385854480  \\ 
WASP-77 & ETD30 & R & Yves Jongen & 1602850422  \\ 
WASP-77 & ETD6 & V & Ferran Grau Horta & 1386541771  \\ 
WASP-93 & ETD21 & R & Mark Salisbury & 1509911995  \\ 
WASP-93 & ETD42 & V & Yves Jongen & 1569913117  \\ 
WASP-93 & ETD49 & R & Josep Gaitan & 1601800673  \\ 
WASP-93 & ETD8 & R & Mark Salisbury & 1475784797  \\ 
WASP-93 & ETD9 & Clear & Marc Deldem & 1475523697  \\ 
XO-2 & ETD71 & R & Ramon Naves & 1295965000  \\ 
XO-2 & ETD77 & R & Thomas Sauer & 1300726462  \\ 
XO-2 & ETD110 & R & Jacques Michelet & 1364137719  \\ 
XO-2 & ETD124 & Clear & Martin Zibar & 1429649013  \\ 
XO-2 & ETD132 & Clear & Trnka J. & 1483125364  \\ 
XO-2 & ETD133 & R &  Wonseok Kang & 1484831673  \\ 
XO-2 & ETD142 & Clear & Stan Shadick & 1522015555  \\ 
XO-2 & ETD143 & I & Marc Bretton & 1523058603  \\ 
XO-2 & ETD154 & Clear &  Joe Garlitz & 1552959381  \\ 
XO-2 & ETD170 & Clear & Manfred Raetz & 1585437743  \\ 
XO-2 & ETD172 & R & Manfred Raetz & 1586597501  \\ 
\end{longtable}
\twocolumn

\onecolumn
\begin{landscape}
\section{List of Information About Light Curves analyzed In This Study}
We present the detailed information about the light curves analyzed in this study in \ref{table_gj1214} for GJ 1214 system. Same table for other systems within this study can be found in online material of the journal with a read-me file describing the columns.

\begin{table}
\caption{Detailed information about GJ 1214 light curves analyzed in this study}
\label{table_gj1214}
\centering
\resizebox{\columnwidth}{!}{\begin{tabular}{ccccccccccccccccc}
\hline
\textbf{Beta}             & \textbf{PNR (ppt)} & \textbf{Source} & \textbf{Depth} & \textbf{T$_{14}$} & \textbf{T$_c$ error}      & \textbf{T$_{ingress/egress}$ error}    & \textbf{Depth error}   & \textbf{T$_{ingress/egress}$}     & \textbf{T$_c$ (BJD\_TDB)}       & \textbf{Filter}         & \textbf{Type}       & \textbf{Discard} & \textbf{Removed} & \textbf{Epoch} & \textbf{ O-C (min.)}                 & Linear Residual (min.)      \\ \hline
1.54002179740703  & 2.90501157258658  & ETD44                  & 0.01918  & 0.039077 & 0.00038267613  & 0.001325629   & 0.00059327354  & 0.010842 & 2455385.3323   & Clear          & ETD        & 10      & 1       &       &                    &                    \\
1.41497140877592  & 1.56266457657402  & ETD48                  & 0.015165 & 0.038365 & 0.00017306575  & 0.00059951733 & 0.00025728618  & 0.006539 & 2455396.395716 & I              & ETD        & 0       & 0       & -255  & -1.75953648984432  & 0.504614573294904  \\
1.95335549554965  & 1.12704441253944  & ETD71                  & 0.014853 & 0.044947 & 0.00066474353  & 0.0023027391  & 0.00075196981  & 0.009814 & 2456420.497064 & Clear          & ETD        & 0       & 0       & 393   & -3.16405445337296  & -0.503744490372933 \\
1.73996684084739  & 1.02451060166468  & ETD86                  & 0.014383 & 0.035416 & 0.00079393994  & 0.0027502886  & 0.0014203748   & 0.004132 & 2457215.438891 & Clear          & ETD        & 1       & 1       & 896   & -5.75239688158035  & -2.78457468612666  \\
1.44406723953075  & 3.01927543583687  & ETD97                  & 0.016271 & 0.038981 & 0.00032347593  & 0.0011205535  & 0.00054037784  & 0.005701 & 2458307.500585 & I              & ETD        & 0       & 0       & 1587  & -2.92512930929661  & 0.465140108076548  \\
1.92712266700976  & 2.17085192544031  & ETD105                 & 0.015684 & 0.037229 & 0.00039431705  & 0.0013659543  & 0.00063640066  & 0.00606  & 2459296.833314 & Clear          & ETD        & 0       & 0       & 2213  & -3.916310146451    & -0.14333166840985  \\
2.19062931990964  & 1.23194238093312  & Kundurthy et al 2011   & 0.014553 & 0.036928 & 0.000090594633 & 0.00031382901 & 0.0001491875   & 0.004873 & 2455307.892618 & sdssr          & Literature & 0       & 0       & -311  & -2.37597115337849  & -0.146056044548218 \\
0.971768878384822 & 4.69936153795429  & Caceres et al 2014     & 0.014276 & 0.03705  & 0.00007518823  & 0.00026045967 & 0.00012155812  & 0.004842 & 2455315.794883 & I              & Literature & 0       & 0       & -306  & -2.02907502651215  & 0.20389686395285   \\
2.22060231209693  & 2.40455477107574  & Harpson et al 2013     & 0.015432 & 0.038471 & 0.00040131617  & 0.0013902     & 0.00064413727  & 0.005629 & 2455315.794995 & sdssi          & Literature & 0       & 0       & -306  & -1.86779513955116  & 0.365176750913834  \\
2.01714726205399  & 1.27922416375966  & de Mooj et al 2012     & 0.014329 & 0.036998 & 0.00029215788  & 0.0010120646  & 0.00044607319  & 0.005613 & 2455342.661299 & r              & Literature & 0       & 0       & -289  & -2.70002894103527  & -0.456663993012198 \\
0.696405871852179 & 7.38398165051281  & de Mooj et al 2012     & 0.014834 & 0.039589 & 0.00023377536  & 0.00080982162 & 0.00031854505  & 0.007352 & 2455342.661793 & K              & Literature & 0       & 0       & -289  & -1.9886688888073   & 0.254696059215776  \\
1.3826047750966   & 1.43953200388388  & Kundurthy et al 2011   & 0.015961 & 0.037895 & 0.00013967494  & 0.00048384818 & 0.00022323204  & 0.006317 & 2455353.72466  & sdssr          & Literature & 0       & 0       & -282  & -1.94077469408512  & 0.306869748226572  \\
1.87808148031968  & 1.36481470615678  & de Mooj et al 2012     & 0.013867 & 0.03667  & 0.00011052198  & 0.00038285939 & 0.00018239978  & 0.004372 & 2455380.590935 & sdssz          & Literature & 0       & 0       & -265  & -2.81476847827435  & -0.556730978404576 \\
2.10949846638068  & 3.46390953487826  & de Mooj et al 2012     & 0.015661 & 0.038892 & 0.00027134763  & 0.00093997577 & 0.00043878893  & 0.005642 & 2455380.591226 & sdssg          & Literature & 0       & 0       & -265  & -2.39572830498219  & -0.137690805112416 \\
2.05667039841214  & 1.47624912995891  & de Mooj et al 2012     & 0.014576 & 0.037102 & 0.00011544754  & 0.00039992201 & 0.00018461769  & 0.00521  & 2455380.591233 & sdssi          & Literature & 0       & 0       & -265  & -2.38564789295197  & -0.127610393082196 \\
2.08973252119274  & 1.73520024460113  & de Mooj et al 2012     & 0.015226 & 0.03786  & 0.00013581942  & 0.00047049229 & 0.00021688722  & 0.005644 & 2455380.591281 & sdssr          & Literature & 0       & 0       & -265  & -2.31652803719044  & -0.058490537320668 \\
2.05904592219586  & 0.534642565539024 & Harpson et al 2013     & 0.015424 & 0.0368   & 0.00021331088  & 0.00073893055 & 0.00035358795  & 0.005539 & 2455383.751738 & I              & Literature & 0       & 0       & -263  & -2.82432921230793  & -0.565068999784269 \\
1.51246622946044  & 1.1554119947922   & Kundurthy et al 2011   & 0.01502  & 0.037032 & 0.000086850798 & 0.00030085999 & 0.00014450977  & 0.005104 & 2455383.752222 & sdssr          & Literature & 0       & 0       & -263  & -2.12736926972866  & 0.131890942795     \\
3.84172941469478  & 0.751229700687564 & Harpson et al 2013     & 0.015354 & 0.037664 & 0.0001518402   & 0.00052598987 & 0.00024378328  & 0.005727 & 2455391.654123 & I              & Literature & 1000    & 1       &       &                    &                    \\
1.97141158071663  & 0.399557185466158 & de Mooj et al 2012     & 0.014209 & 0.036406 & 0.000052885215 & 0.00018319976 & 0.000088523117 & 0.00452  & 2455407.458131 & I              & Literature & 0       & 0       & -248  & -2.36252188682556  & -0.094091329397716 \\
2.27061228061283  & 0.727728845059969 & Harpson et al 2013     & 0.013467 & 0.036102 & 0.00010720389  & 0.00037136516 & 0.0001832348   & 0.00385  & 2455410.618996 & I              & Literature & 0       & 0       & -246  & -2.28280328214169  & -0.013150012059949 \\
0.973337378964746 & 1.22441688249975  & Harpson et al 2013     & 0.014379 & 0.036766 & 0.00020088622  & 0.00069589029 & 0.00034969884  & 0.004189 & 2455415.360198 & sdssi          & Literature & 0       & 0       & -243  & -2.30074591934681  & -0.029258580284235 \\
0.878990098512957 & 9.07091893117274  & de Mooj et al 2012     & 0.015473 & 0.040469 & 0.00024503472  & 0.00084882517 & 0.00029323522  & 0.011569 & 2455426.424175 & K              & Literature & 1       & 1       & -236  & -0.654451623558998 & 1.6213152097922    \\
2.35996325630005  & 0.955226712305003 & Harpson et al 2013     & 0.014877 & 0.038469 & 0.00015059533  & 0.00052167754 & 0.00023111066  & 0.005743 & 2455429.583803 & I              & Literature & 0       & 0       & -234  & -2.35601283609867  & -0.079023290093586 \\
0.822571989863981 & 2.11830531885857  & Gillon et al. 2014     & 0.013214 & 0.038119 & 0.00015918695  & 0.00055143977 & 0.00023098113  & 0.005009 & 2455631.875896 & I+z            & Literature & 0       & 0       & -106  & -1.95851549506187  & 0.396727660792257  \\
1.19576695807761  & 2.34075571461259  & Gillon et al. 2014     & 0.015162 & 0.039121 & 0.0001401945   & 0.000485648   & 0.00021465065  & 0.005905 & 2455650.840331 & I+z            & Literature & 0       & 0       & -94   & -2.56740443408489  & -0.204825002307413 \\
0.889948910909974 & 2.33257626757927  & Gillon et al. 2014     & 0.013247 & 0.036393 & 0.00013005906  & 0.0004505378  & 0.00021279607  & 0.004054 & 2455669.805516 & I+z            & Literature & 0       & 0       & -82   & -2.09629453718662  & 0.273621170514205  \\
0.924076675512996 & 3.6375712201735   & Gillon et al. 2014     & 0.013795 & 0.037471 & 0.00020290215  & 0.00070287368 & 0.00031045503  & 0.005008 & 2455677.707187 & I+z            & Literature & 0       & 0       & -77   & -2.60475888848305  & -0.231786399147492 \\
1.22130279062034  & 2.80072205362528  & Gillon et al. 2014     & 0.014631 & 0.037079 & 0.00013958613  & 0.00048354052 & 0.00023842143  & 0.004505 & 2455680.868054 & I+z            & Literature & 0       & 0       & -75   & -2.52215959131718  & -0.14796438932773  \\
0.692314390641076 & 2.60511774796333  & Gillon et al. 2014     & 0.013419 & 0.037949 & 0.00018892815  & 0.00065446631 & 0.00027064594  & 0.00539  & 2455696.672245 & I+z            & Literature & 0       & 0       & -65   & -2.31652803719044  & 0.063780728068466  \\
1.01307130172199  & 2.85798065969031  & Gillon et al. 2014     & 0.013381 & 0.035668 & 0.00014302326  & 0.00049544709 & 0.00024871441  & 0.003705 & 2455699.832824 & I+z            & Literature & 0       & 0       & -63   & -2.64864921569824  & -0.267117737785447 \\
2.36197219818054  & 1.01691316576869  & Harpson et al 2013     & 0.015037 & 0.037554 & 0.00016333706  & 0.00056581619 & 0.00025973646  & 0.005596 & 2455715.637167 & R              & Literature & 0       & 0       & -53   & -2.22413800656795  & 0.163507034614297  \\
0.911384333866416 & 13.8641033490619  & Caceres et al 2014     & 0.013403 & 0.038567 & 0.00030249204  & 0.0010478632  & 0.00043304365  & 0.005264 & 2455783.594092 & 2.14um\_narrow & Literature & 100     & 1       &       &                    &                    \\
1.12302264232073  & 1.3331571169042   & Narita et al. 2013     & 0.014135 & 0.036622 & 0.00011724402  & 0.00040614518 & 0.00019946084  & 0.004268 & 2455788.335846 & J              & Literature & 0       & 0       & -7    & -2.14165471494198  & 0.274112717279774  \\
1.14779098301767  & 1.33915400192187  & Narita et al. 2013     & 0.014311 & 0.03661  & 0.00011778865  & 0.00040803185 & 0.00019996286  & 0.004414 & 2455788.335853 & H              & Literature & 0       & 0       & -7    & -2.13157497346401  & 0.284192458757741  \\
1.57806255042784  & 1.63289139587076  & Narita et al. 2013     & 0.013572 & 0.036857 & 0.00016212056  & 0.0005616021  & 0.00024613037  & 0.005021 & 2455788.336052 & K              & Literature & 0       & 0       & -7    & -1.84501446783543  & 0.570752964386327  \\
1.82989108161083  & 2.399880960675    & Harpson et al 2013     & 0.015027 & 0.03773  & 0.00038349532  & 0.0013284668  & 0.00057729311  & 0.006351 & 2455799.398034 & sdssz          & Literature & 0       & 0       & 0     & -3.07152025401592  & -0.65147332750555  \\
1.8052699833899   & 2.01483997338865  & Harpson et al 2013     & 0.014445 & 0.037417 & 0.0003181385   & 0.0011020641  & 0.00049451349  & 0.005393 & 2455799.398567 & sdssg          & Literature & 0       & 0       & 0     & -2.30400010943413  & 0.116046817076245  \\
1.91651897753165  & 1.45359009666968  & Harpson et al 2013     & 0.013746 & 0.036933 & 0.00019493565  & 0.00067527691 & 0.00033452789  & 0.003883 & 2455799.398769 & sdssr          & Literature & 0       & 0       & 0     & -2.01311990618706  & 0.406927020323315  \\
2.00010592202447  & 0.561661977317683 & Nascimbeni et al 2015  & 0.013604 & 0.036075 & 0.00004483982  & 0.00015532969 & 0.000077179109 & 0.00388  & 2456015.913815 & R              & Literature & 0       & 0       & 137   & -2.60976925492287  & -0.105966511620937 \\
1.7167103167892   & 1.38303532912177  & Nascimbeni et al 2015  & 0.014552 & 0.036797 & 0.00014755908  & 0.00051115965 & 0.00022948456  & 0.005616 & 2456015.914246 & B              & Literature & 0       & 0       & 137   & -1.9891295582056   & 0.514673185096325  \\
1.66744508440514  & 0.742537137178145 & Nascimbeni et al 2015  & 0.018167 & 0.039255 & 0.0001444947   & 0.00050054431 & 0.0002190867   & 0.009722 & 2456064.906315 & B              & Literature & 0       & 0       & 168   & -2.68093429505825  & -0.158179505621005 \\
1.62895253396986  & 0.581737296919655 & Nascimbeni et al 2015  & 0.013822 & 0.036978 & 0.000050122408 & 0.00017362911 & 0.000079720591 & 0.004676 & 2456064.906354 & R              & Literature & 0       & 0       & 168   & -2.62477487325668  & -0.102020083819438 \\
1.00859192594658  & 2.13477762122242  & 20220512 T80  & 0.012126 & 0.036145 & 0.00022431826  & 0.00077706125 & 0.00035440397  & 0.003622 & 2455701.413994 & iprime         & our        & 0       & 0       & -62   & -1.54679037630558  & 0.835352457934161  \\
1.11617964439914  & 3.51947548154011  & 20200611 T100 & 0.015828 & 0.039198 & 0.00037575773  & 0.0013        & 0.00061163715  & 0.005634 & 2459012.360728 & R              & our        & 0       & 0       & 2033  & -3.51080641150475  & 0.15212792768618   \\
2.08267303466749  & 3.00169245738781  & 20200703 T100 & 0.015081 & 0.037681 & 0.00045417007  & 0.0015732913  & 0.00072708222  & 0.005518 & 2459034.486291 & I              & our        & 0       & 0       & 2047  & -3.66125755012035  & 0.010235777647812  \\
1.97247862626836  & 2.33689387413991  & 20210423 T100 & 0.014749 & 0.037149 & 0.00026050322  & 0.00090240961 & 0.00041434169  & 0.00542  & 2459328.441673 & I              & our        & 0       & 0       & 2233  & -3.53816628456116  & 0.247039320018903  \\
2.00285455916959  & 2.60556008694825  & 20210719 T100 & 0.021252 & 0.050181 & 0.00045961826  & 0.0015921643  & 0.00053107098  & 0.02509  & 2459415.362554 & I              & our        & 10      & 1       &       &                    &                   

\end{tabular}}
\end{table}
\end{landscape}
\bsp	
\label{lastpage}
\end{document}